\documentclass[12pt]{article}
\usepackage{amsmath,amssymb,amsthm,color}
\usepackage{graphicx}
\usepackage{cite}
\usepackage{epsfig}
\usepackage{lineno}
\renewcommand{\baselinestretch}{1.66}

\begin{document}
\title{Optical properties of graphene in magnetic and electric fields \\}
\author{
\small Chiun-Yan Lin$^{a}$, Thi-Nga Do $^{a,*}$, Yao-Kung Huang $^{a}$, Ming-Fa Lin$^{a,*}$ $$\\
\small  $^a$Department of Physics, National Cheng Kung University, Tainan 701, Taiwan \\
 }
\renewcommand{\baselinestretch}{1.66}
\maketitle

\renewcommand{\baselinestretch}{1.66}

\begin{abstract}
Optical properties of graphene are explored by using the generalized tight-binding
model. The main features of spectral structures, the form, frequency, number and intensity,
are greatly enriched by the complex relationship among the interlayer atomic interactions,
the magnetic quantization and the Coulomb potential energy. Absorption spectra have shoulders, asymmetric peaks and logarithmic peaks, coming from the band-edge states of
parabolic dispersions, the constant-energy loops and the saddle points, respectively. The
initial forbidden excitation region is only revealed in even-layer AA stacking systems.
Optical gaps and special structures can be generated by an electric field. The delta-function-like
structures in magneto-optical spectra, which present the single, twin and double peaks,
are associated with the symmetric, asymmetric and splitting Landau-level energy spectra,
respectively. The single peaks due to the non-tilted Dirac cones exhibit the nearly uniform
intensity. The AAB stacking possesses more absorption structures, compared to the
other stackings. The diverse magneto-optical selection rules are mainly determined by the
well-behaved, perturbed and undefined Landau modes. The frequent anti-crossings in the
magnetic- and electric-field-dependent energy spectra lead to the increase of absorption
peaks and the reduced intensities. Part of theoretical calculations are consistent with the
experimental measurements, and the others need further detailed examinations.
\end{abstract}

\par\noindent  * Corresponding author.
{~ Tel:~ +886-6-275-7575.}\\~{{\it E-mail addresses}: l28027034@mail.ncku.edu.tw (T.N. Do), mflin@mail.ncku.edu.tw (M.F. Lin)}


\noindent{Contents}

\pagenumbering{Roman}

\addcontentsline{toc}{chapter}{摘要}

\vskip0.6 truecm

\noindent {1. Introduction} \hfill ............................................................................................................~~~01

\vskip0.5 truecm

\noindent

\noindent {2. The generalized tight-binding model}.........................................................................~~~11

\vskip0.5 truecm

\noindent {3. AA-stacked graphene }
\hfill..............................................................................................~~~15

\noindent {3.1 Electronic and optical properties without external fields }
\hfill........................................~~~16

\noindent {3.1.1 Monolayer}
\hfill........................................................................................................~~~16

\noindent {3.1.2 Multilayer}
\hfill............................................................................................................~~~16

\noindent {3.2 Magneto-electronic and magneto-optical properties }
\hfill..............................................~~~18

\noindent {3.2.1 Monolayer}
\hfill...........................................................................................................~~~18

\noindent {3.2.2 Multilayer}
\hfill............................................................................................................~~~20

\noindent {3.3 Electric field enriched optical properties }
\hfill................................................................~~~24

\noindent {4. AB-stacked graphene }
\hfill..............................................................................................~~~25

\noindent {4.1 Electronic and optical properties without external fields }
\hfill.......................................~~~27

\noindent {4.2 Magneto-electronic and magneto-optical properties }
\hfill...............................................~~~29

\noindent {4.3 Electric field enriched optical properties }
\hfill.................................................................~~~31

\noindent {5. ABC-stacked graphene }
\hfill............................................................................................~~~34

\noindent {5.1 Electronic and optical properties without external fields }
\hfill........................................~~~35

\noindent {5.2 Magneto-electronic and magneto-optical properties }
\hfill................................................~~~37

\noindent {5.3 Electric field enriched optical properties }
\hfill.................................................................~~~43

\noindent {6. AAB-stacked graphene }
\hfill...........................................................................................~~~45

\noindent {6.1 Electronic and optical properties without external fields }
\hfill........................................~~~46

\noindent {6.2 Magneto-electronic and magneto-optical properties }
\hfill............................................~~~48

\noindent {6.3 Electric field enriched optical properties }
\hfill.................................................................~~~55

\noindent {6.4 Comparisons among the different trilayer stackings }
\hfill................................................~~~56

\noindent {7. Sliding bilayer graphene }
\hfill..........................................................................................~~~58

\noindent {7.1 Electronic and optical properties without external fields }
\hfill.......................................~~~60

\noindent {7.2 Magneto-electronic and magneto-optical properties }
\hfill.............................................~~~63

\noindent {7.3 Electric field enriched optical properties }
\hfill...............................................................~~~66

\noindent {8. Concluding remarks}
\hfill..................................................................................................~~~68

\noindent {Acknowledgments}
\hfill.........................................................................................................~~~75

\noindent {References (Ref. 1-Ref. 338)}
\hfill.....................................................................................~~~76

\noindent {Figures (fig. 1-fig. 36)}
\hfill.................................................................................................~~~110

\pagebreak
\renewcommand{\baselinestretch}{2}
\newpage

\vskip 0.6 truecm



\setcounter{page}{1}
\pagenumbering{arabic}

\section{ Introduction}
\vskip 0.3 truecm

Graphene is a 2D material made up of hexagonal carbon lattices\cite{Science306;666,Proc102;10451}.
Since mono- and few-layer graphene sheets were first fabricated in 2004\cite{Science306;666,Proc102;10451}, low-dimensional graphene-related systems have been a great interest to experimental and theoretical studies.
The stacking orders of graphene sheets include the essential sequences of AA\cite{JCP129;234709,PRL102;015501}, AB\cite{PRB77;155426,SurfSci610;53,JAP109;093523,PRB91;155428,PRL109;126801}, ABC\cite{PRB77;155426,SurfSci610;53,JAP109;093523,PRB81;161410,ACSNano6;5680} stackings.
While the AA stacking configuration has only been artificially made from intercalated graphite compounds, the AB and ABC configurations are the common orders in natural graphite, respectively, with their estimated volume fractions: 80 $\%$ and 14 $\%$\cite{PRSLSA106;749,PRSLSA181;101}.
The rest parts $\sim 6\%$ consist of haphazardly stacked graphene sheets, called turbostratic configuration\cite{CARBON32;289,PRB46;4540,PRB81;165105}.
Few-layer graphene desired with a specific stacking configuration can be exfoliated from highly orientated pyrolytic graphite\cite{Science306;666,Proc102;10451}, and chemically and electrochemically reduced from graphene oxide\cite{JPCC118;15455,Carbon45;1558,RSCADV2;1168,ACSNano3;2653,AdvFunMat19;2782,
AdvFunMat20;3050,SciRep6;21282,ACSnano10;2600}.
Nevertheless, chemical vapor deposition method has the advantage of producing large-scale size of high-quality graphene sheets.
Recently, large area of graphene with high mobility and highly symmetric configurations, e.g., AA, AB and ABC, have been found in CVD-grown samples\cite{NanoLett9;30,ACSNano6;8241,JPCC115;11666,Nature457;706,AdvMat26;5488,
SurSci611;67,NatCom4;2096,NanoLett11;3612,NatTech5;574,SCIENCE324;1312,ACSAMI8;4878,
SciRep6;21152,SciRep6;21676}.
The improved quality is adequate for research experiments and industry applications\cite{NatCom3;906,NatTech9;780,NatPho4;611,CARBON100;608,CARBON99;579,
SABC226;342,NANOTEC27;105705,OPTLETT4;816,NanoLett14;6097,Nanoscale5;262}.
In addition, the AAB stacking and intermediate bilayer configurations, with relative layer-layer shifts or twists, are also found and becoming an interesting subjects\cite{Arx;1509.02253,Carbon94;619,PRB84;161406,SciRep4;7509,SurSci601;498,PRB75;235449,
APL102;163111,PRB79;125411,APL107;263101}.
The synthesized graphene generally comprises different kinds of domains due to the tiny difference among the total energies of different configurations.
Infrared spectroscopy, transmission electron microscopy (TEM)\cite{JCP129;234709,PRL102;015501,JAP109;093523,SCIENCE324;1312}, scanning tunneling microscopy (STM) and spectroscopy (STS)\cite{PRL109;126801,
PRB91;155428,PRB77;155426,SurfSci610;53,ACSNano6;5680,NatCom4;2096,NanoLett11;3612,NatTech5;574
}, angle resolved photoemission spectroscopy (ARPES)\cite{PRL98;206802,PRB88;075406,
PRB88;155439,PRL102;056808,Nature3;36,PRB77;155303,PRL103;226803}, and Raman spectroscopy\cite{PRB81;161410,Carbon99;118,Carbon99;212}, are used to visualize stacking domains with a namometric resolution for cases of tri-, tetra-, and pentalayer graphene with typical or intermediate configurations.

Monolayer graphene belongs to a gapless system which possesses extremely high carrier mobility of up to 15,000 cm$^{2}$V$^{-1}$s$^{-1}$\cite{Science306;666}.
P. R. Wallace has discovered that the low-energy spectrum around the corners K and K$'$ in the Brillouin zone is described by isotropically conical dispersions in the framework of the nearest-neighbor tight-binding model\cite{PR71;622}.
The conical band structure and the vertex points at K and K$'$ are known as a Dirac cone and Dirac points, respectively.
The linear-dispersion quasi-particles in graphene behave as relativistic Dirac fermions with an effective speed of light $\simeq c/300$ ($c$ is the speed of light).
Few-layer graphenes with specific stacking configurations, e.g., AA\cite{CPL550;104,JAP114;233701,PRB83;165429}, AB\cite{PRB78;205425,Rep76;056503,PRB75;193402,PRB81;115315,RevModPhys81;109,
JPCM18;5849,NatMat12;887,Nature6;109,NanoLett3;5153,PRB76;201402,PNAS107;14999,
PRL106;126802,PRB92;075442}, ABC\cite{RevModPhys81;109,PRL97;036803,PRB82;035409,PRB84;165404,PRB83;220503,PRB73;245426,
PRB87;155116}, and AAB\cite{Carbon94;619}, are predicted to display unique electronic energy dispersions.
This new material holds a great promise for the development of next-generation electronic and optoelectronic nano-devices, because the electronic and optical properties can be flexibly tuned by the application of external fields
\cite{Arx;1509.02253,Carbon94;619,SciRep4;7509,PRB84;161406,PRL98;206802,
PRB88;075406,PRB88;155439,PRL102;056808,Nature3;36,PRB77;155303,PRL103;226803} and the changes of geometric structures\cite{PRB92;075442,PRL99;216802,Nature459;820,PRB87;165102,NatPhys7;944,Science313;951,Nature7;948,
SCIENCE320;206,NatNanotechnol4;383,Carbon99;118,Carbon99;212,PR71;622,CPL550;104,JAP114;233701,PRB83;165429,
JPCM18;5849,PRB78;205425,Rep76;056503,PRB75;193402,PRL106;126802,NatMat12;887,
Nature6;109,NanoLett3;5153,PRB76;201402,PRB81;115315,PNAS107;14999,
RevModPhys81;109,PRL97;036803,PRB82;035409,PRB84;165404,PRB73;245426,PRB87;155116,
PRB83;220503,PRB73;144427,JPSJ76;024701,PRB74;161403,PRB75;155115,APL98;222101,APL98;263107,PRB79;035421} and dopants\cite{ACSNano5;7517,SciRep6;21311,JMC21;3335,NanoLett9;1752,SSC151;13,STAM16;014803,CARBON98;358,CARBON100;337}.
Stacking configurations are experimentally and theoretically verified to have a great significance on the magneto-electronic and optical properties\cite{SciRep4;7509,PCCP17;26008,Carbon42;2975,PhyE40;1722,
JAP110;013725,PRL96;086805,PRB77;085426,PRB77;155416,PRB83;165443,PRB84;205448,
PRB84;125455,Jnn11;4938,Ann326;721,PRB80;165409,RSCAdv4;56552,PRB90;205434,
JPCC119;10623,Arx;1509.02253,Carbon94;619,PRB84;161406,PRB83;195405,JVSTB28;386,
OptEx22;7473,APL103;041907,JAP108;043509,PRR89;045419,NJP15;015010,SynMet162;800,
PRB83;245418,CARBON69;151,PRB73;245411,PRL98;157402,ACSNano4;1465,PRB77;115313,
Nature4;532,PRB78;235408,NANOLETT14;4548,PRB79;115441,PRL14;126804,APL117;112803,
PRL102;037403,PRL104;176404,PRL111;077402,PRL101;267601,PRL100;087401,
PRB83;125302,PRL102;166401,PRB85;245410,PRL100;087403,SCIREP3;3143,JPSJ17;808,
PTPS176;227,SSC142;123,PRB81;115432,PRB79;125443,ACSNano9;6765,PRL104;067404,
PRL107;036807,ACSNano8;10766,PRB91;235403,APE7;115102,EPL108;27011,PRL98;197403,
PCCP17;15921,RSCAdv5;80410,PRB87;165102,APL97;101905,PRB84;153405}, shedding some light on other intriguing structure-specific physical properties, e.g., the Coulomb excitations and the quantum Hall transport properties\cite{PRB74;085406,PLA352;446,PRB86;125434,ACSnano5;1026,NanoResLett7;134,
PRB89;165407,RSCAdv5;51912,PRB84;035439,PRB85;235444,PRB80;085408,PRB84;115420,
PRB75;115314,JAP109;113721,NatMat6;183,Nature438;197,PRL100;016602,NatTech3;206,SciRep4;4558,
NatTech5;487,PRL95;146801,Nature438;201,SCIENCE315;1379,PRB82;165404,NatPhys2;177,
NatMater7;151,NatPhys7;621,PRL107;126806,PRB88;121302,NatPhys7;953,NanoLett13;1627,
PRX2;011004,Science342;720}.
The essential properties under the previous investigations are rather different from those in other layered materials, conventional metals, semiconductors, and topological insulators\cite{PRL113;266801,NJP14;033003,PRL109;055502,PRB92;174514,Science349;524,PRB89;155316,
APLHo;acc,PRL110;197402,PRB88;085434,PRB85;195440,PRB81;125120,JAP76;1363,RMP54;437}; they are dominated by the interplay between the external fields and the specific geometric symmetries.
In this paper, we thoroughly review the electronic and optical properties of multilayer graphene to explore the response under external fields with respect to the specific AA, AB, ABC, AAB and sliding configurations.

In AA-stacked graphene, all the graphene sheets are identically stacked along the $\widehat{z}$ direction\cite{APL103;041907,PCCP17;26008}.
The low-energy band structure consists of several monolayer-like massless Dirac cones\cite{APL103;041907,CPL550;104,JAP114;233701,PRB83;165429}.
They are exactly located at the two valleys K  and K$'$, while the energies of the Dirac points shift away from the Fermi energy due to the vertically projected geometry.
The absorption spectrum is, in general, a superposition of monolayer-like spectra because only the vertical excitations of intra-Dirac cones are permitted, regardless of external electric and magnetic fields\cite{APL103;041907,APL97;101905}.
However, the stacking effect gives rise to an optical gap for the even-layer configurations, a forbidden optical transition zone resulting from free carriers in Dirac cones.
The middle-frequency spectrum shows symmetric peaks as a result of the saddle-point excitation channels around the M point.
Such peaks are also observed for other stacked graphenes, as well as being a critical characteristic of sp$^{2}$-bonding carbon systems [Refs in page 27].
In a uniform perpendicular magnetic field $B=B_{0}\widehat{z}$, the Landau level (LL) spectrum behaves as several groups of monolayer-like ones\cite{APL103;041907,PCCP17;26008,APL97;101905}.
Responsible for each Dirac cone, the quantized LLs of each group are characterized by a set of single quantum modes and their energies can be described by a square-root dependence on $B_{0}$.
Therefore, the magneto-absorption spectrum consists of several monolayer-like spectra with delta-function-like absorption peaks following the optical selection rule of $\Delta n=\pm1$\cite{APL97;101905,APL103;041907}.
A forbidden transition region and a discontinuous threshold frequency would be revealed for all the bilayer-like intragroup LL transitions, as the Fermi level ($E_{F}=0$) deviates from the Dirac point.
Moreover, the preservation of the optical and electronic properties of the massless Dirac fermions indicates the unchanged spatial symmetry of the chemical environments under external fields.

AB-stacked graphene is a fascinating material, which has attracted intense attention due to the nature of massless- and/or massive-Dirac fermions, depending on whether the numbers of their stacking layers are even or odd\cite{PRB78;205425,Rep76;056503,PRB75;193402,PRB81;115315,PRL106;126802,NatMat12;887,
Nature6;109,NanoLett3;5153,PRB76;201402,RevModPhys81;109,PNAS107;14999,PRB92;075442}.
Of special interest is the electronic structure that can be essentially characterized by a superposition of a monolayer-like and bilayer-like structures.
Optical excitation channels accessible for the monolayer-like and bilayer-like subbands construct the optical spectrum\cite{JAP108;043509,PRR89;045419,NJP15;015010,
APE7;115102,JPCM18;5849,PRB73;144427}.
Under an electric field, the bilayer graphene shows a sharp peak due to new transition channels within the energy gap (up to 100 meV)\cite{PRB74;161403,PRB75;155115,
PRL99;216802,Nature459;820}, and the cases of more than three layers display enhanced semimetal characteristics attributed to the significant overlap of the low-energy subbands\cite{JPSJ76;024701,Nature459;820,NatPhys7;944,NatNanotechnol4;383}. In particular, as a result of the nearly isotropic energy dispersions around the K point, quasi-1D asymmetric peaks are revealed for the vertical excitations between two constant-energy loops.
The electric-field-induced characteristic peaks are also observed in other stacking systems.
On the other hand, a magnetic field gives rise to interesting phenomena of the the Landau quantization in electronic and optical properties for the odd-layer cases of symmetry-broken graphene and for the even-layer cases of inversion-symmetry graphene\cite{PCCP17;26008,PRB77;085426,PRL96;086805,PRB77;155416,PRB83;165443,PRB84;205448,PRB84;125455}.
Massless and massive Dirac quasi-particles, respectively, reveal single- and twin-peak structures in the monolayer-like and bilayer-like spectra\cite{ACSNano4;1465,Arx;1509.02253}.
The corresponding ratio of the peak intensities is estimated to be of about 2:1.
In the presence of an electric field, the magneto-excitation channels around K and K$'$ are separated from each other due to the LL splitting so that the channel numbers become double than those under zero electric field\cite{PCCP17;26008}.
In addition, very few LLs with a special relationship of quantum modes avoid to cross each other during the variation of the magnetic or electric field; they are characterized by a main quantum mode and the side modes under the hybridization of LLs\cite{PCCP17;26008,PRB83;165443,JPSJ17;808}.
Extra optical selection rules are induced for such LLs with multi quantum modes in addition to the ordinary $\Delta n=\pm1$.
At certain regions of the absorption spectra, the effects of the geometry breaking and the anticrossings of inter-group LLs would trigger some special peaks without simple relationships among the intensity, frequency and field strength.

Few-layer ABC-stacked graphene is a semimetal\cite{RevModPhys81;109,PRL97;036803,
PRB82;035409,PRB84;165404,PRB73;245426,PRB87;155116,PRB83;220503}.
The electronic structure is characterized by one pair of partially flat subbands near $E_{F}=0$, and pairs of sombrero-shaped subbands near the energy of the vertical atomic interactions between nearest-neighboring layers\cite{RevModPhys81;109,PRL97;036803,
PRB82;035409,PRB84;165404,PRB73;245426,PRB87;155116,PRB83;220503}.
Contributed by the surface-localized states\cite{PRB73;245426,PRB87;155116,PRB83;220503}, the partially flat subbands give rise to a prominent peak in DOS and plays an important role in the low-frequency optical spectrum\cite{NatPhys7;944,Nature7;948}.
An electric field can separate the two partially flat bands and trigger a new kind of optical channels responsible for a gap transition\cite{NatPhys7;944,Nature7;948}.
Under a magnetic field, intergroup and intragroup excitation channels among well-behaved LLs comprise the main part of the absorption spectrum\cite{PRL104;176404,PCCP17;15921,RSCAdv5;80410,PRB90;205434}, in which the absorption peaks show the twin-peak structure because the LL energy spectrum is asymmetric about $E_{F}=0$\cite{PCCP17;15921,RSCAdv5;80410}.
Moreover, it is interesting to study the quantization effects of any two constant-energy loops in the sombrero-shaped structure, where the quantized energies of the LLs are proportional or inversely proportional to the magnetic field strength\cite{PCCP17;26008,RSCAdv4;56552,PRB90;205434}.
This leads to many anticrossings of intragroup LLs in the ABC stacking configuration.
By varying the electric field strength, one can suppress or enhance the degree of LL anticrossings and change the components of the main mode and side modes, that is, one can induce and modulate optical channels for the progressive LLs during the variation of the field strength.
Furthermore, the absorption spectrum is no longer dominated by the twin-peak structure but rather by the double-peak structure under a sufficiently large electric field, a transition being attributed to the spatial symmetry breaking of the LL subenvelope functions.

AAB-stacked graphene exhibits an extraordinary band structure as compared to AA, AB and ABC systems mainly due to the low symmetry of the stacking geometry\cite{Carbon94;619}.
In the case of trilayer system, AAB-stacked graphene is a narrow-gap semiconductor, which
presents a pair of subbands with strong oscillatory energy dispersions near $E_{F}=0$.
With the increasing energy, pairs of sombrero-shaped and parabolic bands substantially appear.
In the low-frequency region, rich optical excitation channels cause many absorption peaks, some of which show a quasi-1D asymmetric divergence due to the vertical excitations between two constant-energy loops.
The LLs coming from the two valleys K and K$'$ are obviously separated in the mirror symmetry broken system\cite{Carbon94;619}.
The results present certain important differences for the state degeneracy, spacing, quantum number and spatial distribution of the wavefunctions, as compared with the highly symmetric configurations.
Furthermore, the low-symmetric geometry of the AAB configuration implies the easiness of triggering complex patterns of LL anticrossings and the richness of magneto-absorption spectra.
At low energies, the oscillatory subbands can induce triple-degenerate Landau states under the condition of three constant-energy contours.
In the $B_{0}-$dependent LL spectrum, there are many prominent multi-crossings and -anticrossings, i.e., existences of both well-behaved and perturbed subenvelope functions, and sometimes the abnormal ones resulting from severe LL hybridizations.
Accordingly, the intergroup and intragroup LL anticrossings are both shown for various regions, and they even coexist in a narrow region.
Optical characteristics resulting from due to either intragroup or intergroup inter-LL excitations are remarkably different for the K- and K$'$-valley splitting states.
The main reason is based on the phase and the amplitude of LL wavefunctions.


An intermediate bilayer configuration is formed from AA to AB, and then to AA$^{'}$, as the two graphene sheets relatively shift to each other in the armchair direction\cite{SciRep4;7509,JPCC119;10623}.
The two Dirac cones transform into parabolic bands from AA to AB configurations, and for a further layer-layer shift, two Dirac cones tilting to each other by a small angle are restored in the AA$^{'}$ configuration.
It should be noticed that the two Dirac points are located at different wave vectors and energies once the stacking configuration deviates from the AA configuration.
Furthermore, an eye-shaped stateless region is formed near $E_{F}=0$ as a result of a strong hybridization of Dirac cones.
The dramatic transformation between the Dirac-cone structure and the parabolic band structure leads to drastic changes of the optical response under external electric and magnetic fields.
The layer-layer shift has a significant impact on the LLs in the sliding bilayer systems.
In additional to the well-behaved and perturbed LLs stemming from the massless and massive Dirac Fermions in the highly symmetric AA and AB configurations, some undefined LLs appear to be promising in the case of a significant mixture of LLs as the geometry deviates from the typical configurations.
However, there is absence of any regular optical selection rule for such undefined LLs, because their wave functions display extremely irregular oscillations without a dominating quantum mode.
On the other hand, when AA$^{'}$ configuration is built, the LLs transform back to be well-behaved, while additional optical channels and selection rules are accessible for intragroup LL excitations within the tilted Dirac cones.
In cases of different bilayer configurations, it is worthwhile to investigate the considerate shift-induced absorption peaks, which appear in a wide range of the anti-crossing patterns in the field-dependent energy spectrum.


Electronic properties are studied by the use of STM, STS and ARPES\cite{PRL109;126801,
PRB91;155428,PRB77;155426,SurfSci610;53,ACSNano6;5680,NatCom4;2096,NanoLett11;3612,NatTech5;574,
PRL98;206802,PRB88;075406,PRB88;155439,PRL102;056808,Nature3;36,PRB77;155303,PRL103;226803,
PRB81;161410,Carbon99;118,Carbon99;212}.
Isolated Dirac cones are clearly shown in the monolayer graphene\cite{PRB88;075406,PRL102;056808,PRB77;155303,PRL103;226803}.
The stacking effects for the typical configurations, AA, AB and ABC, are also identified by the main characteristics of the band structures.
The former two, respectively, show multiple Dirac-cone dispersions and a superposition of a monolayer-like and bilayer-like dispersions\cite{PRB77;155426}, while the latter is not analogous to AA and AB configurations\cite{PRL98;206802,PRB88;155439}.
On the other hand, the deformation of Dirac cones for the AAB- and sliding bilayer configurations could be verified by using ARPES.
These aforementioned electronic characteristics have also been confirmed in absorption\cite{PRL98;157402,Nature4;532,PRB78;235408,PRB79;115441,PRL102;037403,PRL104;176404,
PRL98;197403}, transmission\cite{PRB89;085426,PRL100;087401}, refection\cite{ACSNano9;6765}, Raman scattering and Rayleigh scattering spectroscopies\cite{PRL107;036807,Carbon99;118,Carbon99;212,PRL97;187401,
NanoLett12;1749,NanoLett11;164,Carbon46;272,ACSNano8;10766,NANOLETT14;4548,
EPL108;27011,NANOLETT13;1541,PRB78;113407,PRB76;201401,JACS133;10394}.
The evidence of Dirac cones has been observed for monolayer graphene in terms of the zero threshold frequency and the linear dependence of the intensity on the frequency.
The low-frequency spectral features are also verified for the AB- and ABC-stacked samples on SiO$_{2}$/Si substrates.
Furthermore, the results provide spectroscopic evidence of a tunable gap in biased AB-stacked bilayer and ABC-stacked trilayer graphenes through the observation of
new gap-induced absorption peaks, with a blue or red shift at different gate voltages\cite{Nature459;820,NatPhys7;944}.
The gap size is compatible to the theoretical calculations based on the tight-binding model and Ab initio theory\cite{PRB73;144427,JPSJ76;024701,PRB74;161403,PRB75;155115,APL98;222101,APL98;263107,PRB79;035421}.
Transport and STS measurements also give the modified electronic properties under the electric field, including the enhanced semi-metal properties of AB-stacked trilayer graphene\cite{PRB87;165102,Nature7;948,SCIENCE320;206,NatNanotechnol4;383}.
The E-field induced excitations rely on the geometric structure.
That is to say, verifications on the spectral structure are worthwhile for further research to identify inter-layer atomic interactions and geometric structures.

The quantized LL energies directly correspond to the prominent DOS peaks measured in STS, where the LLs near $E_{F}=0$ have been identified for monolayer, AB-stacked bilayer and trilayer and ABC-stacked trilayer graphenes\cite{PRL103;206804,PRB94;226403,SCIENCE324;924,NATURE467;185,
NATURE7;245,PRL109;116802,PRL109;176804,NATURE7;649,NATURE6;811,PRL109;116805,NATURE10;815}.
Quantum Hall transport measurements also observe the sequence of the LLs\cite{PRB82;165404,NatMat6;183,Nature438;197,Science342;720,NatTech3;206,
PRL100;016602,SciRep4;4558,NatTech5;487,PRL95;146801,Nature438;201,SCIENCE315;1379,
NatPhys2;177,NatMater7;151,NatPhys7;621,PRB88;121302,PRL107;126806,NatPhys7;953,
NanoLett13;1627}.
Furthermore, the low-energy optical transitions with specified selection rules are verified for the inter-group LLs near $E_{F}=0$ in infrared transmission, absorption and magneto-Roman spectra\cite{PRL98;157402,Nature4;532,PRB78;235408,PRB79;115441,PRL102;037403,
PRL104;176404,PRL98;197403,PRB89;085426,PRL100;087401,ACSNano9;6765,PRL107;036807,
Carbon99;118,Carbon99;212,PRL97;187401,NanoLett12;1749,NanoLett11;164,Carbon46;272,
ACSNano8;10766,NANOLETT14;4548,EPL108;27011,NANOLETT13;1541,PRB78;113407,
PRB76;201401,JACS133;10394}
Spectral features of the equivalent monolayer-like and bilayer-like Landau quantization are, respectively, identified with square-root and linear dependence of the energies on the magnetic field strength in infrared transmission of AB-stacked trilayer graphene, and in magneto-Raman spectroscopy of AB-stacked graphene up to 5 layers\cite{NANOLETT14;4548}.
For ABC-stacked trilayer graphene, while QHE transport measurements show the low-lying LLs\cite{PRB88;121302,PRL107;126806,NatPhys7;953}, the experimental evidence on the inter-LL excitations, distinct from the massless and massive Dirac-fermion excitations, is needed for further verification.
Crossings and anticrossings of LLs enrich the essential magneto-electronic and optical properties\cite{Arx;1509.02253,PRB83;165443,PRB79;125443,PRB90;205434,RSCAdv4;56552,SciRep4;7509,Carbon94;619,
PCCP17;15921,PCCP17;26008}.
The former has been confirmed by the shift of plateaus in QHE measurements due to the increase in the LL degeneracy\cite{NatPhys2;177,
NatMater7;151,NatPhys7;621,PRB88;121302,PRL107;126806,NatPhys7;953,NanoLett13;1627}.
About the latter, one can use optical spectroscopy to identify the non-monotonic changes in the structures, frequencies, intensities and numbers of the absorption peaks\cite{PCCP17;15921}.
This phenomenon is expected to be more significant in the low symmetric AAB-stacked and sliding bilayer graphenes\cite{Arx;1509.02253,SciRep4;7509}.
At a sufficiently low magnetic fields, the spectral features reflect band structures, such as the symmetry, energy dispersions, atomic interactions, as well as the broken mirror symmetry under the electric field.
These experimental results can be used to identify the stacking configurations and the tight-binding parameters.

In this work, the generalized tight-binding model is developed to investigate optical properties of graphene in magnetic and electric fields. The $2p_{z}$-dominated Hamiltonian is built from the subspace spanned by a lot of tight-binding functions in an enlarged unit cell. Its dimension is determined by the commensurate period of the geometric structure and external field. This model can be utilized in cases where many kinds of external fields are applied, e.g., uniform/modulated magnetic fields \cite{Carbon42;2975,PhyE40;1722};\cite{PRB83;195405}, modulated electric fields, and composite fields \cite{JVSTB28;386}. It is suitable for multi-layer graphene \cite{Ann326;721,Jnn11;4938,PRB77;085426,PCCP17;26008}
 and bulk graphites \cite{Carbon43;1424, EPJB60;161}, with arbitrary stacking configurations. Most important interlayer atomic interactions and external fields are simultaneously taken into account without the need for treating either of them as a perturbation term. The Hamiltonian matrix for determining the magneto-electric properties becomes too large for the experimental field strengths. By using an exact diagonalization method, the eigenvalues and eigenfunctions of the Hamiltonian matrix are efficiently solved. The developed model could be further used to understand the other essential material properties by combing the single- and many-particle theories, such as optical spectra and electronic excitations \cite{NanoResLett7;134,CARBON69;151,ACSNano4;1465,PCCP17;15921,RSCAdv5;80410,PRB74;085406,PLA352;446,PRB86;125434,ACSnano5;1026,
 PRB89;165407,RSCAdv5;51912}. On the other hand, the effective-mass approximation is frequently adopted to comprehend the electronic and optical properties. The low-lying energy bands are expanded about the high-symmetry point, and then the magnetic quantization is done for the approximate ones.  The magneto-optical absorption spectra are calculated for the AA- \cite{JAP110;013725} and AB-stacked graphene \cite{PRB77;115313}. Detailed comparisons between these two methods will be made in the later discussions.


This review article broadly covers the fields related to multilayer graphene systems in the forms of AA, AB, ABC, AAB and sliding configurations and discusses many factors affecting their magneto-electronic properties.
Absorption spectra of multilayer graphene systems are discussed in detail under the various stacking configurations and their competition and corporation with the applied electric field and magnetic field.
The main spectral characteristics focus on the prominent structures, and number, intensity and frequency of absorption peaks.
In Chapter 2, we introduce the Peierls tight-binding model under a electric and magnetic fields.
The optical response, used to describe the spectral properties, is calculated based on the Fermi's golden rule.
We start with the monolayer graphene in Chapter 3, and subsequently discuss the AA-stacked trilayer graphene.
The symmetry-preserved multi Dirac-cone structures are responsible for the optical absorption spectra.
Intra-Dirac-cone excitations comprise the whole frequency region of the spectra.
Chapters 4 and 5 are devoted to AB- and ABC-stacked trilayer graphenes.
The excitations of the massless and massive Dirac quasi-particles dominate the former system, while the vertical transitions between any two inter- or intra-group LLs are permitted for the latter systems.
 In particular, the phenomenon of LL anticrossings in the energy spectrum is discussed and explained by a clear physical picture, especially for the energy width of the sombrero-shaped subbands.
This leads to the corresponding non-monotonic dependence of the spectral frequency and intensity on the magnetic field.
In chapters 6 and 7, we discuss the optical response under effects of broken inversion symmetry in AAB-stacked trilayer and sliding bilayer graphenes.
The lowly-symmetric stacking configurations trigger more significant LL anticrossings and cause distinct features for the K- and K$'$-splitting LLs.

In the beginning of each chapter, we present the Peierls Hamiltonian and the spectral function with respect to each system.
We focus on the mechanism of optical excitations responsible for the spectral characteristics based on the LL spectra and wave functions.
The electric-field-enriched optical properties are also discussed.
The optical absorbtion experiments are available for the investigation of the layer and stacking dependence of the magneto-electronic properties.
Meanwhile, a detail comparison among these systems is also presented in Chapter 7. Finally, Chapter 8 contains concluding remarks.

\pagebreak
\renewcommand{\baselinestretch}{2}

\section{  The generalized tight-binding model }

The primitive unit cell of the honeycomb structure of monolayer graphene, marked by the gray diamond, possesses two carbon atoms $A$ (black) and $B$ (red) atoms, shown in Fig. 1(a).
The symbols $\gamma_{0}$ ($\simeq-2.6$ eV) and $b$ ($=1.42$ \AA ) indicate the nearest-neighbor hopping integral and C-C bond length, respectively.
The first Brillouin zone is shown in Fig. 1(b), where
$\Gamma$, M and K are three highly symmetric points.
The low-energy band structure is dominated by the $\pi$-bonding perpendicular to the graphene plane.
In the tight-binding model based on the 2$p_{z}$ orbitals, the Hamiltonian is a $2\times2$ Hermitian matrix.
The Bloch wave function can be described by a linear combination of the two tight-binding functions of $A$ and $B$ sublattices ($\varphi _{A}$ and $\varphi_{B}$)\cite{PR71;622}:
\begin{eqnarray}
\Psi =\varphi _{A}+\lambda \varphi_{B}=\sum\exp (i\mathbf{k}\cdot \mathbf{R}_{A})\chi (%
\mathbf{r}-\mathbf{R}_{A}) +\lambda \exp (i\mathbf{k}\cdot \mathbf{R}_{B})\chi (%
\mathbf{r}-\mathbf{R}_{B}),
\end{eqnarray}
where $\lambda =\pm 1$ indicates the bonding and anti-bonding, and $\chi (\mathbf{r})$ is
the normalized 2$p_{z}$ orbital wave function for an isolated carbon atom.
The $2\times2$ tight-binding Hamiltonian in the nearest-neighbor approximation is expressed as:
\begin{equation}
\left[
\begin{array}{cc}
0 &  \gamma_{0}f(k) \\
\gamma_{0}f(k)^{\ast} &  0%
\end{array}%
\right] \text{,}
\end{equation}%
where $f(k)=e^{ik_{x}b/\sqrt{3}}+2e^{-ik_{x}b/2\sqrt{3}}\cos(\frac{k_{y}b}{2})$, and the site energies of $A$ and $B$ sublattices and the overlap integral between $A$ and $B$ sublattices are set to zero.

Optical response due to an electromagnetic field is described by the absorption spectral function $A(\omega )$.
According to the Fermi Golden Rule, $A(\omega )$ at zero temperature is expressed as:
\begin{eqnarray}
A(\omega )&\propto&\sum_{n^{v},n^{c}}^{}\int_{1stBZ}\frac{d\textbf{k}}{{
2\pi }*{2\pi }}
\left\vert \left\langle \Psi ^{c}_{\textbf{k}}(n^{c})\left\vert \frac{
\widehat{\mathbf{E}}\cdot \mathbf{P}}{m_{e}}\right\vert \Psi ^{v}_{\textbf{k}}(n^{v})\right\rangle \right\vert^{2} \\ \nonumber%
&\times& Im\left\{\frac{{
f[E^{c}_{\textbf{k}}(n^{c})]-f[E^{v}_{\textbf{k}}(n^{v})]}} {E^{c}_{\textbf{k}}(n^{c})-E^{v}_{\textbf{k}}(n^{v}){-\omega }{-\imath \Gamma }}
\right\},
\end{eqnarray}
where $\textbf{E}$ is the electric polarization vector, $\textbf{P}$ is the momentum operator, $f[E_{k_{y}}(n)]$ is the Fermi-Dirac distribution, $m_{e}$ is the electron mass and $\Gamma$ is the phenomenological broadening parameter.
Only vertical transitions ($\Delta k=0$) are available from the initial occupied state $n^{v}$ to the final unoccupied state $n^{c}$ because of the vanishing photon momentum.
The transition intensity is determined by the velocity matrix element $M^{c,v}_{\textbf{k}}(n^{c},n^{v})$, the first term in Eq. 2, which can be evaluated as follows in the gradient approximation
\begin{eqnarray}
M^{c,v}_{\textbf{k}}(n^{c},n^{v})\sim\frac{\partial}{\partial k_{y}}\left\langle \Psi ^{c}_{\textbf{k}}(n^{c})\left\vert H \right\vert \Psi ^{v}_{\textbf{k}}(n^{v})\right\rangle \quad \text{for} \quad     \widehat{E}\|\widehat{y}.
\end{eqnarray}
The approximation is well verified for carbon-related systems, e.g., graphene, carbon nanotubes, graphite, and nanographite ribbons.
When $M^{c,v}_{\textbf{k}}(n^{c},n^{v})$ slowly varies in the Brillouin zone, this term can be left out of the integral in Eq. (3), and $A(\omega )$ is therefore approximately proportional to the joint density of states.

In the presence of a uniform perpendicular magnetic field $\mathbf{B}=B_{0}\widehat{z}$, electronic states are evolved into highly degenerate LLs due to the cyclotron motion of electrons.
The momentum $\mathbf{P}$ changes to $\mathbf{P}-e\mathbf{A}$, and an extra Peierls phase $G_{R}\equiv {\frac{2{\pi }}{{\phi }_{0}}}$ $\int_{\mathbf{R}^{\prime }}^{\mathbf{R}}\mathbf{A}(\mathbf{r})\cdot d\mathbf{r}$ is introduced in the tight-binding wave function, where $\mathbf{A}(r)$ is the vector potential.
This results in the magnetic field induced enlargement of the primitive unit cell, which changes to $2R_{B}\cdot3b^{'}/2$ $\hat{x}$ (armchair direction) in the Landau gauge
A=(0, $B_{0}$x, 0), where $R_{B}=\frac{\phi _{0}/(3\sqrt{3}b^{\prime 2}/2)}{B_{0}}\simeq \frac{79000\text{ T}}{B_{0}}$.
Accordingly, the enlarged rectangle cell includes $4R_{B}$ atoms$\ $($2R_{B}$ $A$ and $2R_{B}$ $B$ atoms).
The Bloch tight-binding wave function is a linear combination of the products between the value of the subenvelope function and the tight-binding function on each sublattice site:
\begin{equation}
\begin{gathered} \left\vert\Psi _{\textbf{k}})\right\rangle
=\sum_{m=1}^{2R_{B}}A_{m}\left\vert \
A_{m\textbf{k}}\right\rangle +B_{m}\left\vert \
B_{m\textbf{k}}\right\rangle,
\end{gathered}
\end{equation}
where $A_{m}$ and $B_{m}$ are the subenvelope functions and $m$ indicates the atom number. The quantum number of each Landau state can be characterized by the numbers of the zero points of the dominating subenvelope wavefunction.
The Hamiltonian can be expressed as a band-symmetric matrix in the bases of $\{|A_{1\mathbf{%
k}}\rangle $, $|B_{2R\mathbf{k}}\rangle $, $|B_{1\mathbf{k}}\rangle $, $%
|A_{2R\mathbf{k}}\rangle $, $|A_{2\mathbf{k}}\rangle $, $|B_{2R-1\mathbf{k}%
}\rangle $, $|B_{2\mathbf{k}}\rangle $, $|A_{2R-1\mathbf{k}}\rangle $, ......%
$|A_{R-1\mathbf{k}}\rangle $, $|B_{R+2\mathbf{k}}\rangle $, $|B_{R-1\mathbf{k%
}}\rangle $, $|A_{R+2\mathbf{k}}\rangle $, $|A_{R\mathbf{k}}\rangle $, $%
|B_{R+1\mathbf{k}}\rangle $, $|B_{R\mathbf{k}}\rangle $; $|A_{R+1\mathbf{k}%
}\rangle \}$; that is,
\begin{equation}
\left[
\begin{array}{cccccccc}
0 & q^{\ast } & p_{1}^{\ast } & 0 & \cdots & 0 & 0 & 0 \\
q & 0 & 0 & p_{2R} & 0 & \cdots & 0 & 0 \\
p_{1} & 0 & 0 & 0 & \ddots & \ddots & \vdots & 0 \\
0 & p_{2R}^{\ast } & 0 & 0 & \ddots & \ddots & 0 & \vdots \\
\vdots & 0 & \ddots & \ddots & \ddots & 0 & p_{R}^{\ast } & 0 \\
0 & \vdots & \ddots & \ddots & 0 & 0 & 0 & p_{R+1} \\
0 & 0 & \cdots & 0 & p_{R} & 0 & 0 & q \\
0 & 0 & 0 & \cdots & 0 & p_{R+1}^{\ast } & q^{\ast } & 0%
\end{array}%
\right] \text{.}
\end{equation}%
The independent elements $p_{m}$ and $q$ are defined as
\begin{eqnarray}
\begin{array}{l}
p_{m} \equiv \gamma _{0}\exp (-ik_{x}b/2)\cdot 2\cos(k_{y}\sqrt{3}b/2+\Delta G_{mm^{\prime }}),  \\
q \equiv \gamma _{0}\exp (ik_{x}b)\text{;}   \\
\Delta G_{mm^{\prime }} =-\pi (m-5/6)/R\text{.}
\end{array}
\end{eqnarray}
Substituting Eq. (5) and the derivative of Eq. (6) into Eq. (4), one obtains
the velocity matrix $M^{c,v}_{\textbf{k}}(n^{c},n^{v})=$
\begin{equation}
\begin{array}{l}
\left[
\begin{array}{cccccccc}
{A_{1}^{c}}^{\ast} &{B_{2R}^{c}}^{\ast} &{B_{1}^{c}}^{\ast} &\hdots&{B_{R+1}^{c}}^{\ast}&{B_{R}^{c}}^{\ast} &{A_{R+1}^{c}}^{\ast} %
\end{array}%
\right]
\left[
\begin{array}{cccccccc}
0 & 0 & u_{1}^{\ast } & 0 & \cdots & 0 & 0 & 0 \\
0 & 0 & 0 & u_{2R} & 0 & \cdots & 0 & 0 \\
u_{1} & 0 & 0 & 0 & \ddots & \ddots & \vdots & 0 \\
0 & u_{2R}^{\ast } & 0 & 0 & \ddots & \ddots & 0 & \vdots \\
\vdots & 0 & \ddots & \ddots & \ddots & 0 & u_{R}^{\ast } & 0 \\
0 & \vdots & \ddots & \ddots & 0 & 0 & 0 & u_{R+1} \\
0 & 0 & \cdots & 0 & u_{R} & 0 & 0 & v \\
0 & 0 & 0 & \cdots & 0 & u_{R+1}^{\ast } & 0 & 0%
\end{array}%
\right]
\left[
\begin{array}{c}
A_{1}^{v} \\
B_{2R}^{v} \\
B_{1}^{v} \\
\vdots  \\
\vdots  \\
B_{R+1}^{v} \\
B_{R}^{v} \\
A_{R+1}^{v} %
\end{array}%
\right]\\
={A_{1}^{c}}^{\ast}B_{1}^{v}u_{1}^{\ast}+{A_{2}^{c}}^{\ast}B_{2}^{v}u_{2}^{\ast}+\cdots+{A_{2R}^{c}}^{\ast}B_{2R}^{v}u_{2R}^{\ast}+
{B_{1}^{c}}^{\ast}A_{1}^{v}u_{1}+{B_{2}^{c}}^{\ast}A_{2}^{v}u_{2}+\cdots+{B_{2R}^{c}}^{\ast}A_{2R}^{v}u_{2R}\\
=\sum_{m=1}^{2R_{B}}{A_{m}^{c}}^{\ast}B_{m}^{v}u_{m}^{\ast}+{B_{m}^{c}}^{\ast}A_{m}^{v}u_{m}
\text{,}
\end{array}
\end{equation}%
where the element $u_{m}$ is given by
\begin{eqnarray}
\begin{array}{l}
u_{m} \equiv -\sqrt{3}b\gamma _{0}\exp (-ik_{x}b/2)\cdot 2\sin(k_{y}\sqrt{3}b/2+\Delta G_{mm^{\prime }})\text{.}
\end{array}
\end{eqnarray}
Considering the transition intensity of optical excitations in Eq. (4), one can move the Peierls phase term out of the summation because it changes slowly in the enlarged unit cell.
In consequence, the transition intensity $M^{c,v}_{\textbf{k}}(n^{c},n^{v})$ is determined by the inner product of the different subenvelope functions of the initial and final states, i.e., it is proportional to $c_{1} A_{m}^{c}B_{m}^{v}+c_{2} B_{m}^{c} A_{m}^{v}$, where $c_{1}$ and $c_{2}$ are constants.
According to the orthogonal relationship of the subenvelope functions, the selection rule is deduced to be $\Delta n=\pm1$.


The generalized tight-binding model, based on the subenvelope functions of the distinct sublattices, is suitable to study the magneto-electronic properties of multilayer graphene with arbitrary stacking.
The effects of external fields and atomic interactions are simultaneously taken into account.
When an electric field is perpendicularly applied to the graphene planes, the electric potentials are added to the site energies of the diagonal elements of the Hamiltonian matrix.
This model provides an exact diagonalization method for a band-like Hamiltonian matrix to efficiently solve the eigenvalues and eigenfunctions of the quantized states.
It could also be utilized to comprehend the essential properties in the nonuniform or composite external fields \cite{OptEx22;7473}. As to absorption spectra, they are closely related to the Landau-level spectrum and the transition matrix elements.
The spectral characteristics, such as the prominent structures, number, intensity and frequency of the absorption peaks, are well depicted.
Moreover, the results are accurate and reliable within a wide-frequency range.


The velocity matrix, determining the transition intensity and the selection rule, deserves a closer examination.
It should also be noted that the velocity matrix needs to be calculated from the wavefunctions of the initial and final states, of which the individual wavefunction is a linear combination of the products between the subenvelop function and the tight-binding function on each site in the magnetically enlarged unit cell.
Using the gradient approximation, one can obtain that the expectation value of the velocity matrix is simplified as the product of three matrices: the subenvelope functions of the initial state and the final state, and the operator $\frac{\partial H}{\partial k}$.
The calculation result is proportional to the sum of the products of the components of the
the initial and final states correlated to the nonvanishing hopping integrals.
It can be deduced that for multilayer graphene, the terms attributed to the in-plane atomic interaction $\gamma_{0}$ dominate the optical excitations, because intra-layer atomic interactions are much stronger than interlayer ones.
Consequently, the available optical transitions satisfy the condition in which the quantum mode needs to be identical for the $A$ ($B$) sublattice of the initial state and the $B$ ($A$) sublattice of the final state on the same layer, according to the orthogonal relationship of the subenvelope functions.
The detailed calculations are given in the following sections, as well as clear physical pictures for explaining the magneto-optical properties.

\section{ AA-stacked graphene}

AA-stacked graphene has the highest symmetry, in which the carbon atoms of one layer are directly above/below those of another one.
It has been artificially synthesized and is macroscopically stable, although AA stacking configuration is absent in natural graphite due to its higher total energy \cite{JCP129;234709}.
The low-lying multiple Dirac cones have attracted intense interest of theoretical and experimental research\cite{PRB83;165429}.
In this chapter, we focus on the optical absorption spectra in a wide energy range; the results are closely related to the electronic band structure and the dipole transition matrix.
In the absence of external fields, the low- and middle-frequency optical properties are, respectively, dominated by the excitations of massless Dirac fermions and saddle-point states.
On the other hand, the magneto-optical properties are dominated by inter-LL transitions that obey a special selection rule derived from the dipole transition moment between the initial and final states.
For all excitation channels, the linear symmetric and asymmetric superpositions of the wave functions in different layers are responsible for the non-zero and zero probability, respectively, for intra- and inter-Dirac-cone transitions, regardless of external fields.
In particular, intensity discontinuities and forbidden transition regions are revealed for certain kinds of excitation channels.
The optical spectra can be further modified by the application of an electric field.

\subsection{ Electronic and optical properties without external fields}
\subsubsection{ Monolayer}

The electronic structure of monolayer graphene is symmetric about $E_{F}=0$, as shown by the gray curve in Fig. 2(a).
In the low-energy region, two Dirac-type linear subbands intersect at $E_{F}=0$, with the Fermi-momentum state located at the K point.
At middle energies, the saddle-point states are located at the M point with the energies of $\pm\alpha_{0}$.
The isotropic conical dispersion measured from the K point is described by $E^{c,v}\simeq3\alpha_{0}b(\mathbf{k}-\mathbf{K})/2$ according to Eq. (2), where the the conical point is defined as the Dirac point and the slope $dE^{c,v}/dk=3\alpha_{0}b/2$ is regarded as the effective speed of light of the Dirac fermions in graphene.

The main characteristics of the electronic properties are reflected in the absorption spectra.
The spectral absorption function $A(\omega)$, Eq. (3) is obtained by a summation over all possible transitions in the first Brillouin zone.
For the excitations of the Dirac fermions, $A(\omega)$ is simplified as a product of the joint density of states and the Fermi velocity.
Therefore, a linear dependence of the spectral intensity on the frequency appears in the low-frequency region, while in the middle-frequency region, a symmetric peak peak in the logarithmically divergent form is revealed at $\omega\simeq2\alpha_{0}$ due to the saddle-point transitions around the M point, as shown by the black curve in Fig. 2(b).
This very strong $\pi$-electronic absorption peak is one of the critical characteristics in the carbon-related  $sp^{2}$ bonding systems, such as graphite \cite{JPSJ70;897}, graphite intercalation compounds \cite{PRB55;13961}, carbon nanotubes \cite{JPSJ68;3806, PRB61;14114 }, and carbon fullerenes \cite{Carbon99;439}.

\subsubsection{ Multilayer}

The highly symmetric AA-stacked graphene is formed by perpendicularly stacking graphene sheets along the $\hat{z}$ direction.
The geometric structure is illustrated on the top right side in Fig. 1(c), where the four representative atomic interactions taken into account in the tight-binding calculations are indicated: one nearest-neighbor intra-layer interaction $\alpha _{0}=2.569$ eV, and three interlayer interactions $\alpha _{1}=0.361$ eV, $\alpha _{2}=0.013$ eV and $\alpha _{3}=-0.032$ eV\cite{PRB46;4531}.
The electronic band structure of an $N$-layer AA stacked graphene comprises $N$ sets of monolayer-like energy bands.
In the trilayer case, three sets of monolayer-like subbands, $S^{c,v}_{1}$, $S^{c,v}_{2}$ and $S^{c,v}_{3}$, are displayed by the thin curves in Fig. 2(a); one of the three sets is mapped to the monolayer case, and the other two are oppositely shifted by energy $\sim\alpha_{1}$.
Note that three low-energy Dirac cones are located at the K point and three saddle points at the M point.
With an increase of the graphene layers, more Dirac cones are induced while the energy difference among Dirac points decreases, as described in the effective-mass model by \cite{JAP110;013725}
\begin{eqnarray}
[2\cos(j+1\pi/(N+1))-2\cos(j\pi/(N+1))]\alpha_{1},
\end{eqnarray}
where $j=1,2,......N-1$.
Recently, STS experiments have conformed the two Dirac cones in AA-stacked bilayer graphene\cite{NatMat12;887}.
The measured slope and location of the Dirac cones are in agreement with theoretical predictions.
The multiple Dirac-cone structure for multilayer cases can be further confirmed by ARPES and STS experiments.

According to the spectral absorption function $A(\omega)$, Eq. (3), we can find that the inter-band excitations are only available within the same pair of energy subbnads.
That is to say, they come from intra-Dirac-cone structures, but not from inter ones.
The reason is attributed to that the Landau wavefunction is the linear symmetric or asymmetric superposition of the subenvelope functions in different layers, which leads to
to non-zero (zero) electric dipole moments for intra-Dirac-cone (inter-Dirac-cone) transitions, regardless of external perpendicular electric and magnetic fields.
However, a forbidden transition zone, as indicated by shadows for the case of trilayer graphene in Fig. 2(a), is expected for the lower and upper Dirac cones because of their respective free electrons and holes according to the Fermi-Dirac distribution.
This would lead to a discontinuity in the optical spectrum.

As a result of Dirac-fermion excitations, AA-stacked trilayer graphene also shows the linearly dependent absorption intensity in the low-frequency spectrum (red curve in Fig. 2(b)). However, it should be noticed that due to the forbidden transition zones around the Fermi energy (shadows in Fig. 2(a)), a shoulder structure occurs near $\omega\backsimeq1$ eV, which is evaluated as a double of the energy difference between $E_{F}=0$ and the lower (higher) Dirac-point energy or the difference between the two Dirac-point energies.
The measured forbidden zone can be used to estimate the value of $\alpha_{1}$ according to Eq. 10. The spectral intensity for $\omega<1$ eV is equal to the intensity of monolayer graphene, because only the middle Dirac cone contributes to the optical transitions.
On the other hand, for $\omega >1$ eV, the inter-band transitions within all the three Dirac cones are permitted.
Therefore, the linear intensity is three times as large as that of monolayer graphene.
In the middle-frequency region $\sim2\alpha_{0}$, instead of one single peak for monolayer graphene, three close peaks are induced as a result of more excitation channels around the saddle points at M point.
An increase of graphene layers gives rise to verified and diversified spectral structures in which more steps and peaks, respectively, appear at low and middle frequencies.
It should be noted that an even-layer AA-stacked graphene possesses an bilayer-like optical gap determined by the spacing between the valence and conduction Dirac points nearest to the Fermi level.
Previous works have shown the layer-dependent spectral structures for odd-layer and even-layer scenarios.
Nevertheless, for certain frequency regions, the cases in both scenarios exhibit the properties identical to those of 3D graphite as the layer number exceeds the critical values.


\subsection{ Magneto-electronic and magneto-optical properties}

\subsubsection{ Monolayer}

In the presence of a magnetic field, the electronic states congregate to form dispersionless LLs. Each LL in monolayer graphene is fourfold degenerate without the consideration of spin degeneracy. For the states at ($k_{x}=0, k_{y}=0$), their corresponding wave functions are localized around four different centers: $1/6$, $2/6$, $4/6$ and $5/6$ positions of the enlarged unit cell. At $B_{0}=40$ T, the LLs and the localized wave functions are shown in Figs. 3(a) and (b), where only the $1/6$-localized Landau states are discussed.
In the LL energy spectrum in Fig. 3(c), the conduction and valence LLs are symmetric about $E_{F}=0$, and characterized by the quantum numbers $n^{c,v}$ corresponding to the numbers of the zero points of subenvelpe functions ($A_{o}^{c,v}$ for $1/6$-localized states). The $n^{c,v}=0$ LL (at $E_{F}=0$) is constant as a function of $B_{0}$, and the $n^{c,v}=1,2,3...$ LLs are counted away from $E_{F}=0$; they follow the simple square-root relationship $|E_{n}^{c,v}|\propto \sqrt{n^{c,v}B_{0}}$ which is responsible for the Dirac-type linear dispersion.

The Landau wavefunctions, characterized by the subenvelope functions on different sublattices, are used to define the quantum numbers of LLs. The wavefunction can be decomposed into odd-indexed and even-indexed components in the magnetically enlarged unit cell as follows
\begin{equation}
|\Psi _{\mathbf{k}}\rangle =\sum\limits_{m=1}^{2R_{B}-1}(A_{o}|A_{m\mathbf{k}}\rangle +B_{o}|B_{m\mathbf{k}}\rangle
)+\sum\limits_{m=1}^{2R_{B}}(A_{e}|A_{m\mathbf{k}}\rangle +B_{e}|B_{m\mathbf{k}}\rangle )\text{,}
\end{equation}%
where $o (e)$ represents an odd (even) integer. The subenvelope functions $A_{o,e}$ ($B_{o,e}$), standing for the probability amplitudes, exhibit even or odd spatial symmetry and can be described by an $n$-th order Hermite polynomial multiplied with a Gaussian function. The amplitudes are identical for the 1/6 and 4/6 localized wave functions, and so are for those 2/6 and 5/6 ones. $A_{o,e}^{c,v}$ of the 1/6 localized state is equal to $B_{o,e}^{c,v}$ of the 2/6 localized state and vice versa.
Moreover, the odd-indexed component has only a phase difference of $\pi$ with respect to the even-indexed component, i. e., $A_{o}^{c,v}=-A_{e}^{c,v}$ and $B_{o}^{c,v}=-B_{e}^{c,v}$. Therefore, discussing the odd-indexed components of the 2/6 localized wave functions is sufficient to comprehend the LL properties.
For a Landau state with the quantum number $n^{c,v}\geq1$, the subenvelope functions for $A_{o}$ and $B_{o}$ sublattices correspond to $n^{c,v}$-th and $(n^{c,v}+1)$-th order Hermite polynomials, respectively. The conduction and valence subenvelope functions are related to each other by the following relationship: $A_{o}^{c}=A_{o}^{v}$ and $B_{o}^{c}=-B_{o}^{v}$. It can be  deduced that a simple linear relationship exists between the subenvelop functions $A_{o}$ and $B_{o}$ of conduction and valence Landau states, i.e., $A_{o}^{v}(n^{v})\propto B_{o}^{c}(n^{c}=n^{v}-1)$ and $B_{o}^{v}(n^{v})\propto
A_{o}^{c}(n^{c}=n^{v}+1)$.
This accounts for the specific selection rule $\Delta n=n^{c}-n^{v}=\pm 1$ for inter-LL transitions because, according to Eqs. (8) and (9), the expectation value of the velocity matrix is nonzero only when $A_{o}^{v}$ $(B_{o}^{v})$ of the initial state and $B_{o}^{c}$ ($A_{o}^{c}$) of the final state have the same mode.

Feature-rich magneto-absorption spectra directly reflect the LL characteristics.
The inter-LL transitions that obey the selection rule $\Delta n=n^{c}-n^{v}=\pm1$ contribute to many symmetric delta-function-like absorption peaks, as shown in Fig. 4(a).
For a peak denoted as $\omega_{n(n+1)}$, the first number with underline and the second number without underline, respectively, indicate the quantum number of the unoccupied valence LL and that of the occupied conduction LL.
The absorption intensity is the same for each peak because of the same degree of degeneracy for each LL.
Due to the band symmetry about $E_{F}=0$, the peak indicated by $\underline{n}(n+1)$ ($n=0,1,2,3...$) results from two cases of transitions: one from $n^{c}-n^{v}=1$ with $n^{v}=n$ and the other from $n^{c}-n^{v}=-1$ with $n^{v}=n+1$.
Besides, the absorption frequency is proportional to $\sqrt{B_{0}}$, and at a fixed magnetic field, the peaks are sequenced according to a proportional relationship $E^{c}-E^{v}\propto(\sqrt{n+1}+\sqrt{n})$ (Fig. 3.4(b)).

\subsubsection{ Multilayer}

For an N-layer AA-stacked graphene, the primitive unit cell under a uniform magnetic field consists of $N\times 4R_{B}$ atoms.
In the tight-binding basis of {$|A_{1\mathbf{k}}^{1}\rangle $, $|B_{1\mathbf{k}}^{1}\rangle $, $|A_{1\mathbf{k}}^{2}\rangle $, $|B_{1\mathbf{k}}^{2}\rangle $,...$|A_{1\mathbf{k}}^{N-1}\rangle $, $|B_{1\mathbf{k}}^{N-1}\rangle $, $|A_{1\mathbf{k}}^{N}\rangle $, $|B_{1\mathbf{k}}^{N}\rangle $,.....$|A_{2\mathbf{k}%
}^{1}\rangle $, $|B_{2\mathbf{k}}^{1}\rangle $, $|A_{2\mathbf{k}}^{2}\rangle
$, $|B_{2\mathbf{k}}^{2}\rangle $,...$|A_{2\mathbf{k}}^{N-1}\rangle $, $|B_{2%
\mathbf{k}}^{N-1}\rangle $, $|A_{2\mathbf{k}}^{N}\rangle $, $|B_{2\mathbf{k}%
}^{N}\rangle $,.....$|A_{2R_{B}\mathbf{k}}^{1}\rangle $, $|B_{2R_{B}\mathbf{k}%
}^{1}\rangle $, $|A_{2R_{B}\mathbf{k}}^{2}\rangle $, $|B_{2R_{B}\mathbf{k}%
}^{2}\rangle $,...$|A_{2R_{B}\mathbf{k}}^{N-1}\rangle $, $|B_{2R_{B}\mathbf{k}%
}^{N-1}\rangle $, $|A_{2R_{B}\mathbf{k}}^{N}\rangle $; $|B_{2R_{B}\mathbf{k}%
}^{N}\rangle $}, the Peierls Hamiltonian is a band-like matrix and can be expressed as
\begin{eqnarray}
\begin{array}{l}
\langle B_{m\mathbf{k}}^{L}|H|A_{m^{\prime }\mathbf{k}}^{L^{\prime }}\rangle
=\alpha _{0}(t_{1k}\delta _{m^{\prime },m}+t_{4k}\delta _{m^{\prime
},m+1})\delta _{L,L^{\prime }}+\alpha _{3}(t_{1k}\delta _{m^{\prime
},m}+t_{4k}\delta _{m^{\prime },m-1})\delta _{L-1,L^{\prime }}\\
+\alpha_{3}(t_{1k}\delta _{m^{\prime },m}+t_{4k}\delta _{m^{\prime },m+1})\delta
_{L+1,L^{\prime }}\text{;}  \label{4.1} \\
\langle A_{m\mathbf{k}}^{L}|H|A_{m^{\prime }\mathbf{k}}^{L^{\prime }}\rangle
=\langle B_{m\mathbf{k}}^{L}|H|B_{m^{\prime }\mathbf{k}}^{L^{\prime
}}\rangle =\alpha _{1}\delta _{L,L^{\prime }+1}+\alpha _{2}\delta
_{L,L^{\prime }+2}\text{,}  \label{4.2} \\
\end{array}
\end{eqnarray}
where the superscript represents the layer number.
The four independent phase terms are
\begin{eqnarray}
\begin{array}{l}
t_{1k} ={\exp }\{i[-(k_{x}b/2)-(\sqrt{3}k_{y}b/2)]+\pi(m-5/6)/R\}\\
+{\exp }\{i[-(k_{x}b/2)+(\sqrt{3}k_{y}b/2)]-\pi (m-5/6)/R\}\text{,} \label{4.5} \\
t_{2k} ={\exp }\{i[-(k_{x}b/2)-(\sqrt{3}k_{y}b/2)]+\pi (m-3/6)/R\}\\
+{\exp }\{i[-(k_{x}b/2)+(\sqrt{3}k_{y}b/2)]-\pi (m-3/6)/R\}\text{,} \label{4.6} \\
t_{3k} ={\exp }\{i[-(k_{x}b/2)-(\sqrt{3}k_{y}b/2)]+\pi (m-1/6)/R\}\\
+{\exp }\{i[-(k_{x}b/2)+(\sqrt{3}k_{y}b/2)]-\pi (m-1/6)/R\}\text{,} \label{4.7} \\
t_{4k} ={\exp }(ik_{x}b)\text{.}  \label{4.8}
\end{array}
\end{eqnarray}

The magnetic quantization of the N pairs of Dirac cones in an N-layer AA-stacked graphene gives rise to N low-lying LL groups, of which the group divisions are based on the energy spectrum and the characteristics of the spatial wave function.
In the case of the AA-stacked trilayer graphene, the three Dirac cones at $B_{0}=40$ T are evolved into three monolayer-like LL patterns, as indicated by black, red and blue curves in Fig. 5.
Each Landau state at a given ($k_x,k_y$) is four-fold degenerate.
The conduction LLs ($E^{c}_{n_{1}^{c}=0,1,2...}, E^{c}_{n_{2}^{c}=0,1,2...}$, and $E^{c}_{n_{3}^{c}=0,1,2...}$) are approximately symmetric to the valence LLs ($E^{v}_{n_{1}^{v}=0,1,2...}, E^{v}_{n_{2}^{v}=0,1,2...}$, and $E^{v}_{n_{3}^{v}=0,1,2...}$) with respect to the zero-mode LL, as a result of the three isotropic Dirac cones.
The zero-mode energies $E^{c,v}_{n_{1}^{c,v}=0}, E^{c,v}_{n_{2}^{c,v}=0}$, and $E^{c,v}_{n_{3}^{c,v}=0}$ are revealed to be the energies of the three Dirac points.
The two subenvelop functions in the first group behave the same relationship of monolayer graphene for each layer, and so does the third group.
It should be noticed that the second group of LLs has the same relationship only in the two outmost layers, while the inner-layer subenvelop functions are empty.
This reflects the fact that the second group of LLs is regarded as a result of monolayer graphene, which is consistently presented when the number of graphene is odd.
It is a visible means to identify the LLs in terms of single-mode wave functions,
since there is an absence of LL couplings for AA stacking configuration.

The energy distributions of three groups of LLs are related to their individual Dirac cones that are primarily separated by the specific interlayer atomic interaction $\alpha_{1}$\cite{PRB83;165429,JAP110;013725,CPL550;104}.
Each group exhibits a simple square-root dependence on the field strength and the quantum number like the monolayer graphene.
The energies of the zero-mode LLs are constant as a function of the field strength; each LL moves towards the individual Dirac-point energy when the magnetic field approaches zero.
The Fermi energy for a constant carrier density oscillates as a function of the magnetic-field strength around the zero-field value of $E_{F}=0$ (the bold wriggling curves around $E=0$ in Fig. 6), leading to an oscillating energy gap and the metal-semiconductor transition.
The magneto-resistance measurements are used to detect the oscillation of the Fermi level for other 2D systems, such as FeSe and AlGaAs/GaAs heterojunctions \cite{PRL62;1173,PRB90;144517,PRB64;201311}.
The Fermi distribution changes with the oscillating Fermi energy, which might cause the change on the threshold channel and frequency for each category of intragroup LL transitions and induce a forbidden region for optical transitions.
Meanwhile, all the LLs directly cross one another, i.e., absence of  LL anticrossing, and no additional selection rule is induced, since each Landau state is characterized by a single-mode harmonic function.

In the presence of a magnetic field, the velocity matrix elements are expressed as follows in the gradient approximation.
\begin{eqnarray}
\begin{array}{l}
M^{c,v}_{\textbf{k}}(n^{c},n^{v})\cong\frac{\partial}{\partial k_{y}}\left\langle \Psi ^{c}_{\textbf{k}}(n^{c})\left\vert H \right\vert \Psi ^{v}_{\textbf{k}}(n^{v})\right\rangle\\ =\sum_{l,l'=1}^{3}\sum_{m,m'=1}^{2R_{B}}(c_{A_{m,k}^{l}}^{*}c_{A^{l'}_{m',k}}
\frac{\partial}{\partial k_{y}}\left\langle A_{m,k}^{l}\left\vert H \right\vert A_{m',k}^{l'}\right\rangle+c_{A_{m,k}^{l}}^{*}c_{B^{l'}_{m',k}}
\frac{\partial}{\partial k_{y}}\left\langle A_{m,k}^{l}\left\vert H \right\vert B_{m',k}^{l'}\right\rangle\\ +c_{B_{m,k}^{l}}^{*}c_{A^{l'}_{m',k}}
\frac{\partial}{\partial k_{y}}\left\langle B_{m,k}^{l}\left\vert H \right\vert A_{m',k}^{l'}\right\rangle
+c_{B_{m,k}^{l}}^{*}c_{B^{l'}_{m',k}}
\frac{\partial}{\partial k_{y}}\left\langle B_{m,k}^{l}\left\vert H \right\vert B_{m',k}^{l'}\right\rangle),
\end{array}
\end{eqnarray}
where they are evaluated from the derivative of Eq. (12), and the initial- and final-state wavefunctions.
Substituting the velocity matrix into the spectral absorption function $A(\omega)$, Eq. (3), one can find that those terms correlated by the in-plane atomic interaction play an important role, and the optical LL transitions are available only within the same group (those between intergroup LLs are forbidden) \cite{APL97;101905}.
Given that each LL behaves like the massless Dirac LL, the selection rules of the intragroup-LL transitions can be ascribed to $\Delta n=\pm 1$.
On the other hand, the prohibition of intergroup-LL transitions can be attributed to that the Landau wavefunctions are the linear symmetric or asymmetric superpositions of the subenvelope functions in different layers, giving rise to a zero electric dipole moment for the intergroup transitions.

The prominent delta-function-like absorption peaks can be classified into N categories, each of which owes to the individual intragroup-LL transitions and satisfies the particular selection rule $\Delta n=\pm1$ (same as the monolayer).
In the case of $N=3$, the first, second and third categories of the magneto absorption peaks are indicated by black, red and blue numbers, respectively, as shown in Fig. 7(a) for $B_{0}$= 40 T.
The distributions of the peak frequency and intensity with respect to each category are based on the Fermi level and the Fermi-Dirac distributions of the conduction and valence LLs.
As a result, the second category corresponds to a monolayer-like spectrum.
Nevertheless, the first and third categories can be regarded as a bilayer-like spectrum \cite{ACSNano4;1465}.
Outside the threshold peaks, $\omega_{1}^{8^{v} 7^{v}}$ and $\omega_{3}^{6^{c} 7^{c}}$, there exists a forbidden transition region within these two categories as a result of free carriers in Dirac cones.
The forbidden region, enclosed by two peaks excited from the Fermi level, $\omega_{1}^{8^{v} 7^{v}}$ and $\omega_{1}^{8^{v} 7^{c}}$; $\omega_{3}^{6^{c} 7^{c}}$, and $\omega_{3}^{6^{v} 7^{c}}$, is evaluated as a double of the excitation frequency between the highest occupied LL and the Fermi level, either for the first or third category.
In general, two kinds of channels can contribute to one excitation peak due to the symmetric distribution of the conduction and valence LLs, e. g., $n_{i}^{c}=n\rightarrow n_{i}^{v}=n+1$ and $n_{i}^{v}=n+1\rightarrow n_{i}^{v}=n$ for $\omega_{i}^{n,n+1}$, where index $n$ equals $0,1,2...$ and $i$ represents a group number.
The same kind of permitted channels from three intragroup categories, i.e., $\omega_{1}^{n,n+1}$, $\omega_{2}^{n,n+1}$ and $\omega_{3}^{n,n+1}$, results in three close peaks, between which the slightly different frequency reflects the distortion of Dirac cones.
However, it should be noticed that for the four peaks neighboring the forbidden transition region, $\omega_{1}^{8^{v} 7^{v}}$, $\omega_{3}^{6^{c} 7^{c}}$, $\omega_{1}^{8^{v} 7^{c}}$ and $\omega_{3}^{6^{v} 7^{c}}$, their intensities are about a half of the others due to only one kind of channels available for the excitations from the Fermi level.

It is useful to elaborate the magnetic quantization of AA-stacked graphene in terms of the numbers and frequencies of the field-dependent absorption peaks in the three categories, as in Fig. 7(b).
During the variation of the magnetic field, the Dirac-fermion excitations behave as the direct consequence of the $\sqrt{B_{0}}$ proportionality of the absorption frequency.
When the Fermi level crosses two LLs in the bilayer-like groups, a discontinuity of the threshold frequency appears accompanied by the change of the threshold channel in the $\omega-B_{0}$ spectrum.
As $B_{0}$ is reduced, the threshold frequencies decrease in the first and third categories and they are dominated by larger quantum numbers, such as $\omega_{1}^{8^{v} 7^{v}}$ and $\omega_{3}^{6^{c} 7^{c}}$ at $B_{0}$=40 T and $\omega_{1}^{7^{v} 6^{v}}$ and $\omega_{3}^{5^{c} 6^{c}}$ at  $B_{0}$=50 T.
The main reason is that the threshold frequency contributed by the excitations from the highest occupied LL is strongly affected by the LL degeneracy which is proportional to the magnetic field strength.
That is to say, at a constant carrier density, the weaker the magnetic field the more the valence LLs occupy.
In addition, the forbidden transition region also depends on the magnetic field strength.
It converges on the frequency approximately at $2\alpha_{1}\simeq 0.73$ eV, corresponding to the difference between two outer Dirac points (Fig. 2).
With an increase of the graphene sheets, we can classify the spectra depending on whether the numbers of the graphene sheets are even or odd.
All the categories of absorption peaks in the even-layer scenario can be regarded as bilayer-like spectra, each with a discontinuous threshold frequency and a forbidden transition region.
Except for a monolayer-like spectrum exclusively shown for the odd-layer scenario, all the others are bilayer-like.

\subsection{ Electric field enriched optical properties}

By means of applying a perpendicular electric field $\textbf{E}=E_{0}\widehat{z}$, one can modify the spectral features, such as those from Dirac fermions, saddle-point states and Landau states.
A uniform perpendicular electric field does not break the linear dispersions but only cause a rigid shift of the Dirac-cone structures with a slight decrease of the linear slope.
The upper and lower Dirac cones shift to opposite directions corresponding to the middle one that is placed at the Fermi level.
The energy shift is increased with an increment of the electric field strength \cite{CPL550;104}.
The spectral intensity is also linearly dependent on the frequency, while it slightly increases due to the slope variation for the Dirac-type linear dispersion, as shown in Fig. 8.
However, the shoulder structure, at the doubled frequency between two Dirac points, exhibits a blue shift in the same magnitude with respect to the enlargement of the forbidden transition zone across the Fermi level.

Responsible for the Dirac-cone shift, the first and third groups of LLs are also rigidly shifted to opposite directions with regard to the second group.
The selection rule is the same as the case without electric field because the quantum modes of Landau states are unchanged and still described by well-behaved Landau wavefunctions.
Consequently, the second category of absorption peaks is almost unchanged under the influence of the electric field, while for the other two categories, the forbidden transition region is enlarged with threshold frequencies and channels dominated by larger quantum numbers.
In short, the applied electric field rarely changes the main characteristics of the absorption spectrum because it doesn't break the spatial symmetry in the AA stacking configuration.

The aforementioned absorption spectral features can be verified by optical spectroscopies, such as the absorption, transmission, reflection, Raman scattering and Rayleigh scattering spectroscopies.
The multiple Dirac-cone dispersion is verified by ARPES in the bilayer case \cite{NATMAT12;3717}, while the optical excitation spectrum for multilayer cases is still absent.
On the other hand, the previous experiments have identified the $\sqrt{B_{0}}$ dependent frequencies in the infrared transmission spectra for monolayer and AB-stacked trilayer graphene (Chapter 4).
Recently it is also observed by magneto-Raman spectroscopy for AB-stacked graphene up to 5 layers.
While the magneto-optical properties of massless-Dirac-fermions have been observed, verifying on the multiple Dirac points in the AA stacking configuration is worthwhile for further research to identify inter-layer atomic interactions and geometric structures.

\section{  AB-stacked graphene}

ABA stacking has been known as the most stable and common stacking configuration among trilayer graphene systems, as predicted by the theoretical studies \cite{PTPS176;227,PRB78;205425,SSC142;123,PRB81;115432} and confirmed from the experimental measurements \cite{Nature7;948,PRB88;155439,NatPhys7;944,PRL98;206802,PRL104;176404}.
It is analogous to the composite of a monolayer and a AB-stacked bilayer graphenes.
The energy band structure is composed of a pair of linear subbands and two pairs of parabolic ones near the K point.
There are three groups of valence and conduction LLs, in which the monolayer- and bilayer-like ones, respectively, present the almost symmetric and asymmetric energy spectrum about the Fermi level.
Each LL is four-fold degenerate because of the extremely small splitting energy, except for the three special doubly-degenerate ones, $n_1^{c, v}=0$ and $n_2^{c, v}=$ 0 $\&$ 1.

Most of LL wavefunctions belong to the well-behaved modes, and the perturbed ones only appear in few anti-crossings.
The zero-field absorption spectrum mainly exhibits two kinds of special structures including discontinuities (shoulders after broadening) at low frequencies and logarithmically divergent peaks at middle frequencies.
There are only five kinds of excitation channels due to the forbidden transitions between the linear and parabolic energy bands.
Up to now, infrared reflection \cite{PRB78;235408,PRL102;037403,PRB79;115441,NANOLETT14;4548}, absorption spectroscopies \cite{Nature459;820,Nature4;532,SCIENCE320;206}, and infrared scattering scanning near-field optical microscopy \cite{ACSNano9;6765} have successfully verified the characteristic structure and the AB stacking domains.

A perpendicular electric field can dramatically change electronic structures and thus diversify optical properties.
It significantly modulates the subband spacings, causes the subband hybridizations, and induces extra band-edge states.
One pair of parabolic subbands near $E_{F}$ become oscillatory ones with three circular constant-energy contours in the energy-wave-vector space.
The second pair might have the sombrero-shaped dispersions with two constant-energy loops.
These will lead to the asymmetric absorption peaks in the square-root divergent form.
Moreover, it is expected that the increase in the band-edge states enhances density of states, creates extra excitation channels and enriches spectral structures.
Besides, a perpendicular magnetic field can generate the plentiful absorption spectra.
Five categories of inter-LL excitations quite differ from one another in intensity, frequency, and structure of absorption peaks.
Moreover, the magneto-absorption spectrum can be diversified by the difference in Coulomb potential energy between graphene layers.
The four-fold degenerate LLs become double-degenerate ones so that more inter-LL channels are excited.
The spectral intensity is lower with the increase of absorption peaks and the appearance of double-peak structures.

\subsection{  Electronic and optical properties without external fields}

The geometric structure of the AB-stacked configuration is shown in Fig. 1(c), where two neighboring layers shift relatively to each other by one bond length $b$ along the armchair direction.
The $A$ atoms on the odd-number layers and the $B$ atoms on the even-number layers have the same x-y projections, while the $B$ ($A$) atoms on the odd- (even-) number layers are located at the hexagon centers of the adjacent layer.
The primitive unit cell consists of $2\times N$\ atoms, where $N$ is the number of layers.
The atomic interactions are derived from the SWMcC model  \cite{CARBON32;289, PRB46;4540}, including one nearest-neighbor intralayer $\gamma _{0}$, three nearest-neighbor interlayer ($\gamma _{1},\gamma _{3},\gamma _{4}$), two next-nearest-neighbor interlayer ($\gamma _{2},\gamma _{5}$), and the chemical environment difference $\gamma_{6}$ between $A$ and $B$ atoms.
Their values are listed as follows: $\gamma _{0}=-3.12$ eV, $\gamma _{1}=0.38$ eV, $\gamma _{2}=-0.021$eV, $\gamma _{3}=0.28$ eV, $\gamma _{4}=0.12$ eV, $\gamma _{5}=-0.003$ eV, and $\gamma _{6}=-0.0366$ eV \cite{PRB46;4540,CARBON32;289}.

The electronic properties in AB-stacked trilayer graphene can be basically regarded as a hybridization of monolayer and bilayer properties, while the interlayer atomic interactions play an important role on the energy dispersions in the vicinity of highly symmetric points, as indicated in Fig 9(a).
The low-energy band structure is characterized by massless- and massive-Dirac fermions; it consists of a pair of linear subbands, labeled as $S_{1}^{c,v}$, and two pairs of parabolic subbands, $S_{2}^{c,v}$ and $S_{3}^{c,v}$, in the vicinity of the K point.
Analytic solutions show that $\gamma_{3}$ and $\gamma_{4}$ modify the low-lying energy dispersions, and $\gamma_{2}$ and $\gamma_{5}$ separate the intersections of the nearby subbands and influence their spacings as well \cite{JPCM18;5849,PRB73;144427}.
Nevertheless, a slight overlap between the linear valence band and the parabolic conduction one indicates a gapless 2D semimetal.

Away from $E_{F}=0$, the two subbands $S_{3}^{c,v}$ exhibit parabolic dispersions and they are separated mainly by $\gamma_{1}$, with a fluctuation of $\gamma_{2}$, $\gamma_{5}$ and $\gamma_{6}$. Furthermore, those parabolic dispersions extend along the KM direction, with three saddle points near the M point.
It is remarkable that the energy dispersions near the K and M points give rise to special structures in the DOS \cite{JPCM18;5849,PRB73;144427}, and dominate the main characteristics of the absorption spectra.
Recently, the theoretical band-structure calculations have been confirmed by STS and ARPES measurements \cite{PRB87;165102,Nature7;948,NatNanotechnol4;383}.

In the absence of an electric field, the absorption spectrum of trilayer graphene mainly exhibits two kinds of special structures: discontinuities at low frequencies and logarithmic divergences at middle frequencies, which, respectively, owes to the vertical  transitions around the K point and those around the M point (Fig. 9(b)).
The $3\times3$ excitation channels are induced between three pairs of conduction and valence subbands; however, on the basis of the symmetry characteristics of the subenvelope functions, the permitted channels are only those completely within the monolayer-like subbands $S_1^{c,v}$ or the bilayer-like subbands $S_2^{c,v}$ and $S_3^{c,v}$.
That is to say, there are only five kinds of channels accessible in the absorption spectra, while noticeably, the optical excitations are forbidden from $S^{c,v}_{1}$ to $S^{c,v}_{2}$ or $S^{c,v}_{3}$.

At low frequencies, the spectral structure is determined by the latter excitation channels, since those between the linear bands only contribute to linearly increasing intensity.
The special shoulder structure at $\omega\sim0.5$ eV is due to the edge-state transitions from $S_2^{v} \rightarrow  S_3^{c}$ and $ S_3^v  \rightarrow S_2^c$, but otherwise the spectral intensity grows with the higher frequency.
In the middle-frequency spectrum, there are five logarithmic peaks around $\omega=2\gamma_{0}$, coming from the saddle points of $S^{c,v}_{1}$, $S^{c,v}_{2}$ and $S^{c,v}_{3}$ subbands.
Recently, infrared reflection spectroscopy and absorption spectroscopy have shown a clear evidence of the characteristic structure at $\omega\sim0.5$ eV \cite{PRB78;235408,PRB79;115441,NANOLETT14;4548}.
Furthermore, the infrared scattering scanning near-field optical microscopy can be used to verify the AB stacking domains based on the specific conductivities \cite{ACSNano9;6765}.
The whole-frequency spectrum deserves a closer experimental examination to verify the whole band structure and to identify the stacking configuration \cite{PRL104;176404}.

\subsection{  Magneto-electronic and magneto-optical properties }

The magnetic Hamiltonian matrix elements of the ABA stacking are expressed as
\begin{eqnarray}
\begin{array}{l}
\langle A_{m}^{L}|H|A_{m^{\prime }}^{L^{\prime }}\rangle =\gamma
_{1}\delta _{m,m^{\prime }}\delta _{L,L^{\prime }\pm 1}+(\gamma _{5}+\gamma
_{6})\delta _{m,m^{\prime }}\delta _{L,L^{\prime }}+{\frac{\gamma _{5}}{2}}%
\delta _{m,m^{\prime }}\delta _{L,L^{\prime }\pm 2},  \label{6.1} \\
\langle B_{m}^{L}|H|B_{m^{\prime }}^{L^{\prime }}\rangle =\gamma
_{3}(t_{2,k}\delta _{m,m^{\prime }}+t_{4,k}^{\ast }\delta _{m,m^{\prime
}-1})\delta _{L,L^{\prime }\pm 1}+\gamma _{2}\delta _{m,m^{\prime }}\delta
_{L,L^{\prime }}+{\frac{\gamma _{2}}{2}}\delta _{m,m^{\prime }}\delta
_{L,L^{\prime }\pm 2},  \label{6.2} \\
\langle A_{m}^{L}|H|B_{m^{\prime }}^{L^{\prime }}\rangle
=\gamma_{0}(t_{4,k}\delta _{m,m^{\prime }}+t_{3,k}^{\ast }\delta _{m,m^{\prime
}+1})\delta _{L,L^{\prime }}+\gamma _{4}(t_{1,k}\delta _{m,m^{\prime
}}+t_{4,k}^{\ast }\delta _{m,m^{\prime }+1})\delta _{L,L^{\prime }\pm 1}%
\text{ for odd }L,  \label{6.3} \\
=\gamma _{0}(t_{1,k}\delta _{m,m^{\prime }}+t_{4,k}^{\ast }\delta
_{m,m^{\prime }+1})\delta _{L,L^{\prime }}+\gamma _{4}(t_{4,k}\delta
_{m,m^{\prime }}+t_{3,k}^{\ast }\delta _{m,m^{\prime }+1})\delta
_{L,L^{\prime }\pm 1}\text{ for even L.}  \label{6.4}
\end{array}
\end{eqnarray}

The ABA stacking exhibits three unique groups of LLs which could be regarded as the superposition of those of monolayer and AB-stacked bilayer graphenes, as indicated in Figs. 10(a)-(g).
The first group of LLs, being quantized from the linear subbands, is initially formed near $E_F$.
In the vicinity of $E_F$, the square-root dependence of LL energies on $n_1^{c,v}$ is absent because of the slightly distorted Dirac-cone structure with an energy spacing of $\sim 20$ meV.
Furthermore, the $n_1^{c,v}=0$ LLs are doubly degenerate and their energies are different at 2/6 $\&$ 1/6 (5/6 $\&$ 4/6) centers.
The $n_1^{c,v} \geq 1$ LLs, which the conduction and valence ones are almost symmetric to each other, are four-fold degenerate, similar to those of monolayer graphene.
The second and third groups come from the two pairs of parabolic subbands.
The $n_2^{c,v}$ LLs are initiated at $E^{c,v} \approx 0$, while the $n_3^{c,v}$ ones are located at (0.57 eV, -0.6 eV).
Each LL has double degeneracy, mainly owing to the destruction of the mirror symmetry about the z-plane.
The energy spacing between the splitting LLs is extremely small ($\approx$ 0.1 meV) so that their absorption peaks will merge together.
Exceptionally, the $n_2^{c, v}=0$ $\&$ 1 LLs exhibit the observable splitting energies of about 10 meV, accompanied with certain important differences in the LL wavefunctions and magneto-absorption spectra at 2/6 and 1/6 centers (5/6 and 4/6 ones).
Moreover, all the LLs possess the well-behaved spatial distributions except for few anti-crossings in the $B_0$-dependent energy spectrum.
It is also noticed that the low-lying LLs of the first and second groups have been identified from the STS \cite{PRB91;155428} and quantum Hall transport measurements \cite{NanoLett13;1627,PRX2;011004}.

There are five categories of absorption peaks in the optical spectrum, as observed in the zero-field available excitation channels (Fig. 11).
The intergroup optical excitations associated with the first LL group are forbidden due to the simple phase relations among six different sublattices.
The first group only survives on ($A^1$, $B^1$, $A^3$, $B^3$) sublattices, since the neighboring-layer atomic interaction do not make contributions to its electronic states.
Furthermore, the $A^1$ $\&$ $A^3$ ($B^1$ $\&$ $B^3$) subenvelope functions are identical for the second and third groups whereas those of the first group possess the opposite phases.
As to the intragroup excitations $n_1^v \rightarrow n_1^c$, they exhibit only single peaks with a predominantly uniform intensity because the velocity matrix elements are nearly constant for the almost symmetric Dirac-cone structure.

The second and third LL groups sharply contrast with the first one in term of intensity, structure, and frequency of intragroup absorption peaks.
The peak intensities are nonuniform and only about half of that of the first group.
The magneto-absorption spectrum presents twin-peak structures with an observable spacing of $\sim$ 10 meV.
The differences in peak structures can account for the lower intensity.
Exceptionally, the $\omega_2^{2^v1^c}$ peak is a single structure because of the requirement of the Fermi-Dirac distribution.
Concerning the two categories of intergroup excitations, $n_2^v \rightarrow n_3^c$ and $n_3^v \rightarrow n_2^c$ display the single peaks.
Such peaks obey the selection rules of $| n_2^{c,v} - n_3^{v,c}| = 2$ and 0, and their intensities are comparable.
In general, the intensity of the intergroup inter-LL excitations is weaker than half of that of the intragroup ones.
This is mainly due to the phase relations between the $A^l$ ($B^l$) subenvelope functions of the initial states and $B^l$ ($A^l$) of the final ones.
For the intragroup excitations, all the subenvelope functions of the $n^v$ initial and $n^c$ final states on the $A^l$ and $B^l$ sublattices have the same (for the second group) or opposite (for the third one) phases, resulting in to the high peak intensities.
On the other hand, the complex phase relations in the wavefunctions remarkably weaken the intergroup inter-LL transitions.
In short, the magneto-optical spectra at 2/6 (5/6) and 1/6 (4/6) centers are almost identical in the main features of absorption peaks, including frequency, structure, and intensity.
Especially, the optical excitations coming from the double-degenerate $n_1^{c, v}=0$ and $n_2^{c, v}=0$ $\&$ 1 LLs are distinct in absorption frequencies at different centers.
Such energy splitting is observable ($\sim$ 10 meV) and very useful in identifying the stacking configurations from the experimental measurements.

Experimentally, magneto-optical spectroscopies have been utilized to observe the inter-LL vertical transitions in the ABA stacking.
For the intragroup excitations of the first LL group, the $\sqrt B_0$-dependence of absorption frequency and the selection rule of $\Delta n=\pm 1$ are revealed in the infrared transmission spectrum \cite{PRL104;067404,PRL98;197403,PRL100;087401} and the  magneto-Raman scattering spectrum \cite{PRL107;036807,ACSNano8;10766,PRB91;235403}.
In particular, the square-root and linear $B_{0}$-dependent frequencies due to the massless and massive Dirac-fermions of few-layer AB-stacked graphenes are confirmed by the magneto-Raman spectroscopy \cite{PRL107;036807,ACSNano8;10766,PRB91;235403}.
However, the other four categories of excitation channels need further experimental verifications.
The experimental examinations on the main features of absorption peaks can distinguish the stacking configurations of trilayer graphenes.


The magnetic-field-dependent absorption frequencies of the first few categories are shown in Fig. 12.
Monolayer-and bilayer-like dependence are, respectively, reflected in the first and second intragroup categories.
For the latter, the tiny energy difference $\backsim$ 10 meV for the twin-peak excitation channels is revealed as corresponding to the electron-hole asymmetry in the LL spectrum\cite{JPCM18;5849,PRB73;144427}.
The splitting energies have been observed for the inter-LL transitions near the K point in bulk graphite and graphite intercalation compounds\cite{PRB84;153405,
PRB85;245410}.
On the other hand, the intergroup excitation channels are shown at higher frequencies.
Under a sufficiently small magnetic field, the onset energies of the intergroup categories correspond to the band-edge states in Fig. 9, which are mainly affected by the interaction, $\gamma_{1}$.
The aforementioned results imply that the optical experiments can be used to determine the interlayer atomic interactions that dominate the electron-hole asymmetry and band dispersions through the observation of the splitting energies in the twin-peak structure and the convergent threshold frequencies of the intergroup categories.


\subsection{ Electric field enriched optical properties}

A perpendicular electric field significantly modulates the low-lying energy dispersions and optical excitation channels.
It changes subband spacings, causes subband hybridizations, and induces extra band-edge states as well (Fig. 13(a)).
Unlike the induction of a significant band gap in bilayer graphene, trilayer graphene has been shown both theoretically \cite{PRB82;035409} and experimentally \cite{PRB88;155439} to behave as a semimetal with weakly oscillating dispersions near the Fermi level.
Attributed to the hybridization of the linear and parabolic subbands, the oscillation width is greatly enhanced as the applied electric field increases.
Especially, the pair of parabolic subbands near $E_F$ become oscillatory ones, similar to the $S_1^{c,v}$ subbands in the AAB stacking (Fig. 20 in Chap. 6, section 1).
In the energy-wave-vector space, each subband has three circular constant-energy contours associated with the concave and convex dispersions.
This will lead to the asymmetric absorption peaks in the square-root divergence form, since such subbands could be regarded as one-dimensional ones. Similar behavior can be observed in the sombrero-shaped bands with two constant-energy loops.
Moreover, it is expected that the increase in the band-edge states of the oscillating subbands enhances the density of states, induces optical excitation channels and diversifies the low-energy optical spectra.

The characteristics of the absorption spectra, such as prominent structures, and number, intensity and frequency of absorption peaks, are significantly changed under the influence of the applied electric field (Fig. 13(b)).
Owing to the subband hybridizations, not only the original transition channels survive, but also the extra channels are created between the first pair and the other two pairs of energy bands.
That is to say, five kinds of effective excitation channels are changed into nine ones under the application of $\mathbf{E}$.
The low-energy optical excitation channels from the field-induced oscillatory subbands give rise to multi-peak structures, due to multiple channels, in the absorption spectra instead of the featureless spectra without external fields.

Accordingly, as the field strength is larger than a moderate value $\sim0.3$ eV/layer, three multi-peak structures appear in the low-frequency region, as shown in Fig. 13(b).
The first, second and third peaks, as denoted by $P_{f}$, $P_{s}$ and $P_{t}$, correspond to the transitions from $S_{1}^{v}\rightarrow S_{1}^{c}$, ($S_{1}^{v}\rightarrow S_{2}^{c}$, $S_{2}^{v}\rightarrow S_{1}^{c}$), $S_{2}^{v}\rightarrow S_{2}^{c}$, respectively.
The number of excitation channels for each multi-peak depends on the number of band-edge states of the oscillatory subbands; that is, it depends on the electric-field strength.
At $E_0=0.3$ eV/layer (red curve), each multi-peak structure consists of one main peak accompanied with one or two weak peaks (like shoulders after broadening).
However, the absorption spectra exhibit more and sharper symmetric peaks with an increment of the field strength, especially for $P_{s}$ which displays a rather strong symmetric peak as $E_{0}$ goes up to 0.5 eV/layer (green curve).
The peak intensity is enhanced by the larger radius of the constant-energy loop (the higher density of states).
Specifically, the symmetric peak structure of $P_s$ is the combination of two opposite-direction asymmetric peaks due to the concave- and convex-downward subbands ($S_1^v \rightarrow S_2^c$, $S_2^v\rightarrow S_1^c$) in Fig. 13(a).
These results are in qualitative agreement with recent transport \cite{PRB82;035409} and optical experiments \cite{NatPhys7;944}.
Furthermore, the excitation energy of the first peak $P_{f}$ is nearly independent on the electric-field strength, which is responsible for the semi-metal properties in the trilayer case of AB stacking.
The other peaks show blue shifts with the increasing field strength.
It should be noticed that the $\mathbf{E}$-induced low-frequency absorption peaks rely on the geometric structure.
By contrast, the characteristics of middle-frequency spectra due to the $\pi$-electronic states around the M point are preserved in the presence of electric field.

The applied perpendicular electric field can dramatically alter the LL characteristics and thus diversify the magneto-absorption spectrum (Fig. 13(c)).
The four-fold degenerate LLs are replaced by the double-degenerate ones under the broken mirror symmetry about the z-plane.
Each group of LLs is split into two subgroups, in which the splitting LLs are characterized by two different localization centers of 1/6 and 2/6 unit cell (4/6 and 5/6 unit cell).
The dominating sublattice, in determining quantum numbers, remains unchanged for each subgroup, i. e., there exist two inversely symmetric sublattices for two splitting subgroups.
As more inter-LL channels are excited, the absorption spectrum shows double peaks, identified as $\alpha$ and $\beta$ types for the inter-LL transitions from the former and the latter subgroups, respectively.
Furthermore, the anti-crossings of the perturbed LLs are presented in the $E_0$-dependent energy spectrum (red arrow), since the cooperation of electric field and certain interlayer atomic interactions can drastically change the charge distributions in the main and side modes (details discussed in Chap 5.2).
The total number of absorption peaks is greatly increased, but their intensities are significantly reduced.
The double and the twin peaks, which, respectively, arise from the splitting and asymmetric LL energy spectra, are simultaneously revealed in the magneto-optical excitations.
The two components in each double peak are associated with 2/6 and 1/6 centers, while those in a twin-peak structure only rely on a specific center.
It should be noticed that, the inter-LL excitations between the first group and the other ones are recovered in the presence of $\mathbf{E}$.
The potential energy differences between graphene layers result in drastic changes of spatial charge distributions; that is, the first group of LLs are contributed by electronic states in each layer.
The existence of the second-layer charge distribution is responsible for the enriched absorption peaks.
The electric field can generate nine categories of magneto-absorption peaks, being absent in the AAA, ABC, and AAB stacking systems.
Also, there exist extra absorption peaks arising from the frequent intra-subgroup LL anti-crossings.
The $\mathbf{E}$-induced intergroup excitation channels and extra peaks are expected to provide more information in identifying the stacking domains from magneto-optical measurements \cite{ACSNano4;1465,PRL98;157402,ACSNano4;1465,PRB77;115313}.

The double-degenerate splitting LL energy spectrum in the ABA stacking has been identified from the experimental measurements on the Hall conductances \cite{PRB77;115313} under a composite field. The $\mathbf{E}$-created diverse  magneto-optical properties could be further verified by the infrared transmission spectroscopy \cite{PRL100;087401} and the Raman scattering spectroscopy \cite{PRL97;187401}, in which they include nine categories of inter-LL excitation channels, the anti-crossing-dependent extra peaks, the simultaneous appearance of the single, twin and double peaks, and the weakened peak intensities.

\section{ ABC-stacked graphene}

The ABC stacking is a stable configuration in natural graphite\cite{PRSLSA106;749,PRSLSA181;101}.
Few-layer ABC-stacked graphene is semi-metallic with two partially flat bands slightly overlapping near the Fermi level\cite{PRB73;245426,PRB83;220503,PRB87;155116}.
The prominent peak of DOS near the Fermi level, coming from the partially flat bands, play an important role in the low-frequency optical spectrum.
In particular, energy bands exhibit sombrero-shaped dispersions near the nearest vertical interlayer hopping energy \cite{RevModPhys81;109,PRL97;036803,PRB84;165404,PRB82;035409}.
The corresponding absorption structures have been observed in the optical spectrum \cite{NanoLett11;164}.
Besides, the middle-frequency spectrum is relevant to the band structure around the M point in the vicinity where the energy dispersions are significantly modified by interlayer atomic interactions.
Under a magnetic field, it is intriguing within the sombrero-shaped band structure that the LLs are inversely sequenced as a function of the field strength, leading to frequently intragroup LL anticrossings accompanied by significant hybridizations of LLs. \cite{PRL104;176404,PCCP17;15921,RSCAdv5;80410,PRB90;205434} In general, the inter-LL transitions satisfy the particular selection rule $\Delta=\pm 1$ if the wavefunctions are well-behaved. However, extra optical selection rules are induced for the LLs with quantum modes in modulo 3, as a result of the LL hybridization. What could be especially needed is to focus on the excitations within the LL anticrossing spectrum, where the diversified quantum modes lead to the considerable peaks whose intensities are associated with the degree of hybridizations among LLs.

\subsection{ Electronic and optical properties without external fields}

ABC-stacked $N$-layer graphene consists of 2$N$ carbon atoms in a primitive unit cell, in Fig. 1(e).
Each of the graphene sheets is shifted by a distance of $b^{'}$ with respect to the lower (upper) neighboring layer along $+(-) \widehat{x}$.
The sublattice $A$ of one layer is situated directly above the center a hexagon in the adjacent lower layer, whereas the sublattice $B$ lies above the lower-layer sublattice $A$.
The intra-layer and the inter-layer atomic interactions in the tight-binding model are illustrated as $\beta_{i}$'s, where $i$=0,1,2,3,4 and 5. $\beta_{0}$ represents the nearest-neighbor intra-layer hopping integral, $\beta_{1}$, $\beta_{3}$ and $\beta_{4}$ correspond to adjacent inter layers, $\beta_{2}$ and $\beta_{5}$ are for next-nearest inter layers.
$\beta_{1}$ and $\beta_{2}$ couple two vertical sites, and $\beta_{3}$, $\beta_{4}$ and $\beta_{5}$ couple non-vertical sites.
Their values are listed as follows: $\beta_{0}=-$3.16 eV, $\beta_{1}$=0.36 eV, $\beta_{2}=-$0.01 eV, $\beta_{3}$=0.32 eV, $\beta_{4}$=0.03 eV, and $\beta_{5}$=0.0065 eV.$^{56}$\cite{PRB46;4540}

The low-energy electronic properties of ABC-stacked trilayer graphene are characterized by three pairs of conduction and valence subbands in Fig. 14(a).
They display partially flat, sombrero-shaped and linear dispersions, as respectively labeled by $S_{1}^{c,v}$ (black), $S_{2}^{c,v}$ (red) and $S_{3}^{c,v}$ (blue).
Being weakly dispersive in the vicinity of $E_{F}=0$, the subbands of the first pair, $S_{1}^{c,v}$, consist of surface localized states and anisotropic around the K point as a result of the trigonal warping\cite{PRB73;245426,PRB83;220503,PRB87;155116}.
The small energy difference between the two localized subbands is approximated as $2\beta_{2}=0.04$ eV.
Moreover, the tight-binding calculations show that such a characteristic of the surface localized states disappears as the wave vector is moving away from the K point.
At $E_{F}=0$, the three crossing points in a circle are identified as Dirac points\cite{PRB80;165409}: one in the K$\Gamma$ line and another two whose rotation angles are relative to the former by $120^{o}$ and $-120^{o}$.
Away from $E_{F}=0$, the sombrero-shaped and linear subbands cross each other near K point at energies $\sim\pm\beta_{1}$ $(=\pm0.36$ eV), where the crossing point corresponds to a local minimum of the former and a local maximum of the latter.
In particular, within the bottom ($\sim$0.34 eV) and tip ($\sim$0.36 eV) of the sombrero-shaped subband, two closed circular loops at a specific energy leads to a doubly degenerate state with different Landau orbitals under a magnetic field.
This would cause unusual characteristics for certain excitation channels.
Interlayer atomic interactions also modify and diversify the band structures around $\beta_{0}$.
It is mainly caused by $\beta_{3}$ and $\beta_{5}$ that some saddle-point states are distributed away from the M point over a small energy range and some are shifted along the $\Gamma$M direction.
Recently, the theoretical band structure is experimentally confirmed by STS and ARPES \cite{PRB87;165102,Nature7;948,NatNanotechnol4;383}.

The optical absorption spectrum correlates with the particular energy spectrum in both regions of $\omega<2\beta_{1}$ around the K point and of $\omega\sim2\beta_{0}$ around the M point according to Eq. (3), as shown in Fig. 14(b).
The low-frequency spectrum is dominated by the excitations from the former region, mainly affected by the interlayer atomic interactions $\beta_{1}$ and $\beta_{2}$.
There are one obvious asymmetric peak at $\omega\simeq0.35$ eV, one observable peak at $\omega\simeq0.68$ eV and one shoulder at $\omega\simeq0.72$ eV, respectively, due to the vertical transitions between the surface-localized and sombrero-shaped subbands, between the valence and conduction sombrero-shaped subbands, and between two linear subbands. Specifically, the latter two structures, with a small frequency difference,   might be merged together in optical measurements.
On the other hand, $3\times3$ kinds of excitations channels from the saddle points in the latter region induce several obvious peaks and shoulder structures for the middle-frequency spectrum $\sim2\beta_{0}$, in which the interlayer atomic interactions $\beta_{3}$ and $\beta_{5}$ play an important role.
There appears a prominent symmetric peak at $\omega\simeq$6.21 eV, as a result of transitions between the saddle points of the subbands $S_{1}^{c,v}$.
The two peaks at $\omega\simeq$6.28 eV and $\omega\simeq6.46$ eV are responsible for the saddle-point transitions of pairs of subbands $S_{1}^{c,v}$ and $S_{3}^{c,v}$.
Moreover, the characteristic peak at $\omega\simeq$6.32 eV is associated with the saddle point of $S_{2}^{v}$ \cite{NanoLett11;164}.
Recently, infrared reflection spectroscopy and absorption spectroscopy have verify the low-frequency optical properties, showing a clear evidence of two characteristic peaks associated with the partially flat and sombrero-shaped subbands.
Based on the specific infrared conductivities, infrared scattering scanning near-field optical microscopy can distinguish the ABC stacking domains with nano-scale resolution from other domains.
Furthermore, the whole-frequency spectrum deserves a closer examination to verify the band structure over a wide energy region and the measured features are useful to identify the stacking configuration. \cite{PRL104;176404}


\subsection{ Magneto-electronic and magneto-optical properties}

The $4NR_{B}$ Peierls tight-binding functions based on the periodical atoms in the rectangular unit cell are expressed as: $\{|A_{1\textbf{k}}^{1}\rangle, |B_{1\textbf{k}}^{1}\rangle, |A_{1\textbf{k}}^{2}\rangle, |B_{1\textbf{k}}^{2}\rangle...|A_{1\textbf{k}}^{N}\rangle, |B_{1\textbf{k}}^{N}\rangle.....|A_{2R_{B}\textbf{k}}^{N}\rangle, |B_{2R_{B}1\textbf{k}}^{N}\rangle\}$.

The non-zero elements of the Peierls tight-binding Hamiltonian matrix are given by

\begin{eqnarray}
\begin{array}{l}
\langle B_{m\mathbf{k}}^{L}|H|A_{m^{\prime }\mathbf{k}}^{L^{\prime }}\rangle\\
=\beta _{0}(t_{1,k}\delta _{m,m^{\prime }}+t_{4,k}\delta _{m,m^{\prime
}-1})\delta _{L,L^{\prime }}+\beta _{3}(t_{3,k}^{\ast }\delta _{m,m^{\prime
}+1}+t_{4,k}^{\ast }\delta _{m,m^{\prime }})\delta _{L,L^{\prime }-1}\text{
for $L=3l-2$}, \\
\langle B_{m\mathbf{k}}^{L}|H|A_{m^{\prime }\mathbf{k}}^{L^{\prime }}\rangle\\
=\beta _{0}(t_{3,k}\delta _{m,m^{\prime }+1}+t_{4,k}\delta _{m,m^{\prime
}})\delta _{L,L^{\prime }}+\beta _{3}(t_{2,k}^{\ast }\delta _{m,m^{\prime
}}+t_{4,k}^{\ast }\delta _{m,m^{\prime }-1})\delta _{L,L^{\prime }-1}\text{
for }L\text{$=3l-1$}, \\
\langle B_{m\mathbf{k}}^{L}|H|A_{m^{\prime }\mathbf{k}}^{L^{\prime }}\rangle\\
=\beta _{0}(t_{2,k}\delta _{m,m^{\prime }}+t_{4,k}\delta _{m,m^{\prime
}-1})\delta _{L,L^{\prime }}+\beta _{3}(t_{1,k}^{\ast }\delta _{m,m^{\prime
}}+t_{4,k}^{\ast }\delta _{m,m^{\prime }-1})\delta _{L,L^{\prime }-1}\text{
for }L\text{$=3l$}, \\
\langle B_{m\mathbf{k}}^{L}|H|A_{m^{\prime }\mathbf{k}}^{L^{\prime }}\rangle
=\beta _{2}\delta _{m,m^{\prime }}\delta _{L,L^{\prime }-2}+\beta
_{1}\delta _{m,m^{\prime }}\delta _{L,L^{\prime }-1}\text{ for any L},\\
\langle B_{m\mathbf{k}}^{L}|H|B_{m^{\prime }\mathbf{k}}^{L^{\prime }}\rangle
=\langle A_{m\mathbf{k}}^{L+2}|H|A_{m^{\prime }\mathbf{k}}^{L^{\prime}+2}\rangle
=\beta _{4}(t_{1,k}\delta _{m,m^{\prime }}+t_{4,k}\delta
_{m,m^{\prime }-1})\delta _{L,L^{\prime }-1}\text{ for $L=3l-2$}, \\
\langle A_{m\mathbf{k}}^{L}|H|A_{m^{\prime }\mathbf{k}}^{L^{\prime }}\rangle
=\langle B_{m\mathbf{k}}^{L+1}|H|B_{m^{\prime }\mathbf{k}}^{L^{\prime}+1}\rangle
=\beta _{4}(t_{3,k}\delta _{m,m^{\prime }+1}+t_{4,k}\delta
_{m,m^{\prime }})\delta _{L,L^{\prime }-1}\text{ for $L=3l-2$}, \\
\langle A_{m\mathbf{k}}^{L}|H|B_{m^{\prime }\mathbf{k}}^{L^{\prime }}\rangle
=\langle B_{m\mathbf{k}}^{L^{\prime }+2}|H|B_{m^{\prime }\mathbf{k}%
}^{L+2}\rangle=\langle A_{m\mathbf{k}}^{L^{\prime }+1}|H|A_{m^{\prime }%
\mathbf{k}}^{L+1}\rangle\\
=\beta _{5}(t_{3,k}\delta _{m,m^{\prime
}+1}+t_{4,k}\delta _{m,m^{\prime }})\delta _{L,L^{\prime }-2}
\text{ for $L=3l-2,$} \\
\langle A_{m\mathbf{k}}^{L}|H|A_{m^{\prime }\mathbf{k}}^{L^{\prime }}\rangle
=\langle B_{m\mathbf{k}}^{L+1}|H|B_{m^{\prime }\mathbf{k}}^{L^{\prime}+1}\rangle\\
=\beta _{4}(t_{2,k}\delta _{m,m^{\prime }}+t_{4,k}\delta
_{m,m^{\prime }-1})\delta _{L,L^{\prime }-1}\text{ for $L=3l-1$}, \\
\langle B_{m\mathbf{k}}^{L}|H|B_{m^{\prime }\mathbf{k}}^{L^{\prime }}\rangle
=\langle A_{m\mathbf{k}}^{L+2}|H|A_{m^{\prime }\mathbf{k}}^{L^{\prime
}+2}\rangle =\langle A_{m\mathbf{k}}^{L^{\prime }+1}|H|B_{m\mathbf{k}%
}^{L+1}\rangle \\
=\beta _{5}(t_{2,k}\delta _{m,m^{\prime }}+t_{4,k}\delta
_{m,m^{\prime }-1})\delta _{L,L^{\prime }+2}\text{ for $L=3l$}, \\
\langle A_{m\mathbf{k}}^{L}|H|A_{m^{\prime }\mathbf{k}}^{L^{\prime }}\rangle
=\langle B_{m\mathbf{k}}^{L+1}|H|B_{m^{\prime }\mathbf{k}}^{L^{\prime
}+1}\rangle=\langle A_{m\mathbf{k}}^{L^{\prime }+2}|H|B_{m^{\prime }\mathbf{%
k}}^{L+2}\rangle \\
=\beta _{5}(t_{1,k}\delta _{m,m^{\prime }}+t_{4,k}\delta
_{m,m^{\prime }-1})\delta _{L,L^{\prime }+2}\text{ for $L=3l;$} \\
\end{array}
\end{eqnarray}

The magnetically quantized LLs of ABC-stacked trilayer graphene can be divided into three groups according to three zero-field subbands and the spatial characteristics of the wave functions, as indicated by black, red, and blue colors in Fig. 15 at $B_{0}=40$ T.
The valence and conduction LLs are asymmetric with respect to $E_{F}=0$ due to interlayer atomic interactions.
Each level is four-fold degenerate under the inversion symmetry.
The $A^{l}_{o}$-subenvelop functions of the $1/6-$ ($4/6-$) localized states ($l=1,2;3$) have out of phase with $B^{l}_{o}$-subenvelop functions of the $2/6-$ ($5/6-$) localized states.
The wavefunctions at $2/6$ unit cell are illustrated.
As a result of being quantized from the partially flat subband, the first group of LLs are initiated from $E_{F}=0$, in the vicinity of which the first three LLs, coming form the Dirac points, are slightly separated by the interlayer hoppings, while the other higher-order LLs are distributed away from $E_{F}=0$\cite{Ann326;721,PRB80;165409,PRB90;205434}.
The dominating sublattice in the first group corresponds to the outmost $B^{1}$ sites of the surface-localized states.
Similarly, the corresponding outmost $A^{3}$ sites can be used to define the quantum numbers of the other degenerate states localized at $1/6$ and $4/6$ unit cell.
The quantum numbers of LLs are counted in ascending order from $E_{F}=0$ as evidenced by the dominating subenvelope function $B^{1}_{o}$.
Meanwhile, resulting from the sombrero-shaped and linear subbands, the LLs in the second and third groups are initiated from the intersection energy $\sim\beta_{1}$ of the subbands, and their quantum numbers are, respectively, characterized by the dominating subenvelop functions $B^{3}_{o}$ and  $B^{2}_{o}$.

The main characteristics of the LL spectrum are relevant to the zero-field band structure, as shown in Fig. 16.
The onset energy for each group is the same as the band-state-state energy of the respective subband near the K point.
The level spacings are determined by the DOS, that is to say, LLs are densely packed with an increment of the DOS.
Accordingly, the intensive formation of the LLs near $\pm\beta_{1}$ is attributed to the crossover of the subbands in Fig. 14.
The first and third groups present the monotonically $B_{0}-$dependent energies.
Near $E_{F}=0$, the three LLs insensitive to the field strength reflect the characteristics of Dirac points.
They are confined in a small energy spacing between the two partially flat subbands, with the separated energies mainly determined by the specific interlayer atomic interaction $\gamma_{4}$.
This sequence of the first group of LLs was confirmed by the measurements of the Hall conductance \cite{PRX2;011004}.
However, a complicated pattern of a non-monotonic energy dependence is shown in the second group of LLs, which are responsible for the particular magnetic quantization rule relating to the sombrero-shaped energy subbands.
A considerable amount of inverted LLs is observed for the field strength under certain critical fields that correspond to the local minima in the spectrum.

The relationship of the inverted LLs and the critical magnetic field is dominated by the interplay between the magnetic degeneracy and the total states enclosed in the sombrero surface, as shown in Fig. 17(a).
Two kinds of $B_{0}$-dependent energies correspond to two kinds of degenerate Landau orbitals, i.e., the proportional and the inversely proportional dependences, for the respective outer and inner circular constant-energy loops on the surface of the sombrero-shaped subbands.
A weak magnetic field aggregates electronic states near the vertex of the sombrero-shaped subband.
Therefore, as more states on the inner surface are magnetically quantized, the LLs display the inversely proportional energy dependence on the field strength.
In contrast, the magnetic quantification associated to the outer surface is introduced to the $B_{0}$-proportional LLs.
It can be deduced that at a critical field the whole inner-surface states are exactly quantized to the inverted LLs, and the inverse of the critical field strength exhibits a periodicity with the total numbers of the inverted LLs because of the constant enclosed states. Apparently,
the particular band structures characterize the magneto-absorption spectrum.

As a signature of the ABC-stacked graphene, the magnetic-field-dependent spectrum exhibits many anticrossings of the second group of LLs, as shown in Fig. 17(a). The reason is attributed to Wigner-von Neumann non-crossing rule, which indicates that two multi-mode LLs avoid to cross each other once they simultaneously have certain identical modes.
From a qualitative perturbation analysis, the three factors, $\beta_{2}$, $\beta_{3}$ and $\beta_{5}$, are deduced to cause the hybridization among LLs.
Their subenvelope functions have quantum-number difference by the multiple of 3\cite{PRB83;165443,RSCAdv4;56552,PRB90;205434,JPSJ17;808}.
In principle, for Landau states away from the anticrossing regions in Fig. 17(b), the subenvelope functions for all sublattices are well-behaved and described by the Hermite-functions.
For example of $n_{2}^{c}=3$, the quantum modes for sublattices in the order of $\{A_{o}^{1}, B_{o}^{1}, A_{o}^{2}, B_{o}^{2}, A_{o}^{3}, B_{o}^{3}\}$ are, respectively, 4, 5, 3, 4, 2 and 3.
It should be noticed that the two subenvelop functions established on two sublattices in the same layer, e. g., $A_{o}^{1}$ and $B_{o}^{1}$, and those on the vertical sublattices between adjacent layers, e.g., $A_{o}^{1}$ and $B_{o}^{2}$, are, respectively, characterized by the quantum-mode differences of 1 and 0.

At the region of LL anticrossings, the subenvelop functions are perturbed due to a significant hybridization with $n_{2}^{c}=0$ LL according to the difference of 3 in their quantum modes, as evidenced by the continuous change for the components of the main and side modes in the anticrossing region.
Nevertheless, away from this region, the functions convert to well-defined again.
In addition to the well-behaved LLs, the perturbed LLs are expected to enrich and diversify the magneto-optical properties satisfying the extra selection rules of modulo 3.
What should be especially needed is to focus on the small region under the magnetic fields weaker than the critical strength, where the perturbed-LL transitions lead to a considerable amount of extra peaks satisfying the optical selection rules of modulo 3.
Furthermore, their intensities are strongly increased by the degree of the hybridization among the LLs.

Feature-rich optical absorption spectrum is presented in ABC-stacked trilayer graphene, as attributed to the particular LL spectrum.
One can obtain the spectral absorption function $A(\omega)$, Eq. (3), by substituting
the velocity matrix evaluated from the wavefunctions and the derivative of the Hamiltonian over all possible transitions in the first Brillouin zone.
It is shown that occupied Landau states in any of the three valence groups can be excited to the three conduction groups, once the excitation channels satisfy the generalized selection rules (see Chapter 2).
The absorption peaks can be classified into nine categories resulting from three intragroup and six intergroup LL transitions, as shown in Fig. 18 (a) for $B_{0}=40$ T, where the black, red, and green color numbers, respectively, indicate the excitations from the first, second, and third LL groups.
In general, many of the absorption peaks exhibit the twin-peak structures as a result of the two kinds of excitation channels with a tiny energy difference in the asymmetric LL spectrum.
However, for some categories, the threshold channels exclusively from the three Dirac-point related LLs, $n_{1}^{v}=0, n_{1}^{c,v}=1$, and $n_{1}^{c}=2$, give rise to the single-peak structures according to the Fermi-Dirac distribution, i. e., $\omega_{1,1}^{3^v2^c}$, $\omega_{1,3}^{0^v0^c}$, and $\omega_{2,1}^{1^v2^c}$.
The frequency spacing of twin peaks may be observable, providing a way to verify the asymmetry between valence and conduction LLs.
It should be noticed that the magneto-absorption spectrum exhibits some unusual absorption peaks due to the transitions between the perturbed LLs in the region of the LL anticrossings, as indicated by the dashed ellipse in Fig. 16.
Induced by the side modes of the hybridized LL ($n^{v}_{3}=0$), such extra peaks satisfies the special selection rules of modulo 3.
Moreover, the induced peak intensity varies with the degree of the hybridization.
For example, considering the transitions from the $n^{v}_{3}=0$ LL, one can find an extra twin peak near 0.96 eV that includes the $n^{v}_{3}=0\rightarrow n^{c}_{2}=3$ and $n^{v}_{2}=3\rightarrow n^{c}_{3}=0$ excitations, in addition to the regular twin peak $\backsim0.78$ eV resulting from $n^{v}_{3}=0\rightarrow n^{c}_{2}=0$ and $n^{v}_{2}=0\rightarrow n^{c}_{3}=0$ excitations.
On the other hand, the intensities of the regular absorption peaks in ABC-stacked graphene are non-regular with respect to the photon frequency, in contrast to the uniformity in monolayer graphene.
The main reason is attributed to the relationship of the phase and amplitude between the subenvelope functions, depending strongly on the stacking configurations.

The frequency dependence of the low-lying LL excitations on the field strength deserves a closer examination for exploring the spectral characteristics of ABC-stacked graphene.
Neither the intergroup nor the intragroup categories can be described as a simple square-root or linear formula referred to as the behavior of monolayer or bilayer graphene, as shown in Fig. 18(b).
The single-peak channels, resulting from the specific excitations from the three Dirac-point-related LLs, determine the threshold frequencies of the absorption categories associated with the first group of LLs.
The twin-peak spacings, due to the asymmetry in the LL energy spectrum, are more obvious at lower frequencies.
At the field strength of the anticrossing point ($B_{0}\simeq20$ T), a blue or red shift of the twin peak also accompanies the intensity variation, because the two anticrossing LLs, $n^{c}_{2}=0$ and $n^{c}_{3}=1$, are reversed.
The extra peak has higher intensity when the side mode gets stronger with an increment of $B_{0}$, and the decreasing intensity reflects the gradual disappearance of the side mode after $B_{0}$ passes through the strongest anticrossing point \cite{ACSNano4;1465}.
In the first intragroup category, the absorption frequencies grow quickly with an increment of the field strength, while they approach the threshold value of $\omega_{32}$ $\simeq2\beta_{2}$ at a sufficiently low field strength, corresponding to the energy spacing between the partial flat bands along the $K\Gamma$ direction.
On the other hand, the intergroup absorption peaks appear in between the intragroup ones.
At certain frequency and magnetic-field ranges, the intergroup categories exhibit non-monotonic ( e.g., $\omega_{21}^{1^{v} 0^{v}}$; $0.30$ eV $\lesssim \omega\lesssim0.35$ eV and $B_{0}\simeq20$ T) or non-continuous ($0.30$ eV $\lesssim \omega\lesssim0.5$ eV and $B_{0}\simeq20, 55; 70$ T's) dependence of frequency on $B_{0}$, the peculiar phenomena closely correlating with the particular energy spectrum that includes inverted and anticrossing LLs.
Furthermore, both the dependences can also be observed for those higher-frequency excitation channels associated with the second group of LLs.

\subsection{ Electric field enriched optical properties}

In ABC-stacked trilayer graphene, the application of a perpendicular electric field significantly leads to the drastic changes in the energy dispersion, band gap, band spacing, anticrossing and band-edge states shown in Fig. 19(a)\cite{NatPhys7;944}, as well as, enriches absorption spectra with different peak structures in Fig 19(b).
As a result of the special surface-localized states near the Fermi level, the energy gap is proportional to the potential difference between two outmost layers under a moderate electric field.
The first absorption peak directly reflects the optical gap transitions and is used to determine the value of $E_{g}$, as observed in the AB-stacked bilayer graphene.
The second peak, associated with the partially flat subband and the distorted sombrero-shaped subband, exhibits an observable blue shift.
By contrast, the spectral characteristics of AA- and AB-stacked trilayer graphene remain similar under the electric field.
In AB-stacked trilayer graphene, $\mathbf{E}$ field increases the overlap between valence and conduction bands, and enhances the optical conductivity; in AA-stacked trilayer graphene, it critically changes the Dirac-point energies, Fermi momenta and Fermi velocity, and modulates the forbidden optical transition regions.

The four-fold degenerate LLs are split into two-fold degenerate LLs because of the $\mathbf{E}$-induced destruction of the mirror symmetry in the $xy-$plane and the non-equivalence of the two sublattices in identical projections.
Each group of LLs becomes two subgroups, in which the splitting LLs are characterized by two different localization centers of 1/6 and 2/6 unit cell (4/6 and 5/6 unit cell).
The dominating sublattice, in determining quantum numbers, remains unchanged for each subgroup, i.e., two inversely symmetric sublattices for two split subgroups.
As more inter-LL channels are excited, e. g. $B_{0}=40$ T and $E_{0}=0.3$ eV/layer, the absorption spectrum shows double peaks, identified as $\alpha$ and $\beta$ types for the inter-LL transitions from the former and the latter subgroups, respectively (blue and purple curves in Fig. 19(c)).
At a moderate applied electric field $\sim0.06$ eV/layer, coexistence of the twin- and double-peak structures are displayed, as shown in Fig. 3 of \cite{RSCAdv5;80410}.
However, with a further increment of $V_{g}$, a lot of prominent twin-peak absorption peaks gradually change into double-peak ones, a transition being attributed to the spatial symmetry breaking of the subenvelope functions.
In other words, the two peaks in a double-peak structure are ascribed by the inter-LL transitions from the distinct centers of 2/6 and 4/6 (different valleys).

The $\alpha$- and $\beta$-type absorption peaks display dramatic changes because of the breaking of the stacking symmetry that causes sharp contrast of the amplitudes and phases between their non-equivalent subenvelope functions (detailed discussions in \cite{RSCAdv5;80410}).
Specially, the threshold peak transforms from single-peak to double-peak structure (indicated by the black arrow), owing to the Fermi-Dirac distribution and the generalized selection rule (chapter 2).
The electric field also induces intragroup LL anticrossings and thus extra double peaks that exhibit abnormal electric-field-dependent absorption frequencies.
However, such peaks can be suppressed or enhanced by varying the electric field strength\cite{RSCAdv5;80410}.
The induced peak intensity is determined by the weight of the side mode in the progressive LL. The aforementioned optical characteristics can be verified by optical spectroscopy.

The electric-field-dependent low-frequency absorption spectrum of ABC-stacked trilayer graphene has been observed from samples on SiO$_{2}$/Si substrates \cite{RSCAdv5;80410}.
The results provide spectroscopic evidence of gap opening and the blue (or red) shift of low-frequency absorption peaks at different gate voltages.
The measured gap is compatible to that from transport and STM measurements \cite{PRB87;165102} and consistent with the theoretical calculations based on the tight-binding model over a wide energy range\cite{PRB79;035421,JPCM18;5849,PRB74;161403}.
While the sequence of the first group of LLs was confirmed by the Hall transport measurements\cite{PRL107;126806,PRB88;121302,NatPhys7;953}, the further experimental evidence on the magneto-optical spectrum is needed for verifying the inter-LL excitations in ABC-stacked graphene, which display different characteristics than the massless and massive Dirac fermions in terms of the intensity, number, frequency and selection rule of the spectral peaks. Also, the features due to the electric-field induced symmetry breaking that have not yet been observed deserve further experimental examination.
The optical spectrum over a wide frequency range can be used to verify the total of nine categories consisting of single and twin peaks, i.e., three intragroup and six intergroup categories. The peak intensities are different, and in general, the intragroup ones are higher than the intergroup ones. In addition, as a result of the LL anticrossing, the extra peaks exhibit a small range of non-continuous intensities and frequencies during the variation of the magnetic field (Fig. 18 (b)), in which the strongest intensity and blue and red shifts occur at the anticrossing point. At a sufficiently low magnetic fields, the characteristic peaks with non-monotonic frequencies reflect the magnetic quantization of the sombrero-shaped subbands, and the first threshold single channel can be used to determine the vertical next-nearest interlayer atomic interaction $\beta_{2}$.

\section{  AAB-stacked graphene}

The AAB stacking, the second trilayer stacking characteristic, is one of the important stacking configurations.
This system has been successfully synthesized by different experimental methods, such as,
electrostatic-manipulation STM \cite{PRB48;17427,PRB86;085428,PRB50;1839}, mechanical cleavage \cite{SurSci601;498,PRB75;235449}, liquid phase exfoliation of natural graphite \cite{APL102;163111}, and the growth on silicon carbide substrate \cite{PRB79;125411}
or metal surface \cite{APL107;263101}.
Specifically, the stacking configuration can be continuously changed by the STM tip.
The first-principle calculations show that the ground state energies of the four stacking configurations are -55.832866 eV for AAA stacking, -55.857749 eV for AAB stacking, -55.862386 eV for ABA stacking; -55.864039 eV for ABC stacking.
There exist only the small energy differences among the four typical stackings, and the AAB stacking is more stable than the AAA one.
As for the trilayer AAB stacking, only a few theoretical studies are focused on the low-energy electronic \cite{Carbon94;619,PRB75;235449} and optical properties \cite{Arx;1509.02253}.
Nine interlayer hopping integrals are required to characterize the unique band structure (Chap. 6.1).
This clearly illustrates that stacking configurations dominate interlayer atomic interactions and thus essential physical properties.

The electronic structure consists of three pairs of valence and conduction bands corresponding to, oscillatory, sombrero-shaped, and parabolic ones.
As to the oscillatory pair, each subband has three circular constant-energy contours in the energy-wave-vector space.
It is first revealed in the AAB stacking, implying that it is exclusive among the trilayer systems.
Similar characteristic for the sombrero-shaped one with two constant-energy loops.
The unusual energy dispersions will induce the complex magnetic quantization because of the multi-constant-energy loops.
This will be reflected in the extraordinary features of the magneto-optical spectra.
Without the external fields, any excitation channels between the valence and conduction bands are available; furthermore, the asymmetric and symmetric peaks are, respectively, shown in the low- and middle-frequency absorption spectrum.
The magneto-absorption spectra are largely enriched by three intragroup and six intergroup inter-LL optical excitations.
The special relations in phases and amplitudes of LL wavefunctions result in significant difference in intensities between intra- and inter-group transition peaks.
The absorption spectra exhibit the single- or twin-peak structures determined by the quantum modes, LL energy spectra and Fermion distribution.
Moreover, the splitting LLs arising from the destruction of inversion symmetry can diversify the magneto-absorption spectra.
These characteristics are absent in the other stacking systems.
In general, the excitations related to the LLs with well-behaved wavefunctions satisfy the general optical selection rule.
Nevertheless, many extra single peaks are caused by sufficiently strong side modes during the LL anti-crossings.
The important differences among four typical stacking trilayer graphenes mostly lie in the number of excitation categories and the main features of absorption peaks.

 \subsection{ Electronic and optical properties without external fields}

For the AAB stacking, the first two layers are arranged as the AA-stacked configuration while the third one is the shift of the first (or the second) layer by a distance of $b$ along the armchair direction.
The zero-field Hamiltonian built from the six tight-binding functions of the $2p_{z}$ orbitals is dominated by the intralayer and the interlayer atomic interactions $\gamma_i^{'}s$.
In order to simulate the unusual energy dispersions, the 10 atomic hopping integrals are used in the calculations (Fig. 1(f)).
$\gamma_{0}=-2.569  $ eV represents the nearest-neighbor intralayer atomic interaction, ($ \gamma_{1}=-0.263$ eV, $\gamma_{2}=0.032 $ eV) are related to the interlayer atomic interactions between the first and second layers; ($ \gamma_{3}=-0.413$ eV, $\gamma_{4}=-0.177 $ eV, $ \gamma_{5}=-0.119$  eV) denote those between the second and third layers, ($ \gamma_{6}=-0.013  $ eV, $ \gamma_{7}=-0.0177 $ eV, and $ \gamma_{8}=-0.0119  $ eV) correspond to those between the first and third layers, and $ \gamma_{9}=-0.012 $ eV accounts for the difference in the chemical environment of A and B atoms.
Only $\gamma_1$, $\gamma_3$, and $\gamma_6$ belong to the vertical interlayer atomic interactions.

The AAB stacking possesses a unique electronic structure, in which three pairs of conduction and valence bands have special energy dispersions (Fig. 20(a)).
The occupied valence bands are somewhat asymmetric to the unoccupied conduction ones about $E_F=0$ because of the interlayer atomic interactions.
In the vicinity of $E_F$, the first pair of subbands $S_1^{c,v}$ exhibit strong oscillatory energy dispersions.
The conduction and valence states start from the extreme points of about $\pm 4$ meV at the K point.
Along the KM or the K$\Gamma$ direction, they reach the second ones of $\sim \pm$ 58 meV, further exhibit the oscillatory variations at the third ones with $\sim \pm$ 4 meV, and then change monotonously.
There exist three constant-energy contours within $\pm$ 58 meV and a narrow energy gap of $ E_g\sim 8$ meV, being absent in other stacking configurations.
Such peculiar characteristics can be verified by ARPES \cite{PRL103;226803}.
The circular loop closest to the K point has a vanishing radius  at the band-edge states so that it does not contribute any special structure in DOS.
This is similar to that in a Dirac-cone structure.
Specifically, the other two circular loops could be regarded as the 1D energy bands near the band-edge states; that is, they can display the square-root asymmetric peaks in DOS and spectral structures.
These will induce very complicated LL energy spectra so that the low-energy expansion of the effective-mass approximation is impossible and can not deal with the magnetic quantization.
Concerning the sombrero-shaped bands $S_2^{c,v}$, the energy difference between the two extreme points is quite narrow, only about 5 meV.
Away from $E_F$, the monotonic parabolic bands $S_3^{c,v}$ possess the band-edge states at about $\pm 0.49$ eV.
The specific electronic structures distinguish the AAB stacking from the other ones.
Moreover, the generalized tight-binding model is consistent with the first-principle calculations in energy dispersions (dashed curves)\cite{Carbon94;619}.

The unique energy dispersions can enrich optical properties.
There are nine kinds of excitation channels between any valence and conduction bands, in which the plentiful structures are revealed in absorption spectrum (Fig. 20(b)).
At lower frequency, the two asymmetric peak structures appear at $\omega \approx 0.01$ eV and 0.11 eV.
This is mainly owing to the fact that each oscillatory subband has three circular constant-energy loops in energy-wave-vector space which are related to concave and convex dispersions near the band-edge states, respectively (Fig. 20).

The optical threshold frequency is determined by the narrow energy gap.
The latter corresponds to the oscillation bandwidth of the first pair of energy subbands near the K point.
A very predominant sharp peak, which arises from the band-edge states of energy subbands ($S_1^{v}$, $S_2^{c}$) $\&$ ($S_2^{v}$, $S_1^{c}$), displays at $\omega \approx 0.32$ eV.
Its strong intensity is due to the very high DOS associated with three valleys in each oscillatory subband.
As to the weak peak at $\omega \approx 0.525$ eV, it is contributed by the merged excitation channels of ($S_2^{v}$ $\rightarrow$ $S_2^{c}$, $S_1^{v}$ $\rightarrow$ $S_3^{c}$, $S_3^{v}$ $\rightarrow$ $S_1^{c}$).
Moreover, the frequency difference between the optical transitions of $S_2^{v}$ $\rightarrow$ $S_3^{c}$ and $S_3^{v}$ $\rightarrow$ $S_2^{c}$ in the asymmetric band structures further leads to two distinguishable peaks at $\omega \approx 0.67$ eV and 0.79 eV, respectively.
Especially, the excitations of $S_3^{v}$ $\rightarrow$ $S_3^{c}$ do not exhibit the special structure because of the extremely small velocity matrix elements.
For the middle frequencies, there are many predominant peaks due to the saddle points.
The absorption spectrum of the AAB stacking presents more spectral structures at the low- and middle-frequency absorption spectrum compared with those of the other stackings.
The nine interlayer atomic interactions can account for the configuration-created spectral structures.

 \subsection{  Magneto-electronic and magneto-optical properties }

The magnetic Hamiltonian, with $12R_{B} \times 12R_{B} $ matrix elements, are characterized by the non-vanishing ones:

\begin {equation*}
\langle B_{j}^{1} |H|A_{i}^{1}\rangle = \gamma_{0} \sum  { \frac{1}{N} exp [ i\vec{k}.(\vec{R}_{A_{i}^{'}}- \vec{R}_{{B}_{j}^{'}}   )] exp[\frac{2i\pi}{\phi_0} (G_{  \vec{R}_{{B}_{j}^{'}}} -  G_{  \vec{R}_{{B}_{j}^{'}}})] }\\
\end{equation*}
\begin{equation}
=\gamma_{0}t_{1,i}\sigma_{i,j} + \gamma_{0}q\sigma_{i,j+1},
\end {equation}
 \begin{equation}
\langle B_{j}^{2} |H|A_{i}^{2}\rangle
=\gamma_{0}t_{1,i}\sigma_{i,j} + \gamma_{0}q\sigma_{i,j+1},
\end{equation}

 \begin{equation}
\langle B_{j}^{3} |H|A_{i}^{3}\rangle
=\gamma_{0}t_{3,i}\sigma_{i,j-1} + \gamma_{0}q\sigma_{i,j},
\end {equation}

 \begin{equation}
\langle A_{j}^{2} |H|A_{i}^{2}\rangle =\langle A_{j}^{3} |H|A_{i}^{3}\rangle
=\gamma_{9}\sigma_{i,j},
\end {equation}

 \begin{equation}
\langle A_{j}^{2} |H|A_{i}^{1}\rangle =\langle B_{j}^{2} |H|B_{i}^{1}\rangle =\gamma_{1}\sigma_{i,j},
\end {equation}

 \begin{equation}
\langle B_{j}^{1} |H|A_{i}^{2}\rangle =\langle A_{j}^{1} |H|B_{i}^{2}\rangle =\gamma_{2} t_{1,i}\sigma_{i,j},
\end {equation}

 \begin{equation}
\langle A_{j}^{3} |H|A_{i}^{2}\rangle =\gamma_{3}\sigma_{i,j},
\end {equation}

 \begin{equation}
\langle B_{j}^{3} |H|B_{i}^{2}\rangle =\gamma_{5}t_{2,i}\sigma_{i,j},
\end {equation}

 \begin{equation}
\langle B_{j}^{2} |H|A_{i}^{3}\rangle
=\gamma_{4}t_{1,i}\sigma_{i,j} + \gamma_{4}q\sigma_{i,j+1},
\end {equation}

 \begin{equation}
\langle A_{j}^{2} |H|B_{i}^{3}\rangle
=\gamma_{4}t_{3,i}\sigma_{i,j-1} + \gamma_{4}q\sigma_{i,j},
\end {equation}

 \begin{equation}
\langle A_{j}^{3} |H|A_{i}^{1}\rangle
=\gamma_{6}\sigma_{i,j},
\end {equation}

 \begin{equation}
\langle B_{j}^{3} |H|B_{i}^{1}\rangle
=\gamma_{8}t_{2,i}\sigma_{i,j} + \gamma_{8}q\sigma_{i,j+1},
\end {equation}

 \begin{equation}
\langle B_{j}^{1} |H|A_{i}^{3}\rangle
=\gamma_{7}t_{1,i}\sigma_{i,j} + \gamma_{7}q\sigma_{i,j+1},
\end {equation}

 \begin{equation}
\langle A_{j}^{1} |H|B_{i}^{3}\rangle
=\gamma_{7}t_{3,i}\sigma_{i,j-1} + \gamma_{7}q\sigma_{i,j},
\end {equation}
where the four independent phase terms are \\
$t_{1,j} = exp \{ i[ -(k_{x}b/2) - (\sqrt {3}k_{y}b/2) +\pi\phi (j-1+ 1/6) ] \}    \\
+ exp \{ i[ -(k_{x}b/2) + (\sqrt {3}k_{y}b/2) -\pi\phi (j-1+ 1/6) ] \}, $  \\
$t_{2,j} = exp \{ i[ -(k_{x}b/2) - (\sqrt {3}k_{y}b/2) +\pi\phi (j-1+ 3/6) ] \}    \\
+ exp \{ i[ -(k_{x}b/2) + (\sqrt {3}k_{y}b/2) -\pi\phi (j-1+ 3/6) ] \}, $  \\
$t_{3,j} = exp \{ i[ -(k_{x}b/2) - (\sqrt {3}k_{y}b/2) +\pi\phi (j-1+ 5/6) ] \}    \\
+ exp \{ i[ -(k_{x}b/2) + (\sqrt {3}k_{y}b/2) -\pi\phi (j-1+ 5/6) ] \}; $  \\
$q=exp\{ ik_{x}b\}$.\\

The destruction of the inversion symmetry in AAB stacking leads to certain important differences between the LL wavefunctions localized at 2/6 (5/6) and 1/6 (4/6) centers in terms of LL energy spacings, quantum numbers, spatial distributions, and anti-crossings.
At 2/6 center, three groups of unoccupied conduction and occupied valence LLs are initiated at 0 (0), -0.23 (0.24) and -0.58 (0.55) eV (Fig. 21).
All the LL wavefunctions are almost well-behaved in the spatial distributions where the main modes are much stronger than the side ones.
Their quantum numbers are, respectively, determined by the $A^1$, $B^2$ and $A^2$ sublattices with the dominating subenvelope functions.
For the first group, the ordering of quantum number is monolayer-like, in which the middle $n_1^{c,v}=0$ one is located right at $E_F$.
Dissimilarly, $n_2^{c,v}$ begins from $n_2^{c,v} = 1$, then $n_2^{c,v} = 0$, $n_2^{c,v} = 2$, $n_2^{c,v} = 3$, and so on.
As for $n_3^{c,v}$, it normally increases from zero for the higher conduction states, and the deeper valence states.
On the other hand, at 1/6 center, the quantum numbers of the first LL group could be determined by $B^3$ sublattice while those of the other groups are based on the same sublattices as at 2/6 center.
However, the quantum numbers of the low-lying LLs are not easy to identify from the non-well-behaved spatial distributions, since the complex LL anti-crossings cause the subenvelope functions to oscillate abnormally.
Remarkably, the quantum numbers are out of normal ordering, different to those of monolayer graphene.
The LL splitting in AAB stacking is expected to diversify the magneto-absorption spectra.

The AAB stacking configuration induces the special relationships between different subenvelope functions in amplitudes and phases.
The amplitudes of the six subenvelope functions, which depend on the interlayer atomic interactions, are non-equivalent.
Explicitly, at 2/6 center, the amplitude of $A^3$ sublattice is much weaker than those of the others in the first LL group (Fig. 21).
Similar results are also revealed for $A^2$ and $B^2$ sublattices in the second group; $B^3$ sublattice of the low-lying LLs in the third group.
Moreover, the $A^1$ and $A^2$ ($B^1$ and $B^2$) sublattices have similar subenvelope functions with the same number of zero points, mainly owing to the same (x,y) projections.
Concerning the phases in the first and second LL groups, those of the $A^1$ and $A^2$ sublattices are identical for all the conduction states but have the distinct signs for the valence ones in the first and second LL groups.
The opposite is true for the third LL group.
These phase relationships can also be observed for the $B^1$ and $B^2$ sublattices in the first and third groups but opposite in the second group.
The unique features of the subenvelope functions can greatly enrich the absorption spectra.


The AAB stacking reveals the feature-rich $B_0$-dependent LL energy spectrum.
Three groups of LL energies exhibit monotonic and entangled $B_0$-dependencies.
The conduction and valence LL energies monotonously approach to the initial energies in the decrease of field strength from $B_0=60$ T (Fig. 22).
However, for the first and second groups, the unusual $n^{cv}$ ordering and $B_0$-dependence occur below the critical magnetic fields.
There exist the oscillatory $B_0$-dependent energy spectrum, directly reflecting the strong competition in the magnetic quantization of multi-contour energy loops (Fig. 20).
Furthermore, a simple relationship between LL energy spacing and $B_0$ or $n_1^{c,v}$ is absent.
Specifically, the diverse anticrossing and crossing phenomena are revealed in distinct ($E^{c,v}, B_0$)-regions including the intra-group LL anticrossings/crossings from the first and the second groups (at low $B_0$), and the inter-group LL anticrossings/crossings between any two groups (at $B_0 >$ 80 T; not shown).
The main modes of the two anticrossing LLs have a difference of $\Delta n = \pm 3i$ ($i$ is an integer).
These special features are expected to induce the unique magneto-absorption frequency spectra.


Three intragroup and six intergroup optical excitations are clearly revealed in the spectra as indicated in Fig. 23.
They are in sharp contrast with each other in the peak structures, spectral intensities, and energy spacings in twin peaks.
The peak structures are determined by the quantum modes, LL energy spectra and Fermion distribution.
In general, the intragroup excitations present very weak twin peaks.
On the contrary, the intergroup absorption peaks are single peaks with monolayer-like strong intensities.
Exceptionally, the transitions between the first and third LL groups exhibit twin peaks in which one peak is very weak while another is rather strong.
Regarding the intensities of absorption peaks, the intergroup optical excitations are much stronger than the intragroup ones.
The amplitude and phase relationship between the $A^l$ ($B^l$) subenvelope function of the initial state and the $B^l$ ($A^l$) one of the final state is responsible for the differences in spectral intensities.
For example, the intragroup peak $\omega_1^{1^v2^c}$ at $\omega\approx 0.06$ eV is about 1/15 the intergroup one $\omega_{21}^{1^v1^c}$ at $\omega\approx 0.25$ eV.
For the former, the $|B^1\rangle$ of the $n_1^v=1$ initial state and the $\langle A^1|$ of the $n_1^c=2$ final state have the opposite phases, while $|B^2\rangle$ and $\langle A^2|$ possess the same phases.
The distinct signs in $\langle A^1 | B^1 \rangle$ and $\langle A^2 | B^2 \rangle$ greatly reduce the velocity matrix elements and thus the peak intensity.
On the other hand, strong peak intensity in the latter reflects the fact that all the dominating subenvelope functions of the $n_2^v=1$ initial and $n_1^c=1$ final states on the sublattices, ($B^1$, $A^1$), ($B^2$, $A^2$); ($A^3$, $B^3$), have the similar phase relationships.

At 2/6 center, the intragroup inter-LL excitations possess a lot of low-intensity absorption peaks which mostly belong to the twin-peak structure (Fig. 24).
Exclusively, the threshold peak of $\omega_1^{0^v 1^c}$ ($\omega \approx 0.01$ eV) is a single peak with ignorable intensity due to the extremely weak subenvelope functions of $n_1^v=0$ LL in the $B^1$, $A^2$ and $A^3$ sublattices.
Distinctively, the threshold peaks of ($\omega_2^{0^v 1^c}$, $\omega_2^{1^v 0^c}$) and ($\omega_3^{0^v 1^c}$, $\omega_3^{1^v 0^c}$) are twin peaks in which the former are relatively strong while the latter are comparable with the other intragroup excitation peaks.
This is because all the subenvelope functions of the $n_2^v=0$ (1) initial and $n_2^c=1$ (0) final states on the $A^l$ and $B^l$ sublattices have the same (opposite) phases.
In general, three categories of intragroup inter-LL excitations exhibit similar characteristics, including peak intensity, twin-peak structure, and selection rule ($\Delta n = \pm1$).

For the intergroup inter-LL excitations, beside the asymmetry of LL energy spectra, the quantum modes are responsible for the peak structures.
Specifically, the absorption peaks associated with the second LL group are single ones.
Concerning the $n_1^v \rightarrow n_2^c$ $\&$ $n_2^v \rightarrow n_1^c$ transitions, the initial and final states of available excitations satisfy $n_1^{c,v} = n_2^{v,c}$.
Remarkably, the threshold peak of $\omega_{1,2}^{0^v 0^c}$ and $\omega_{2,1}^{0^v 0^c}$ are merged together at $\omega = 0.265$ eV and become a higher-intensity single one.
As to the $n_2^v \rightarrow n_3^c$ and $n_3^v \rightarrow n_2^c$ excitations, the single peaks mainly come from the inter-LL transitions with $| n_2^{c,v} - n_3^{v,c}| = 2$ and 0, in which the former are much higher than the latter as a result of the special phase relations between the subenvelop functions.
On the other hand, the $n_3^{v}\rightarrow n_1^{c}$ and $n_1^{v}\rightarrow n_3^{c}$ inter-LL transitions satisfy the selection rule of $| n_1^{c,v} - n_3^{v,c}| = 1$, identical to that of the intragroup ones.
Their absorption spectra present the twin-peak structures except that the threshold peak of $\omega_{3,1}^{0^v 1^c}$ is a single form, since the $\omega_{3,1}^{1^v 0^c}$ transition is forbidden by the Fermi-Dirac distribution function.
The  intensities of twin peaks quite differ from each other.
One of them is comparable to the intergroup peaks, while another is similar to the intragroup ones.

The splitting of LLs results in critical differences between 1/6- and 2/6-center magneto-absorption spectra in terms of intensity, frequency, and structure of peaks.
These arise from the abnormal ordering of quantum number and the significant side modes during the low-lying LL anti-crossings.
The threshold peak $\omega_1^{0^v1^c}$ associated with the 1/6 center is quite prominent at $\omega \approx 0.15$ eV.
There are a lot of low-frequency single peaks when the magnetic field corresponds to the range of LL anti-crossings \cite{Carbon94;619}.
For example, the $n_1^{c,v}=0$ LL also has the non-negligible side mode of three zero points, being responsible for the anti-crossing of the $n_1^{c,v}=0$ and $n_1^{c,v}=3$ LLs and the appearance of the single peaks $\omega_1^{0^v2^c}$ and $\omega_1^{0^v4^c}$.
Furthermore, many pairs of LLs in the anti-crossings such as ($n_1^{c,v}=1$, $n_1^{c,v}=4$), ($n_1^{c,v}=2$, $n_1^{c,v}=5$), and so on, present several extra single peaks.
Consequently, similar characteristics also occur in the initial absorption peaks of the intergroup transitions between the first LL group and others.
Different from the absorption spectrum at 2/6 center, the $n_2^v \rightarrow n_3^c$ and $n_3^v \rightarrow n_2^c$ excitations only survive for the inter-LL transitions with $| n_2^{c,v} - n_3^{v,c}| = 0$.
In short, the LL anti-crossings play an important role in distinguishing the magneto-absorption spectra due to the splitting LLs.

The optical-absorption peak frequencies of AAB stacking exhibit the plentiful $B_0$-dependent spectrum.
Both intra- and inter-group excitation frequencies have non-monotonic $B_0$-dependence except for the $3^v \rightarrow 3^c$ intragroup excitation category.
As $B_0$ gradually decreases from 60 T, the absorption frequencies normally decline.
However, they start to grow after certain critical magnetic fields, such as $\underline{2}1$ at $B_0 \approx$ 26 T and $\color{red}{\underline{1}} \color{black}{1}$ at $B_0 \approx$ 38 T (Figs. 25(a) and 25(b)).
This is associated with the abnormal field-dependent LL energy spectrum magnetically quantized from the oscillatory and the sombrero-shaped energy bands.
It should be noticed that the LL anticrossing phenomena are also presented in the discontinuous absorption frequencies at certain $B_0$'s, e.g., ($\underline{2}1$, $\underline{1}2$) at $B_0 \approx$ 18 T and ($\color{black}{\underline{1}} \color{red}{1}$, $\color{red}{\underline{1}} \color{black}{1}$) at $B_0 \approx$ 25 T (Figs. 25(a) and 25(b)).
The unusual absorption frequency spectrum is easily observed at low field strength, as indicated by more absorption peaks.
Specifically, the excitation channels of the threshold peaks are dramatically changed from $\underline{0}1$ to $\underline{0}2$, $\underline{4}0$, $\underline{5}0$, and others, being attributed to the frequent intra-group LL anticrossings (Fig. 22).
In regard to the inter-group transitions, the initial peaks, $\color{black}{\underline{0}} \color{red}{0}$ and $\color{red}{\underline{0}} \color{black}{0}$, are forbidden for $B_0$, respectively, smaller and larger than 17 T, by the requirement of the Fermi-Dirac distribution.
Furthermore, they have weakly dispersive relations between $\omega_a$ and $B_0$, since the LL energies of $n_1^{v,c}=0$ and $n_2^{c,v}=0$ are almost initiated at the specific energies.
Significantly, the quantum number ordering plays an important role in the $B_0$-dependence.
The inter-group threshold peak, in the range of 28 T $<B_0<$ 51 T, is not the initial one but replaced by the second excitation channel $\color{red}{\underline{1}} \color{black}{1}$.
This characteristic comes from the unusual $n_2^{c,v}$ ordering of the second LL group, which is assigned $n_2^{v,c}$ = 1, 0, 2, 3, and so on (Fig. 22).
Concerning the twin-peak spacing arising from the asymmetry of valence and conduction LL energy spectra, it remarkably depends on the field strength.
At certain ranges of $B_0$, twin peaks might be merged into a single structure.
This feature is clearly illustrated for both intra- and inter-group inter-LL excitations, e.g, ($\underline{1}2$, $\underline{2}1$) for $B_0 <$ 24 T, ($\color{red}{\underline{3}} \color{black}{3}$, $\color{black}{\underline{3}} \color{red}{3}$) for $B_0 <$ 14 T, and so on.
The stacking-diversified $B_0$-dependent absorption frequencies are worthy of detailed experimental examinations\cite{PRL98;157402,Nature4;532,PRB78;235408,PRB79;115441,PRL102;037403,
PRL104;176404,PRL98;197403,PRB89;085426,PRL100;087401,ACSNano9;6765,PRL107;036807,
Carbon99;118,Carbon99;212,PRL97;187401,NanoLett12;1749,NanoLett11;164,Carbon46;272,
ACSNano8;10766,NANOLETT14;4548,EPL108;27011,NANOLETT13;1541,PRB78;113407,
PRB76;201401,JACS133;10394}.


\subsection{  Electric field enriched optical properties}

The electronic structure of AAB stacking is drastically altered in the presence of $\mathbf{E}$.
The Coulomb potential difference between graphene layers has a strong effect on band gap as illustrated in Fig. 26(a).
As to the oscillatory and sombrero-shaped energy bands, their oscillation widths are greatly widened.
The band-edge states are getting away from the K point.
This induces the larger constant-energy loops, indicating the higher DOS.
The energy spacings of concave and convex energy dispersions become large.
The parabolic subbands might change into linear ones when the electric field is sufficiently large.
Remarkably, the asymmetry between the valence and conduction subbands is largely amplified by the close cooperation relationship of the complex interlayer atomic interactions and the electric field.
Furthermore, their constant-energy loops (or the band-edge states) correspond to the different wave vectors.

The frequency, intensity and number of spectral structures are very sensitive to the changes in electric-field strength, as shown in Fig. 26(b).
With the increase of $E_0$, the optical gap grows monotonously, and then reach a saturated value at higher fields ($E_0 \ge$ 0.3 eV/layer; blue and green curves).
The intensity of the threshold peak is greatly enhanced, a result of the $E_0$-enlarged constant-energy loops.
The similar result is also revealed in the second peak.
There exists an extra weaker peak between these two prominent ones, reflecting the distinct band-edge states in the valence and conduction bands.
Moreover, the original merged absorption peaks might be clearly separated, such as the splitting from one peak into two ones  related to $S_1^v \rightarrow S_2^c$ $\&$ $S_2^v \rightarrow S_1^c$.
This is caused by the $E_0$-enhanced asymmetric energy spectrum.

The double degeneracy of LLs in AAB stacking keeps the same even under an electric field, i.e., the doubly-degenerate LLs are  characterized by the 1/6 and 2/6 localization centers in a composite field.
However, the electric field can amplify the asymmetry of LL energy spectra, alter the spatial distributions in distinct sublattices and generate more frequent anti-crossings, leading to the drastic changes in the frequency, intensity and number of  magneto-absorption peaks.
There are more observable symmetric peaks as a result of the non-merged inter-LL excitation channels and the side-mode-induced ones, as shown in Fig. 26(c) at $E_0=0.1$ eV/layer and $B_0=$ 40 T for the 2/6 case.

The electronic and optical properties of the AAB stacking are easily modulated by the external fields.
The experimental observations on them are absent up to now.
The unique $E_0$-dependence of the band-edge state energies could be directly verified through the STS \cite{PRB91;155428} and ARPES measurements \cite{PRL98;206802}.
STS is useful in examining the $B_0$-dependent splitting LL energy spectra with the strong asymmetry about $E_F$, the frequent  intra- and inter-group anti-crossings, and the complex $E_0$-dependence.
Moreover, many spectral structures at low and middle frequency, the magneto-absorption peaks with distinct structures, the diverse selection rules, and the $E_0$-enriched optical spectra need further examinations using various optical spectroscopies \cite{PRL98;206802}.

\subsection{ Comparisons among the different trilayer stackings}

The stacking configurations can diversify energy dispersions and the optical properties of trilayer graphenes.
As for the low-lying band structure, there are critical differences among four stacking configurations.
The AAA- and ABA-stacked trilayer graphenes possess special band structures with three pairs of linear bands, and a pair of linear bands and two pairs of parabolic bands, respectively.
Their electronic structures can be, respectively, regarded as the superposition of monolayer graphene, and monolayer and bilayer graphenes.
In contrast, the ABC and AAB stackings exhibit unique energy bands, being absent in monolayer and bilayer systems.
The former has pairs of the partially flat, sombrero-shaped and linear bands, and the latter possesses pairs of the oscillatory, sombrero-shaped, and parabolic ones.
Remarkably, only the AAB stacking is a narrow direct-gap semiconductor, while the others are gapless 2D semimetals.

The diverse energy dispersions are directly reflected in the rich optical spectrum.
For the AAA and ABA stackings, only the intra-Dirac-cone optical excitations can survive as a result of the linearly symmetric or anti-symmetric superposition of the wavefunctions in different layers.
There is solely a shoulder structure in the low-frequency absorption spectra.
As to the former, it is displayed right after a forbidden excitation region under the specific fermion distributions due to the almost symmetric upper and lower Dirac cones.
However, that of the latter comes from the distinct parabolic bands.
In the middle-frequency spectrum, these two systems, respectively, present three and five $\pi$-electronic absorption peaks
coming from the saddle points of three pairs of subbands.
Regarding the ABC and AAB stackings, excitation channels between any valence and conduction bands are available.
The threshold frequency of the former is zero while that of the latter is determined by the narrow energy gap.
The low-frequency spectra exhibit peak and shoulder structures, being different from the above-mentioned two stacking systems.
The AAB stacking presents the most plentiful spectral structures due to the complex interlayer atomic interactions.
Under an electric field,  the spectral characteristics of the  AAA stacking remain similar, but the opposite is true for the other ones. The ABA stacking has the $3\times3$ excitation channels, as observed in the ABC and AAB ones. More special absorption structures are created in these three systems; furthermore, an optical gap is opened for the ABC stacking.

The magneto-absorption spectra are greatly enriched by stacking configurations.
The AAA, ABA, ABC and AAB stackings sharply contrast with one another in the excitation category, structure, intensity, and number of absorption peaks.
This is mainly determined by the special relationships among the subenvelope functions on the six sublattices, the Fermi-Dirac distribution, the symmetric or asymmetric LL energy spectra, and the LL anti-crossings.
The AAA and ABA stackings, only present, respectively, three and five categories of absorption peaks.
The linearly symmetric and anti-symmetric superposition of LL wavefunctions in different layers can account for the absence of inter-group excitations related to the Dirac-cone LLs.
On the other hand, the ABC- and AAB-stacked trilayer graphenes possess nine categories of absorption peaks.
Regarding the spectral form and intensity, the intra-Dirac-cone LL excitations in the AAA and ABA stackings are mainly present in the single peaks with a predominantly uniform intensity, owing to the almost symmetric LL spectra and same velocity matrix elements.
Specifically, the ABA stacking also has the  bilayer-like magneto-absorption spectrum in the twin-peak structure.
The ABC stacking exhibits the twin-peak structures except for the initial absorption peaks.
In general, the intragroup excitations are stronger than the intergroup ones which is in sharp contrast with the AAB stacking.
The AAB stacking displays more complex magneto-absorption spectra with many single and twin peaks, reflecting the complex relationship between the interlayer atomic interactions and the magnetic quantization.
The intensity difference between the intra- and inter-group excitation peaks is more obvious.
Especially, the LL anti-crossings and the splitting LLs at different localization centers significantly increase the number of absorption peaks.
The critical differences among the four stacking systems can provide the useful information in identifying the specific stacking configurations by the experimental measurements \cite{PRL98;206802}.

The magneto-absorption spectra can be diversified by an electric field, depending on the stacking configurations.
The main features of optical properties in AAA stacking are hardly changed under a composite external field while those of the other three are greatly enriched.
The obvious changes of magneto-absorption peaks lie in frequency and intensity.
In the ABA and ABC stackings, extra peaks and double peaks are generated, respectively, by the LL anti-crossings and splitting LLs.
Especially, the former displays more inter-LL excitation categories which is absent in the other stacking systems.

\section{ Sliding bilayer graphene}

In addition to the high-symmetry stackings, the low-symmetry ones deserve systematic researches \cite{PRB89;085426,NATMAT12;3717,NANOLETT13;1541,PRB84;045404,PRL109;236604,PRB88;035428,PRB78;113407,APL101;083101,PRB84;155410,SREP5;17490,PRL99;256802,CARBON99;432,SREP5;10985,JPCC116;8271}. A sliding bilayer graphene (BLG), with various misalignment configurations, is chosen for the detailed discussions about the specific essential properties. During the shift between two graphene layers, the first-principle calculations show that the variation of the total ground state energy is small among the configuration-dependent BLGs \cite{JPCC119;10623}. Moreover, some recent experimental studies have successfully synthesized asymmetric stackings using specific techniques\cite{ACSNANO7;1718}. Stacking boundaries in which one of the graphene layer is shifted with respect to another one are successfully grown using the CVD method \cite{PNAS110;11256}. The sliding of graphene flakes on a graphene substrate can also be initiated by the STM tip to overcome the weak van der Waals interactions \cite{ACSNANO7;1718}. In addition, micrometer-size graphite flakes can slide spontaneously after stirred by a STM tip \cite{PRL108;205503,PRB82;155460}.

The generalized tight-binding model is suitable for studying the magnetic quantization in the high- and low-symmetry BGLs. The variation in stacking configuration can induce a dramatic transformation between the Dirac-cone structure and the parabolic band structure \cite{JPCC119;10623,PRB88;115409,SciRep4;7509,PRB75;201404,RSCADV4;63779}, and thus create unusual absorption spectra in terms of optical gap and low- and middle-frequency special structures. The low-frequency optical properties are significantly modified by an electric field. The low-frequency optical properties are significantly modified by an electric field. Moreover, the LLs of a sliding BLG exhibit the drastic changes in the spatial symmetry, initial formation energy, intergroup anti-crossings, state degeneracy and semiconductor-metal transition. There are three kinds of LLs: the well-behaved, perturbed and undefined LLs, being characterized by a specific mode, a main mode plus side modes, and a disordered mode, respectively. Such LLs are clearly revealed in diverse magneto-optical selection rules. Specially, the undefined LLs frequently present intergroup anti-crossings in the B$_{0}$-dependent energy spectra, and exhibit a large number of absorption peaks in the absence of selection rule. The diverse selection rules, the special B$_{0}$-dependent energy spectra, and the unusual zero-field absorption spectra are closely related to stacking configurations; therefore, the experimental measurements on them can be used to resolve the subangstrom misalignment stackings of BLG.

\subsection{  Electronic and optical properties without external fields}

The distinct stacking configurations of BLG can be tuned by the relative shift between two layers. When a shift happens along the armchair direction in the range of 0 $\le \delta \le $12b/8,  stacking configurations alter among the high-symmetry ones: AA ($\delta$ = 0) $\longrightarrow$AB ($\delta$ =b) $\longrightarrow$ AA$'$ ($\delta$ =1.5b) (Fig. 27).
The C atoms in the AA$'$ stacking have the distinct x-y projections, while they experience the same chemical envirnment, as observed in the AA stacking. Four carbon atoms are included in a primitive unit cell. For all the stacking configurations \cite{JPCC119;10623,JAP73;494,PRL70;1858}, the low-energy electronic properties are dominated by the 2p$_{z}$ orbitals, as being indicated by the agreement between the tight-binding model and the first-principle calculations. There exist the complicated interlayer atomic interactions during the variation of shift. The position-dependent empirical formula is chosen to characterize the intra- and inter-layer atomic interactions:

\begin{equation}
-\gamma_{ij}=
\gamma_{0}
e^{-\frac{d-b}{\varrho}}
[1-(\frac{\textbf{d}\cdot\hat{z}}{d})^{2}]
+\gamma_{1}
e^{-\frac{d-d_{0}}{\varrho}}
(\frac{\textbf{d}\cdot \hat{z}}{d})^{2}
\end{equation}
$\gamma_{ij}$ is the atomic interaction between lattice sites i and j, depending on the distance and the angle of two 2p$_{z}$ orbitals.
$\gamma_{0}$$=$ $-2.7$ eV is the nearest-neighbor hopping integral in the same layer, $\gamma_{1}$$=$ $0.48$ eV the interlayer vertical hopping integral, $\textbf{d}$ the position vector connecting two lattice points, $d_{0}$$=$3.35 $\AA{}$ the interlayer distance, and $\varrho$$=$0.184$b$ the decay length. Equation (31) has been acceptable to describe the interlayer atomic interactions in carbon-related systems, e.g., multi-walled carbon nanotubes \cite{JAP73;494,PRL70;1858}.

The sliding BLG exhibits the dramatic changes in band structures. The different stacking systems possess the overlap of valance and conduction bands (Figs. 28(a)-28(f)); that is, they are semi-metallic systems with a low free carrier density.
The space-inversion symmetry keeps the same in the presence of shift so that each valley in the first Brillouin zone is identical\cite{PRB81;115315}. The K valley is sufficient for a model study.
With a small shift ($\delta$= b/8 in Fig. 28(b)), two pairs of isotropic Dirac cones in the AA stacking (Fig. 28(a)) are changed into the distorted energy dispersions. An eye-shaped stateless region is created near the Fermi level. Band structure is symmetric about the k$_{x}$-axis, since the y$\longrightarrow$$-$y and space-inversion symmetries are unbroken. A pair of band contact points remain as the Fermi-momentum states. The Dirac-cone structures gradually become smoother in the further increase of $\delta$. After the serious distortion, the Dirac points are fully separated near $\delta$ $\cong$ 6$b$/8 (Fig. 28(c)).

A further shift in the range of b $\leq$ $\delta$ $\leq$ 1.5b also leads to the special transformation of energy dispersions. Two pairs of parabolic bands are formed in the AB stacking of $\delta$ =b (Fig. 28(d)), as verified by ARPES \cite{PRB77;195403,PRB88;075406}. The first pair of parabolic energy dispersion nearest to E$_{F}$ is seriously distorted along $\pm$$\hat{k}_{y}$ directions, as shown for $\delta$ =11b/8 in Fig. 28(e). An eye-shaped stateless region is created with two distinct Dirac points at different energies; furthermore, two neighboring conduction (valance) bands have strong hybridization. Finally, such energy bands are transformed into two pairs of isotropic Dirac-cone structures at $\delta$ =1.5b (Fig. 28(f)). Specially, two Dirac cones are slightly tilted, and the Dirac points are deviated from the K point (Fig. 29) \cite{SciRep4;7509}.
The AA$'$ and AA stackings, respectively, possess the non-vertical and vertical Dirac-cone structures. This will induce the critical differences in magnetic quantization and magnetic-optical properties. The above-mentioned dramatic transformation of energy bands is also confirmed by the first-principle calculations \cite{JPCC119;10623}.

The two-dimensional electronic structures, as shown in Figs. 30(a)-30(f) which are depicted along high symmetry points, can also provide the useful information on the electronic properties.
The low-lying isotropic energy dispersions near $E_{F}$ are transformed between isotropic and anisotropic characteristics (insets). The AA stacking presents the linear isotropic dispersions near the K point (Fig. 30(a)). During the variation of shift, they change into the anisotropic distorted ones at $\delta$ = b/8 $\&$ 6b/8 (Figs. 30(b) $\&$ 30(c)) and become the parabolic isotropic ones at $\delta$ = b (Fig. 30(d)). And then, energy dispersions are anisotropic in the range of b $\le$ $\delta$ $\le$ 1.5b (Figs. 30(e) and 30(f)). Except for the AA and AB stackings, other systems present the distorted energy dispersions and highly anisotropic 2D band structures. Regardless of the semi-metallic property of the sliding BLG, tiny energy spacings might exist along K $\to$ M and K $\to$ M$'$ (Figs. 30(b) and 30(c)). Concerning the middle-energy electronic states,  the M and M$'$ points remain as the saddle points in the energy-wave-vector space; that is, they will make much contribution to DOS and absorption spectrum. Their energies become split except for the AA and AB stackings. The 3-fold rotation symmetry is destroyed, while the space and y $\to$ $-$y inversion symmetries are invariant. This accounts for the distinct energies between the two M points and the four M$'$ points in the first Brillouin zone. The splitting energy is observable with a wide range of $\approx$ 0.1$-$0.6 eV, when the sliding BLG significantly deviates from the AA or AB stacking.

Up to now, ARPES measurements have confirmed two vertical Dirac cones of the AA stacking and two pairs of parabolic bands of the AB stacking \cite{PRB77;195403,PRB88;075406}. The feature-rich electronic properties of configuration-dependent BLGs, including the transformation between Dirac-cone structures and parabolic bands, the tilted Dirac cones, the stateless arc-shaped region, the distorted energy dispersions, and the spitting of middle-energy states, can be further verified with ARPES \cite{PRB77;195403,PRB88;075406,PRL105;136804,PRB92;195148}. This is one way to identify the tiny misalignment stackings related to various BLG configurations, as suggested by the experimental examinations on the predicted phonon spectra \cite{ACSNANO7;7151}.

The unusual electronic structues can diversify the absorption spectra. The AA stacking reveals a forbidden excitation zone before the threshold frequency, since the spatial wave-function symmetries lead to the zero velocity matrix elements for the inter-Dirac-cone excitations. Two shoulders are presented in absorption spectra, as indicated in Fig. 31(a).
Their frequencies of 0.64 $\&$ 0.73 eV's are, respectively, proportional to the Fermi velocities of the upper and lower Dirac cones; therefore, the optical measurements on the initial absorption structures could be used to identify the Dirac-cone Fermi velocity. The optical gap vanishes in a small shift ($\delta$ = b/8 in Fig. 31(b)), being replaced by a prominent peak at $\sim$ 0.1 eV and a forbidden region with a width of $\sim$0.4 eV. Such features are associated with the saddle points and the vacant arc-shaped region in the distorted energy bands (Fig. 30(b)). Specifically, the $\delta$ =6b/8 stacking exhibits the gapless and featureless absorption spectrum as a result of the quite different wave vectors in the saddle points of valence and conduction states (Fig. 31(c)). A $\sim$0.3-eV shoulder structure, which comes from the band-edge state excitations between the first and the second pairs of parabolic bands, appears in the AB stacking (Fig. 31(d)). This has been verified by optical measurements on absorption \cite{Nature459;820,PSSB247;2931}, reflection \cite{PRB78;235408,PRB80;165406,PRB79;115441,NANOLETT14;4548}, transmission spectra \cite{PRB78;235408,PRB78;235408,PRL14;126804,APL117;112803} and Raman scattering \cite{PRL97;187401, PRB76;201401}. Moreover, an optical gap is recovered in the large-shift BLGs, e.g., $\delta$ = 11b/8 in Fig. 31(e) and $\delta$ = 12b/8 in Fig. 31(f).
As for the middle-frequency optical spectra, the AA, small-$\delta$ and AB stackings (Figs. 31(a), 31(b); 31(d)),  have two very strong absorption peaks, while the other stackings present four or three ones (Figs. 31(c), 31(e); 31(f)). This directly reflects the existence or destruction of the three-fold rotation symmetry. Obviously, the experimental measurements \cite{PRL97;187401,PRB76;201401,PRB91;045418} on the low- and middle-frequency absorption structures are an efficient method to identify the various stacking configurations in the sliding BLG.

\subsection{  Magneto-electronic and magneto-optical properties}

Three types of LLs, the well-behaved, perturbed, and undefined ones, are presented in the different stacking systems. The main characteristics are identified from the spatial distributions of wave functions and the $B_0$-dependent energy spectra. Both AA and AA$'$ stackings belong to the first type. The first and second types could be found in most stacking systems, such as the AB, ABC and AAB ones, and the sliding BLGs with a small shift. The quantum numbers in the sliding systems are defined by the $B^{1}$ sublattice for the localization center of 2/6. Moreover, the third type is revealed in the specific stacking systems corresponding to the dramatic changes of energy bands. In addition, the degeneracy of LL states remains the same for any shifts.

The LLs of AA and AA$'$ stackings exhibit the regular spatial distributions with the symmetry or anti-symmetry about the localization centers and the normal zero points. Their energies are proportional to $\sqrt{n^{c,v}B_{0}}$, as revealed in the $B_0$-dependent energy spectrum ($\delta$ =b in Fig. 33(d)). However, the important difference between them lies in the localiztion centers. The AA stacking has a fixed localization center of 2/6 (Fig. 32(a)).
As to the AA$'$ stacking, the first and second groups of LL wave functions deviate from the 2/6 position along opposite directions ($n^{c,v}_{1}$ $\&$ $n^{c,v}_{2}$ in Fig. 32(f)).
The main reason is that the $k_{y}$ component is different for the current Dirac points and the K points, and the deviation of Dirac-cone centers grows with the increasing energy (Fig. 29) \cite{SciRep4;7509}. The tilted Dirac-cone structures can create the new magneto-optical selection rules.

The AB-stacked BLG has a high-symmetry configuration, while its Hamiltonain can not be exactly diagonalized \cite{PRL96;086805}. The interlayer atomic interactions in the magnetic quantizations could be treated as perturbations. The LL wave functions are presented with the normal quantum numbers, as shown in Fig. 32(d) at $B_{0}$= 40 T. But when magnetic field is very high ($B_{0}$ $\gtrsim$ 60 T; in the inset of Fig. 33(c)), the $B_0$-dependent electronic spectra show that some perturbed LLs in the anti-crossing regions are related to the distinct groups of LLs.
The second type of LLs is frequently revealed in the sliding BLG with a small shift measured from the AA, AA$'$, and AB stackings, as indicated by the anti-crossings in Fig. 32(a). The low-lying LLs of $\delta$= b/8 exhibit the distorted wave functions composed of the main and side modes (Fig. 32(b)). Such LLs come from the low-energy distorted electronic structure (Fig. 28(b)).

With a large shift deviated from the high-symmetry stacking configurations, the LL wave functions present the irregular oscillations without the normal zero points, e.g., $\delta$= 6b/8 and 11b/8 in Figs. 32(c) and 32(e), respectively. They can not be defined as a single-mode oscillator; that is, the main mode is absent for any magnetic-field strength.
Moreover, the $B_{0}$-dependent energy spectrum  indicates that each LL in the second group displays serious anti-crossings with all the LLs in the first one, and vice versa (Fig. 33(b)); This clearly illustrates that such LLs consist of various zero-point modes or that they could be verified as the undefined LLs. They arise from the fully distorted energy bands (Fig. 28(c)). The K and K$'$ points do not belong to the critical points; therefore, the low-energy expansion about them is not suitable for the magnetic quantization. It is impossible to comprehend the undefined modes from the effective-mass approximation. The above-mentioned three types of LLs have the distinct energy spectra, in which their main characteristics can be directly examined by the STS measurements \cite{PRB94;226403,PRL103;206804,PRL109;116802,PRL109;176804}.

The diverse spatial distributions of three kinds of LLs can enrich the magneto-optical absorption spectra \cite{SciRep4;7509}. Absorption spectra of the sliding BLG are revealed in the twin- or single-peak structures, depending on the difference of excitation frequency in the asymmetric LL energies. The well-behaved LLs in the AA and AA$'$ stackings exhibit two categories of intragroup excitations ($n_{1}^{c}$$\rightarrow$$n_{1}^{v}$ $\&$ $n_{2}^{c}$$\rightarrow$$n_{2}^{v}$). The absorption peaks of the former satisfy a specific selection rule of $\Delta$ n =$\pm$1 (Fig. 34(a)), while those of the latter are discribed by $\Delta$n = $\pm$1, 0, and $\pm$2 (Fig. 34(f)).
The new selection rules of $\Delta$ n = 0 $\&$ $\pm2$, which correspond to a lot of observable absorption peaks (blue and green triangles), are associated with the tilted Dirac-cone structures in the AA$'$ stacking. They are mainly determined by the relations between the initial state on the A$^{l}$ (B$^{l}$) sublattice and the final state on the B$^{l}$ (A$^{l}$) one. The valence- and conduction-state subenvelop functions have the different localization centers so that the excitation channels of $\Delta$ n= 0 $\&$ $\pm$2 gradually become obvious at higher frequency. Such rules are never predicted to exist in the other condensed-matter systems with the well-behaved LLs.

On the other hand, four categories of inter-LLs excitations, two intragroup and two intergroup ones, are revealed in the other stacking systems. The AB stacking, with the well-behaved LLs at $B_{0}$ $\lesssim$ 60 T, exhibits many prominent absorption peaks of $\Delta$ n =$\pm$1, as indicated in Fig. 34(d). Specifically, the absorption frequencies due to the inter-LL excitations of the first group present the linear $B_{0}$-dependence (being identified from Fig. 33(c)). As for the sliding BLGs slightly deviated from the normal stackings, most of the absorption peaks agree with $\Delta$ n =$\pm$1 ($\delta$ = b/8 in Fig. 34(b)). However, some peaks have the extra selection rules of $\Delta$ n =0 $\&$ $\pm$2 (blue and green triangles), whose intensities are dominated by the relative strength of the main and side modes in the perturbed LLs (Fig. 32(b)). Specially, the undefined LLs show a large number of absorption peaks without any optical selection rules. This means that all the occupied LLs could be effectively excited to any unoccupied ones, i.e., all the inter-LL excitation channels are effective under the requirement of the Fermi-Dirac distribution. In general, such peaks are relatively low for an easy observation, since the velocity matrix elements are significantly reduced by the random spatial distributions of the disordered quantum modes. The above-mentioned unique magneto-optical properties, the new selection rules, the extra ones, and the absence of selection rule, are worthy of a detailed examination using the optical spectroscopy \cite{NANOLETT14;4548,EPL108;27011,PRL14;126804,APL117;112803}.

\subsection{  Electric field enriched optical properties}

Electronic structures are significantly altered by the perpendicular electric field, such as Dirac-cone structures, saddle points, band gaps, and energy dispersions. All the stacking systems have the y$\rightarrow$$-$y symmetry ($\&$ time reversal symmetry); that is, electronic states are identical in the K and K$'$ valleys. Only the AA stacking presents the similar isotropic Dirac-cone structures \cite{CPL550;104}, since the $\textbf{r}$$\rightarrow$$-$$\textbf{r}$ and y$\rightarrow$$-$y inversion symmetries remain the same. But for the AA$'$ stacking (Fig. 35(e)), the electric potential difference between two layers induces the splitting of Dirac Dirac points (four points) and the seriously distorted Dirac-cone structure with extra saddle points.
Energy gaps are generated in most of the bilayer stacking systems except for those close to AA and AA$'$ ones, as shown in Figs. 35(a) and 35(c) ($\delta$=6b/8 and b). Specifically, the AB stacking possesses the weakly dispersive energy bands with a deep pocket near the Fermi level (the slightly distorted sombrero-shaped bands). The Dirac-cone structures and energy gaps are directly reflected in optical gaps. The electric field causes each stacking system to exhibit a threshold frequency (Figs. 35(b), 35(d); 35(f)), which is rather different from the zero-field case. The dependence of optical gap on the field strength is monotonous, i.e., it gradually grows as in Figs. 35(b) $\&$ 35(d) and declines as in Fig. 35(f) in the increasing field strength. Moreover, there exist extra special structures in absorption spectrum. For the $\delta$=6b/8 stacking, an extra absorption peak appears at a higher frequency ($\sim$0.35 eV), mainly owing to the nearly identical wave vectors in the valence and conduction saddle points (red arrows). The AB stacking has two prominent asymmetric peaks, in which the lower- and higher-frequency ones arise from the optical excitations of the intra- and inter-pair energy bands, respectively. The high DOS from the first pair of energy bands accounts for the strong absorption intensity. The $\bf{E}$-induced saddle points in the AA$'$ stacking can create a quite strong peak structure.

In the presence of a composite electric and magnetic field, the x$\rightarrow$$-$x and y$\rightarrow$$-$y inversion symmetries are conserved in the AA and AA$'$ stackings, while the former is absent in the other BLG stackings. Upon the condition of the splitting, the magnetic quantization of K and K$'$ valleys indicates that only these two high-symmetry stackings possess the four-fold degenerate LLs.
The splitting LLs exist in the other BLG's. Moreover, the seriously distorted and the weakly dispersive energy dispersions are magnetically quantized into the multi-mode LLs. There are more perturbed and undefined LLs, so that the LL-anticorssings occur frequently in the E$_{0}$- and B$_{0}$-dependent energy spectra. Apparently, the destructions in the state degeneracy and the well-behaved mode induce more complicated magneto-absorption structures, as shown in Figs. 36(a)-36(c).
The number of absorption peaks is largely enhanced. Most of them present the lower intensity except those whose LLs are associated with the high-DOS energy dispersions (0.2 eV$\le\omega\le$ 0.4 eV). The original magneto-selection rules can survive in a composite field.

Optical measurements under external fields are made only for the AB-stacked BLG by using the absorption \cite{Nature459;820,PSSB247;2931}, transmission \cite{PRB78;235408,PRL14;126804,APL117;112803}, reflection \cite{NANOLETT14;4548,PRB80;165406,PRB79;115441,PRB78;235408} and Raman scattering \cite{PRB76;201401,PRB78;113407,PRL97;187401} spectroscopies. They agree with a number of theoretical predictions including the low-frequency shoulder structure at zero fields \cite{PRB73;144427,PNAS107;14999,RSCADV4;63779,PRL102;256405,ACSNano9;6765}, the $\textbf{E}$-induced threshold frequency and two prominent peaks \cite{Nature459;820,PSSB247;2931}, the magneto-optical selection rule of $\Delta$ n=$\pm$1 for the inter-LL excitations of the first group \cite{NANOLETT14;4548}, and the linearly $B_{0}$-dependent absorption frequency. The other predicted results, which are the forbidden excitation region induced by the Fermi-Dirac distribution, the $\textbf{E}$-created optical gaps and special structures in any bilayer stackings, the diverse selection rules due to three kinds of LLs, and the drastic changes of the spectral features in a composite field, can be further examined by optical spectroscopies.

\section{ Concluding remarks}

A systematic review has been made on optical properties of graphene. The generalized tight-binding mode and the gradient approximation are used to explore the rich absorption spectra. They are suitable for studying the low-symmetry stacking configurations and the complex external fields. The rich features due to the stacking configurations and the external fields are investigated in detail. The AAA, ABA, ABC and AAB stackings, and the various stackings in BLG's are taken into account. The zero-, electric-, magnetic and composite fields are included in the calculations. The presented results are compared with those from other theoretical calculations and validated by the experimental measurements, while most of predictions require further experimental verifications. The developed theoretical framework can promote the future studies on other 2D materials\cite{PRL113;266801,NJP14;033003,PRL109;055502,PRB92;174514,Science349;524,PRB89;155316,
APLHo;acc,PRL110;197402,PRB88;085434,PRB85;195440,PRB81;125120,JAP76;1363,RMP54;437}.

The high-symmetry AA-stacked trilayer graphene presents three vertical Dirac-Dirac structures with separated Dirac points. Only the intra-Dirac-cone optical excitations can survive because of the linearly symmetric or anti-symmetric superposition of the tight-binding functions in different layers. The spectral intensity due to the middle Dirac cone linearly depends on excitation  frequency, as observed in monolayer graphene. In addition, an even-layer AA stacking possesses an optical gap determined by the first shoulder structure. The almost symmetric upper and lower Dirac cones, with free holes and electrons respectively, result in a forbidden excitation region under the specific fermion distributions. Their Dirac points shift away from the Fermi level in the presence of electric field and thus create the enlarged vacant region or the blue shift of shoulder structure. On the other hand, three obvious symmetric peaks, which come from the saddle points near the M point, are revealed in  the middle-frequency absorption spectrum. Three groups of the well-behaved LLs only exhibit three categories of intra-group optical excitations, in which each prominent delta-function-like peak is characterized by the $\Delta$n=$\pm$1 selection rule. Most spectral structures belong to single peaks with an almost uniform intensity, reflecting the approximately symmetric LL energy spectrum and the close Fermi velocities. Any peak frequency is characterized by the $\sqrt{B_{0}}$ dependence. As for the bilayer-like intra-group excitations, the first inter-LL channel changes discontinuously at certain magnetic fields, so that its frequency has a discontinuous $B_{0}$-dependence. Specially, the main features of optical and electronic properties keep similar in a composite external field, a result of the conservation of spatial symmetries.

The AB-stacked trilayer graphene possesses a slightly separated Dirac-cone structure and two pairs of parabolic bands, a superposition of monolayer- and bilayer-like energy bands. The optical excitations are vanishing between the former and the latter. Only an observable shoulder structure due to the distinct parabolic bands appears in the low-frequency absorption spectrum, while there exist five symmetric peaks at middle frequency. Any excitation channel between valence and conduction bands becomes available for a sufficiently high electric field. More low-frequency special structures are generated in absorption spectra, in which the asymmetric peaks are related to the $E_{0}$-dependent oscillatory and sombrero-shaped bands with circular constant-energy contours. However, the optical threshold frequency keeps zero, i.e., the semimetal-semiconductor transition is absent during the variation of electric field. The similar behavior could also be observed in the AB stackings of N$>$3. In general, the LL splitting energies are negligible so that the magneto-absorption spectra are almost the same for the distinct localization centers. One category and the other four categories of magneto-optical excitations, which, respectively, come from monolayer- and bilayer-like LLs, exhibit many single- and twin-peak structures. Such absorption peaks due to the well-behaved modes obey the specific selection rules. The former is much stronger than the latter; furthermore, a simple relationship between the intensity and absorption frequency is absent for the latter. Especially, the certain initial excitations associated with the double-degenerate LLs of (n$_1^{c,v}$=0; n$_2^{c,v}$=0 $\&$1) can survive and are distinguishable in single-peak absorption frequencies. The magneto-optical spectra are greatly enriched by the electric field because of the splitting LLs and the intra-group LL anti-crossings. Five categories of inter-LL excitations are changed into nine ones. The twin peaks and the double peaks can occur simultaneously, reflecting the asymmetric and the splitting LL energy spectra, respectively. Moreover, the frequent anti-crossings of the perturbed LLs in the $E_{0}$-dependent energy spectrum lead to the increase of peaks and the reduced intensities.

The ABC-stacked trilayer graphene has three pairs of unusual energy bands, namely the partially flat (surface-localized), sombrero-shaped and linear ones. Excitation channels between any valence and conduction bands are available. The threshold frequency is zero  and the first absorption peak corresponds to the band-edge-state excitations between the surface-localized and sombrero-shaped energy bands. There are four obvious peaks in the middle-frequency absorption spectrum. Optical gap (energy gap), being characterized by the threshold peak from the separated surface-localized bands, is created by an electric field. The original peak exhibits an obvious blue shift. Three groups of LLs can generate nine categories of absorption peaks, including six inter-group and three intra-group ones. Remarkably, the intergroup LL anti-crossings frequently emerge in the B$_{0}$-dependent energy spectrum and thus induce extra absorption peaks. Many of the prominent absorption peaks associated with the well-behaved LLs are presented under the specific selection rules. As to these two kinds of absorption peaks, their B$_{0}$ dependence is rather strong. The magneto-optical properties are diversified by an electric field. With the increased electric-field strength, many outstanding inter-LL absorption peaks gradually change from the twin-peak structures into the double-peak ones, as a result of the non-equivalence condition imposed on the two sublattices with identical projections. In particular, a single threshold peak becomes a double-peak structure, owing to the Fermi-Dirac distribution. The extra peaks are generated by the intragroup LL anticrossings, and the abnormal E$_{0}$-dependent absorption frequencies are simultaneously revealed.

The AAB-stacked trilayer graphene, without the z-plane mirror symmetry, sharply contrasts to the other highly symmetric stacking systems. It is a narrow-gap semiconductor, while the others are semimetals. Electronic and optical properties are enriched by the various interlayer atomic interactions. The unique band structure includes the oscillatory, sombrero-shaped and parabolic energy dispersions. The 3$\times$3 excitation channels present more special structures in the low- and middle-frequency optical spectrum, compared to those of the ABC stacking. As for the first asymmetric absorption peak, its frequency is energy gap between the valence and conduction oscillatory bands; furthermore, both frequency and intensity are significantly altered by an electric field. The LLs are split by the broken mirror symmetry, a phenomenon being absent in the other stacking systems. The intragroup and intergroup anti-crossings occur frequently because of the complex relationship between the magnetic quantization and the interlayer atomic interactions. There are nine categories of absorption peaks, in which the inter-group LL excitations are much higher than the intra-group ones, depending on the phases and amplitudes of LL wavefunctions. The magneto-absorption spectra exhibit the single- or twin-peak structures which are dominated by the quantum modes, LL energy spectra and Fermi-Dirac distribution. The splitting LLs, with different localization centers, can generate very distinct absorption spectra including extra single peaks during the anticrossings. A lot of strong peaks, which arise from the well-behaved LLs, can be characterized by the specific selection rules. The intensity and frequency of magneto-absorption peaks are strongly affected by an electric field.

Various stacking configurations can diversify electronic structures and absorption spectra in the sliding BLG. The vertical or non-vertical Dirac cones transform into parabolic bands, accompanied by the separation of cone structures, creation of an arc-shaped stateless region, distorted energy dispersions, extra low-energy critical points, and splitting of middle-energy states. All the stacking systems belong to gapless semimetals, while an optical gap is only present in the AA and AA$'$ stackings because of the intra-cone excitations. The low-frequency shoulders or peaks strongly depend on stacking configuration; furthermore, there are three or four middle-frequency strong peaks in the absence of three-fold rotation symmetry. An electric field causes the semi-metallic BLGs
 to become semiconductors except for AA and AA$'$ ones; therefore, optical gaps are revealed in any absorption spectra. It also leads to extra low-frequency peaks related to the seriously distorted or weakly dispersive energy bands. As for magneto-optical spectra, the diverse selection rules are dominated by three kinds of LLs with the normal or irregular spatial distributions. A lot of strong absorption peaks due to the well-behaved LLs obey the specific selection rules, such as ($\Delta$n=$\pm$1; $\pm$1; $\pm$1, 0, $\pm$2) for the AA, AB and AA$'$ stackings. Two novel selection rules, $\Delta$n=0 and $\pm$2, are verified to arise from the tilted Dirac cones in the AA$'$ stacking. The anti-crossing perturbated LLs induce extra selection rules ($\Delta$n=0$\&$$\pm$2) in the intergroup optical excitations, but the heights of the absorption peaks are relatively low. More peculiarly, the undefined LLs create a large number of absorption peaks in the absence of a specific selection rule. Magneto-optical spectra in a composite field are more complicated, such as the drastic changes in the number and intensity of absorption peaks.

Up to now, the experimental observations are consistent with part of theoretical calculations on the optical properties in external fields. As to the AB-stacked bilayer graphene, the $\sim$0.3-eV shoulder structure under zero field, the $\textbf{E}$-induced semimetal-semiconductor transition and two low-frequency asymmetric peaks, the two strong $\pi$-electronic absorption peaks at middle frequency, the $\Delta$n=$\pm$1 selection rule for the first group of LLs, and the linear $B_{0}$-dependence of the inter-LL excitation energies, have been confirmed by absorption, transmission, reflection, and/or Raman scattering spectroscopies. The similar verifications performed on the trilayer ABA stacking contain one shoulder at $\sim$0.5 eV, the gapless behavior unaffected by electric field, the $\textbf{E}$-generated low-frequency multi-peak structures, several $\pi$-electronic peaks, and the monolayer- and bilayer-like inter-LL absorption frequencies. Moreover, the identified spectral features in the ABC-stacked trilayer graphene cover two low-frequency characteristic peaks and gap opening under electric field. On the other side, a plenty of important predicted results require further experimental examinations, including the shoulders, asymmetric peaks and symmetric ones in zero-field optical spectra, the initial forbidden excitation region in even-layer AA stacking systems, the more spectral structures in lower-symmetry stacking configurations, the $\textbf{E}$-created optical gaps and special structures, the significant dependence of the prominent $\pi$-electronic peaks on stacking configuration and layer number, the diverse magneto-optical selection rules due to three kinds of LLs, the delta-function-like inter-LL absorption peaks with the single, twin and double forms, the almost uniform single peaks related to the non-tilted Dirac cones, the LL-antirossing-induced drastic changes in number and intensity of peaks, and the greatly diversified magneto-absorption spectra under a composite field. The experimental measurements on the frequency and form of spectral structures can provide the critical information in identifying energy bands and LL energy spectra, such as, the linear, parabolic, sombrero-shaped and oscillatory dispersions, and the symmetric, asymmetric and splitting LL energy spectra. They are very useful in understanding the complex relationship among the stacking configurations, the magnetic quantization, and the Coulomb potential energy. Also, they could be used to distinguish the geometric structures of the distinct stacking domains.

In addition to the stacking configurations, the different dimensionalities can create the diverse properties of carbon-related systems. The 1D graphene nanoribbons \cite{SCI319;1229, JACS132;10034, NAT458;872, NAT458;877, ACSNANO4;1362, ACSNANO6;2020, APL105;023101, JACS133;10394, C77;1031, JACS69;3529, PRB76;045418, OPTE19;23350, PRB84;085458, NANOTECH18;495401, CARBON44;508, PCCP18;7573}, 2D graphenes and 3D graphites \cite{JCP129;234709, PR112;698, PR103;1586, ADVPHYS30;139, PRL102;166401, PRSLSA181;101, NJP12;083060, PCCP13;6036, CARBON54;268, CPC189;60, JPSJ40;761, PRB77;115313, PRB83;121201, NJP15;053032, JPHYS27;125602} quite differ from one another in the electronic and optical properties. Concerning the first systems, the finite-size quantum confinement and edge structure dominate the essential properties. A lot of asymmetric absorption peaks, which come from the 1D parabolic energy bands in the absence of external fields, obey the armchair- or zigzag-dependent selection rule \cite{JACS69;3529, PRB76;045418, OPTE19;23350, PRB84;085458}. The spatial distributions of the standing waves are responsible for this unusual rule, since they are mainly determined by the distinct boundary structures. When the ribbon width is longer than the magnetic length, each 1D Landau subband consists of the dispersionless quasi-LL and parabolic dispersion. The almost well-behaved quasi-LLs are localized at the ribbon center, and they exhibit prominent symmetric peaks with a selection rule similar to that in monolayer graphene \cite{NANOTECH18;495401}. The edge- and magneto-dependent selection rules will be destroyed by a sufficiently high transverse electric field \cite{CARBON44;508, PCCP18;7573}. On the other side, the periodically interlayer atomic interactions play a critical role in the 3D graphites. The AA- \cite{JCP129;234709}, AB- \cite{PR112;698, PR103;1586, ADVPHYS30;139, PRL102;166401} and ABC-stacked \cite{PRSLSA181;101} graphites, respectively, have one pair, two pairs and one pair of 3D valence and conduction bands. Their low-frequency absorption spectra, respectively, show a shoulder structure after the interband forbidden region, a concave-downward one and a shoulder one. Specifically, optical gaps are vanishing in the latter two systems \cite{NJP12;083060, PCCP13;6036}. The 3D magnetically quantized states form the Landau subbands (LSs) composed of 2D dispersionless LLs and 1D k$_{z}$-dependent energy dispersions . The number of LS group is, respectively, one, two and one for the simple hexagonal, Bernal and rhombohedral graphites. Many LSs in the AA-stacked graphite cross the Fermi level and their bandwidths are $\sim$1 eV, reflecting the strongest interlayer atomic interactions among three systems. Such LSs generate a plenty of twin structures of opposite-direction asymmetric peaks \cite{CARBON54;268, CPC189;60}. The LSs of the AB-stacked graphite possess $\sim$0.2-eV band widths, and only two of them intersect with E$_{F}$ \cite{JPSJ40;761,PRB77;115313, PRB83;121201}. The LS energy spectrum corresponding to the K and H points (k$_{z}$=0 and 1) is, respectively, almost asymmetric and symmetric about E$_{F}$, so that the magneto-absorption spectrum presents the twin structures of same-direction asymmetric peaks and the single symmetric peaks \cite{PRL102;166401, PRB83;121201}. As for the ABC-stacked graphite, the LS bandwidths are narrower than 10 meV, and only one is located at E$_{F}$ \cite{NJP15;053032}. These result in a lot of single absorption peaks with symmetric forms, being attributed to the merged excitations from the K and H points \cite{JPHYS27;125602}. It is also noticed that the LS energies of the AA- and ABC-stacked graphites \cite{CARBON54;268, NJP15;053032} are characterized by the similar
square-root dependence on the quantum number and the magnetic-field strength. A review study on the electronic and optical properties has been finished for the 1D graphene nanoribbons \cite{PCCP18;7573}, and it will be extended to the 3D graphites.

Two-dimensional materials have attracted much attention since the discovery of graphene in 2004. The layered group-IV condensed-matter systems with nano-scaled thickness and hexagonal symmetry are very suitable for studying the novel physical, chemical and material phenomena \cite{APL96;183102, APL97;223109, PRL108;155501, ACSNANO7;4414, NATCOM5;3383, NANOLETT15;2510, NATMAT14;1020} and have a high potential for future technological applications \cite{ACSNANO4;803, NATNANOTECH10;227, NATCOM4;1500}. Silicene and germanene have been synthesized on metallic substrates \cite{APL96;183102, APL97;223109, PRL108;155501, ACSNANO7;4414, NATCOM5;3383, NANOLETT15;2510}. Recently, tinene is successfully fabricated on a substrate of bismuth telluride \cite{NATMAT14;1020}. Silicene, germanene and tinene possess the buckled structure with a mixed sp$^{2}$-sp$^{3}$ orbital hybridization rather than a sp$^{2}$ one. Their spin-orbital interactions are sufficiently strong so that they can create an observable energy gap. The sp$^{3}$ orbital bondings and the spin-orbital interactions can dramatically change electronic structures; therefore, they are expected to create the rich magnetic quantization.

The generalized tight-binding model could be further modified to comprehend the diverse quantization phenomena in various condensed-matter systems. Recently, a new Hamiltonian is proposed to explore the unusual magnetic properties of monolayer tinene \cite{ARXIV1512;02129}. The magnetic quantization of the multi-orbital-, spin- and configuration-dependent Hamiltonian is solved by using the generalized tight-binding model. This should be very helpful for the further development of the critical Hamiltonian models in the mainstream materials. The Hamiltonian dimension of monolayer tinene is eight times large as that of monolayer graphene. The buckled structure causes the matrix elements to be complex at any wave vectors, so that the Hamiltonian diagonalization becomes difficult. Tinene is predicted to have two groups of low-lying LLs dominated by the 5p$_{z}$ and (5p$_{x}$,5p$_{y}$) orbitals, indicating the critical roles of multi-orbital bondings and spin-orbital coupling. The important differences between tinene and graphene in the main characteristics of LLs include orbital components, spin configurations, localization centers, state degeneracy, and magnetic- and electric-field dependencies. Other quantized properties are worthy of detailed investigations, such as optical spectra, Coulomb excitations and transport properties of layered  group-IV systems.

\par\noindent {\bf Acknowledgments}

This work was supported in part by the National Science Council of Taiwan,
the Republic of China, under Grant Nos. NSC 98-2112-M-006-013-MY4 and NSC 99-2112-M-165-001-MY3.

\newpage
\renewcommand{\baselinestretch}{0.2}

\begin{figure}
\centering
\includegraphics[width=0.9\linewidth]{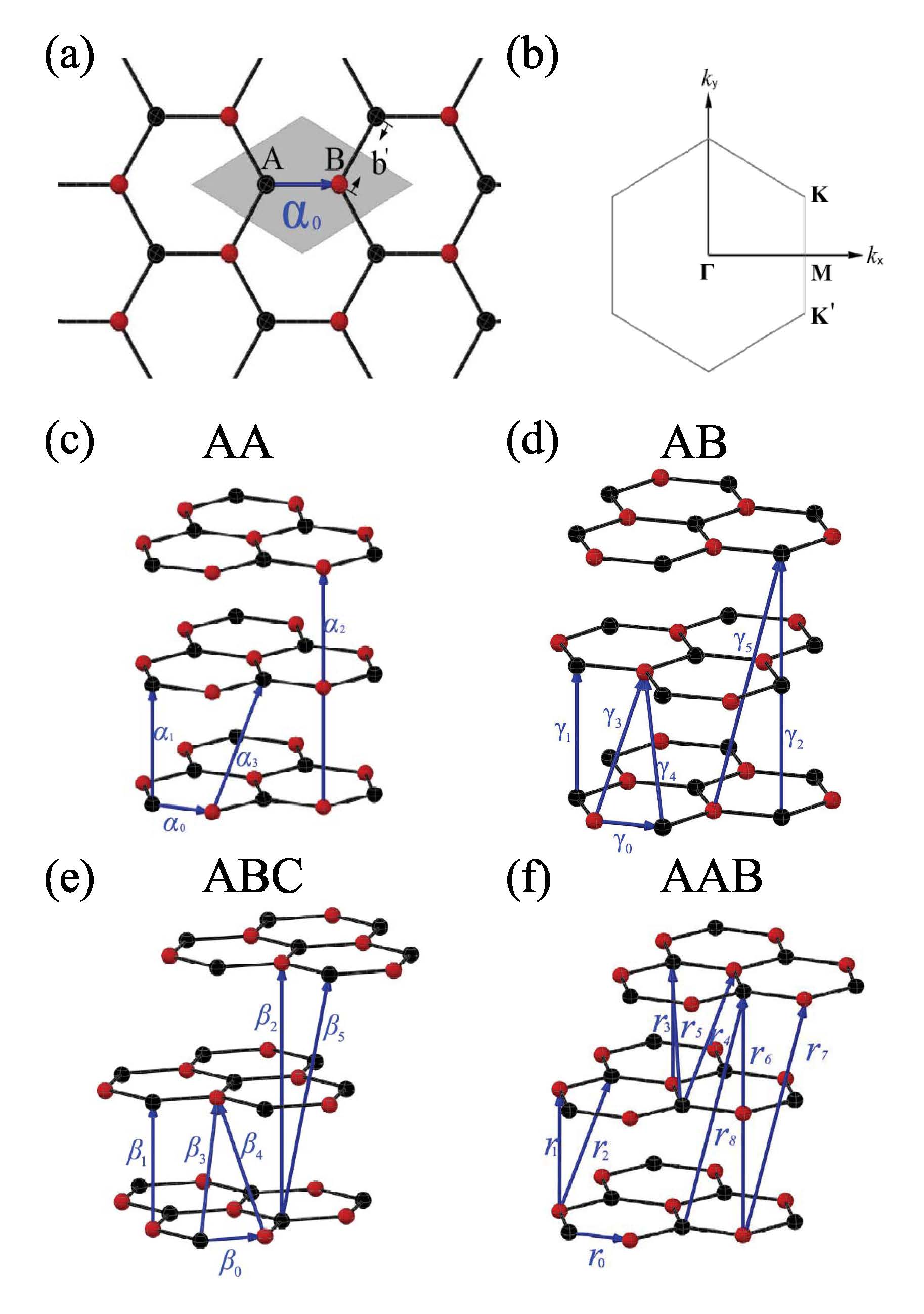}
\caption{(a) Honeycomb lattice structure of monolayer graphene. Sublattices A and B are shown by black and red colors, respectively. The primitive unit cell is depicted by a gray diamond, where $\alpha$ indicates the nearest-neighbor hopping integral and $b^{'}$ represents the C-C bond length. (b). The Brillouin zone of honeycomb lattice and some highly symmetric points. Geometric structures of (c) AA-, (d) AB-, (e) ABC- and (f) AAB-stacked graphenes. Inter- and intra-layer atomic interactions are indicated by blue arrows.}
\label{fig:graph}
\end{figure}

\begin{figure}
\centering
\includegraphics[width=0.9\linewidth]{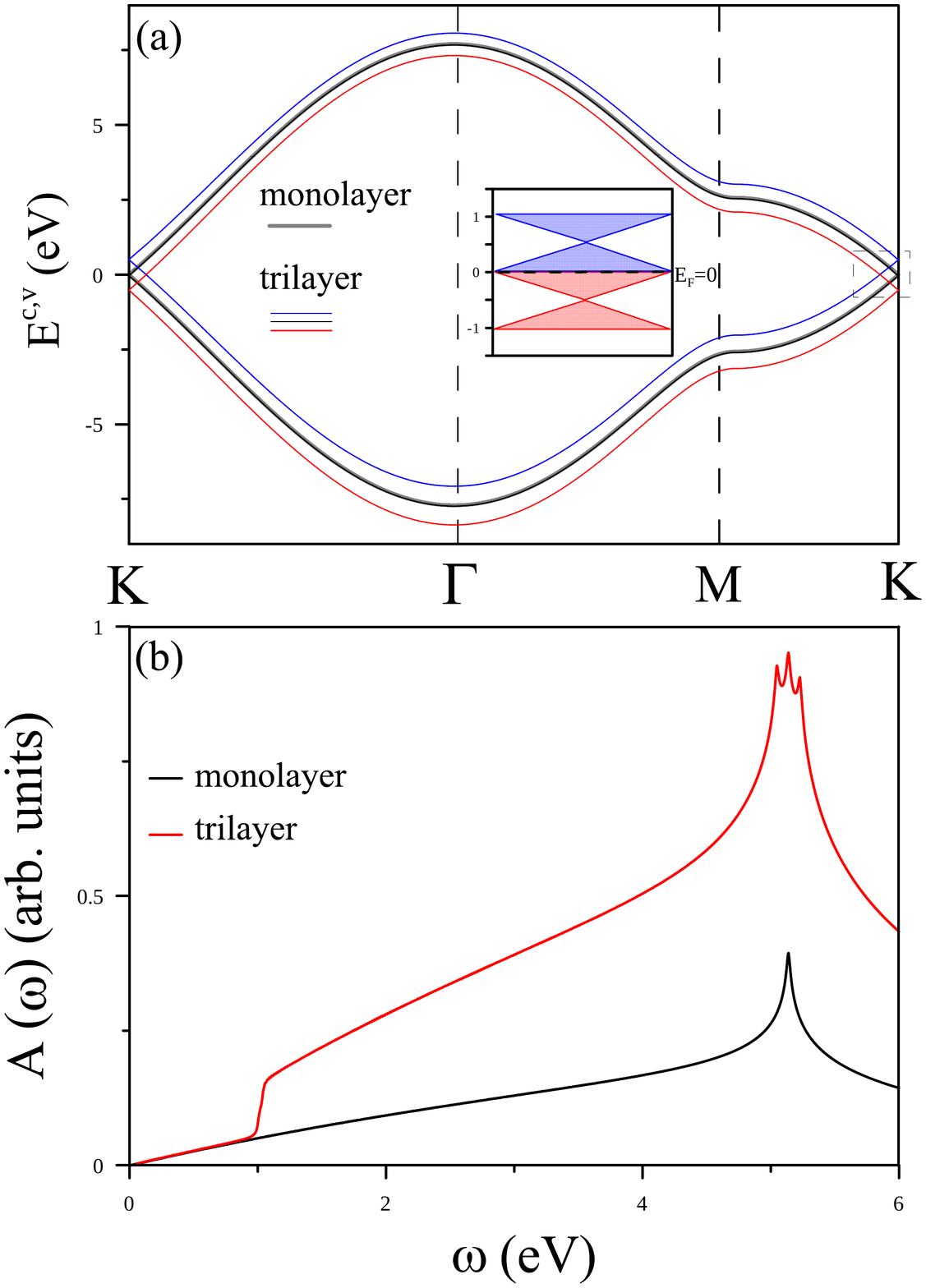}
\caption{(a) Zero-field band structures and (b) absorption spectra of monolayer and AA-stacked trilayer graphenes. The shadows inset in (a) indicate the forbidden optical transition zone in the case of trilayer graphene.}
\label{fig:graph}
\end{figure}

\begin{figure}
\centering
\includegraphics[width=0.9\linewidth]{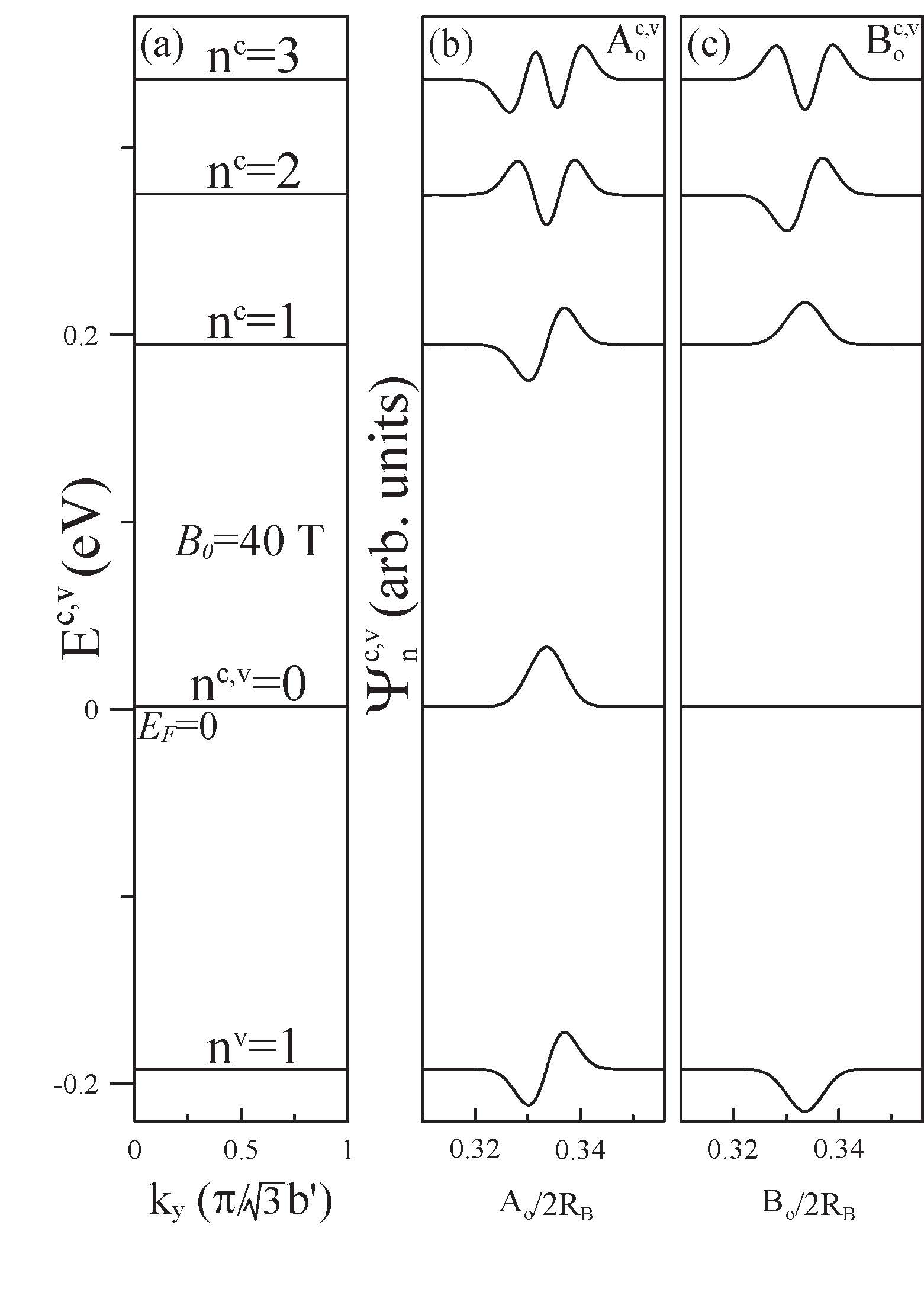}
\caption{Low-lying Landau levels of monolayer graphene at $B_{0}$ = 40 T.
The quantum numbers of conduction and valence Landau levels $n_{c,v}$ = 0, 1, 2, 3. . . are counted from the Fermi level, $E_{F}$ = 0. (b and c) The subenvelope functions of two sublattices A and B are plotted in the unit of 2 $R_{B}$ according to the Landau levels in (a).}
\label{fig:graph}
\end{figure}

\begin{figure}
\centering
\includegraphics[width=0.9\linewidth]{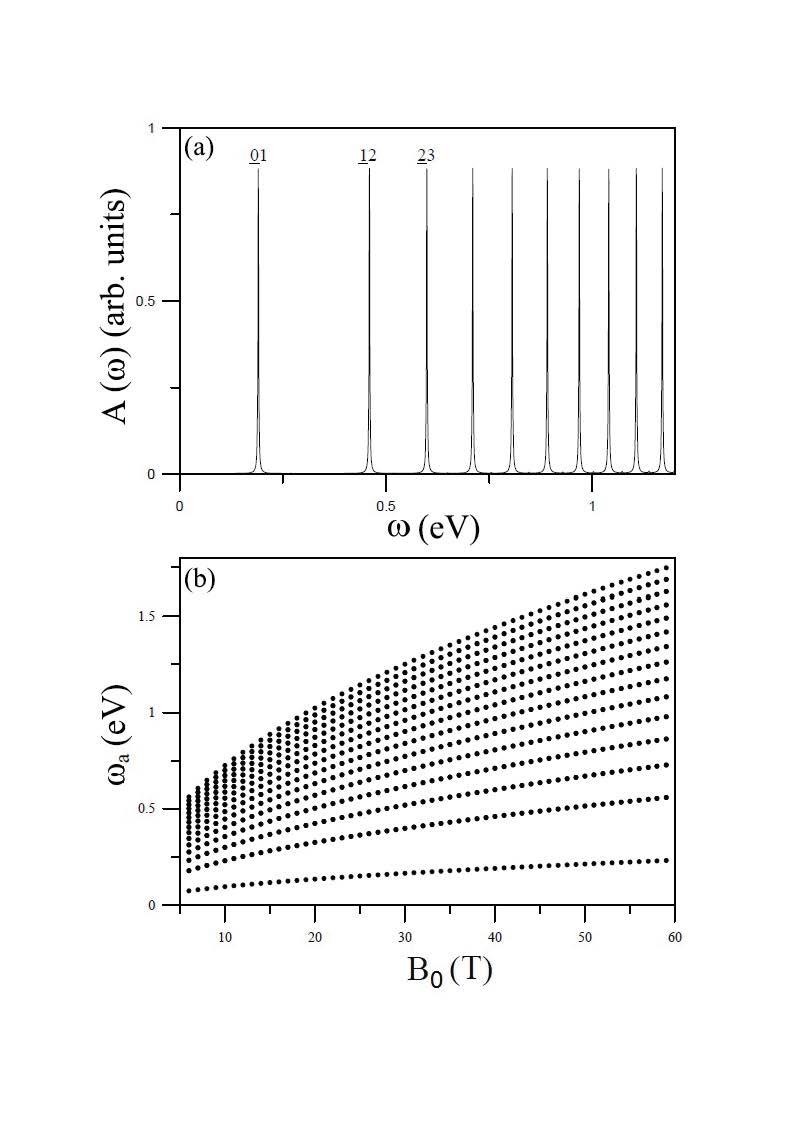}
\caption{(a) Magneto-absorption spectrum of monolayer graphene at $B_{0}=40$ T.
(b) The $\sqrt{B_{0}}$-dependent magneto-absorption frequencies.}
\label{fig:graph}
\end{figure}

\begin{figure}
\centering
\includegraphics[width=0.9\linewidth]{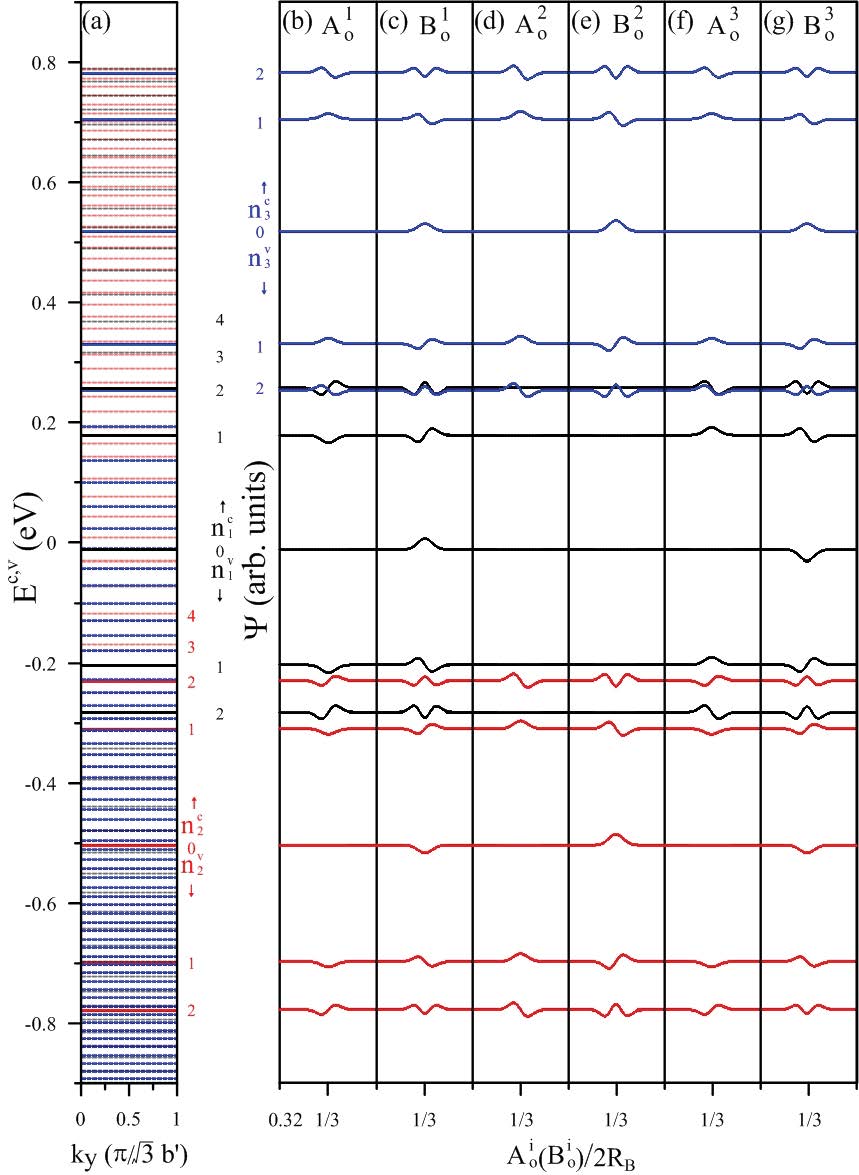}
\caption{(a) Landau levels and (b) six subenvelope functions for the 2/6 localization Landau states of AA-stacked trilayer graphene under $B_{0}$ = 40 T. The black, red, and blue colors, respectively, indicate the first, second, and third groups of Landau levels.}
\label{fig:graph}
\end{figure}

\begin{figure}
\centering
\includegraphics[width=0.9\linewidth]{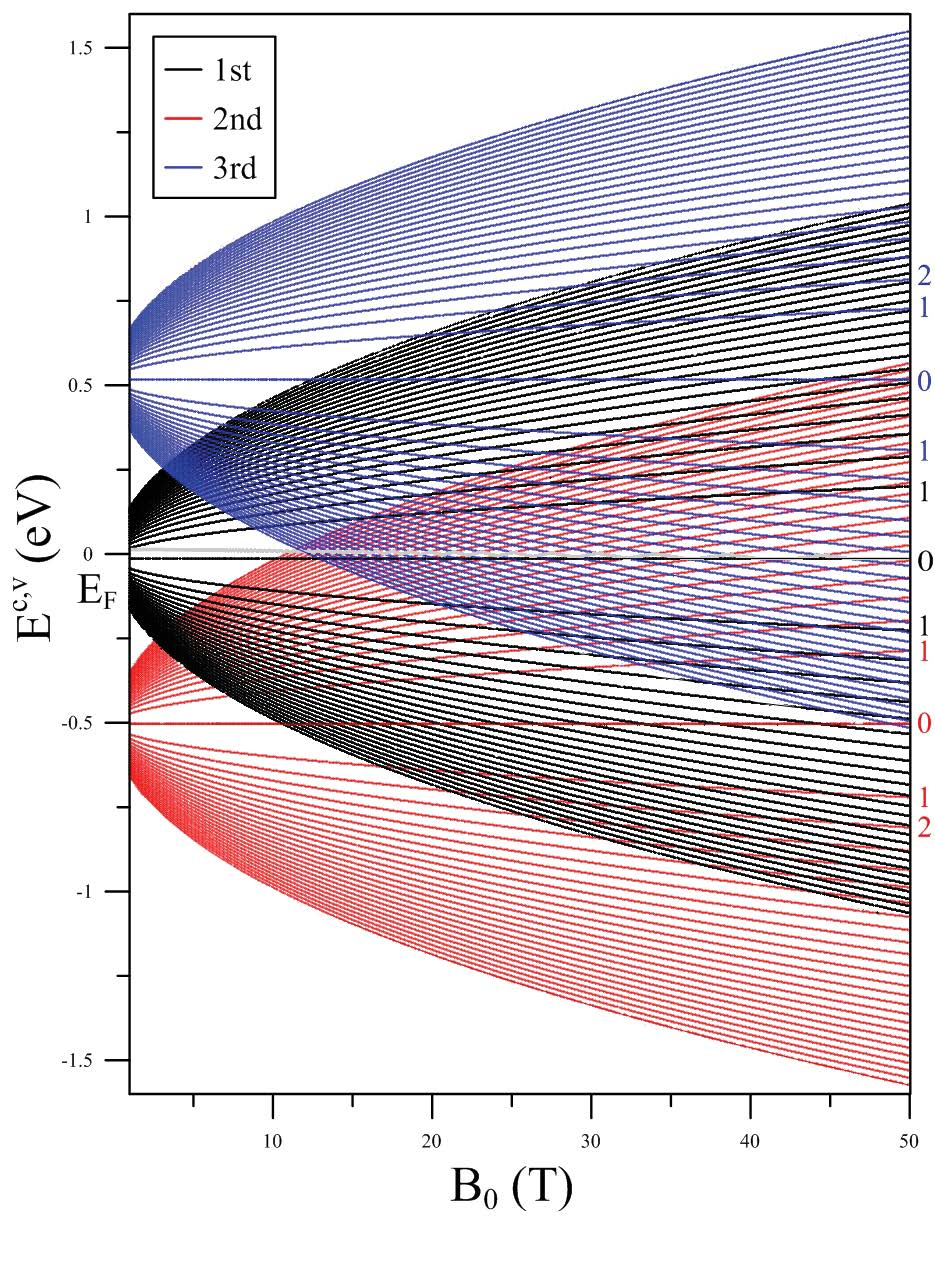}
\caption{The $B_0$-dependent LL energy spectrum of AA-stacked trilayer graphene.
The three monolayer-like LL patterns, as indicated by black, red and blue colors, are evolved from the three isolated Dirac cones in Fig. 2 (a).}
\label{fig:graph}
\end{figure}

\begin{figure}
\centering
\includegraphics[width=0.9\linewidth]{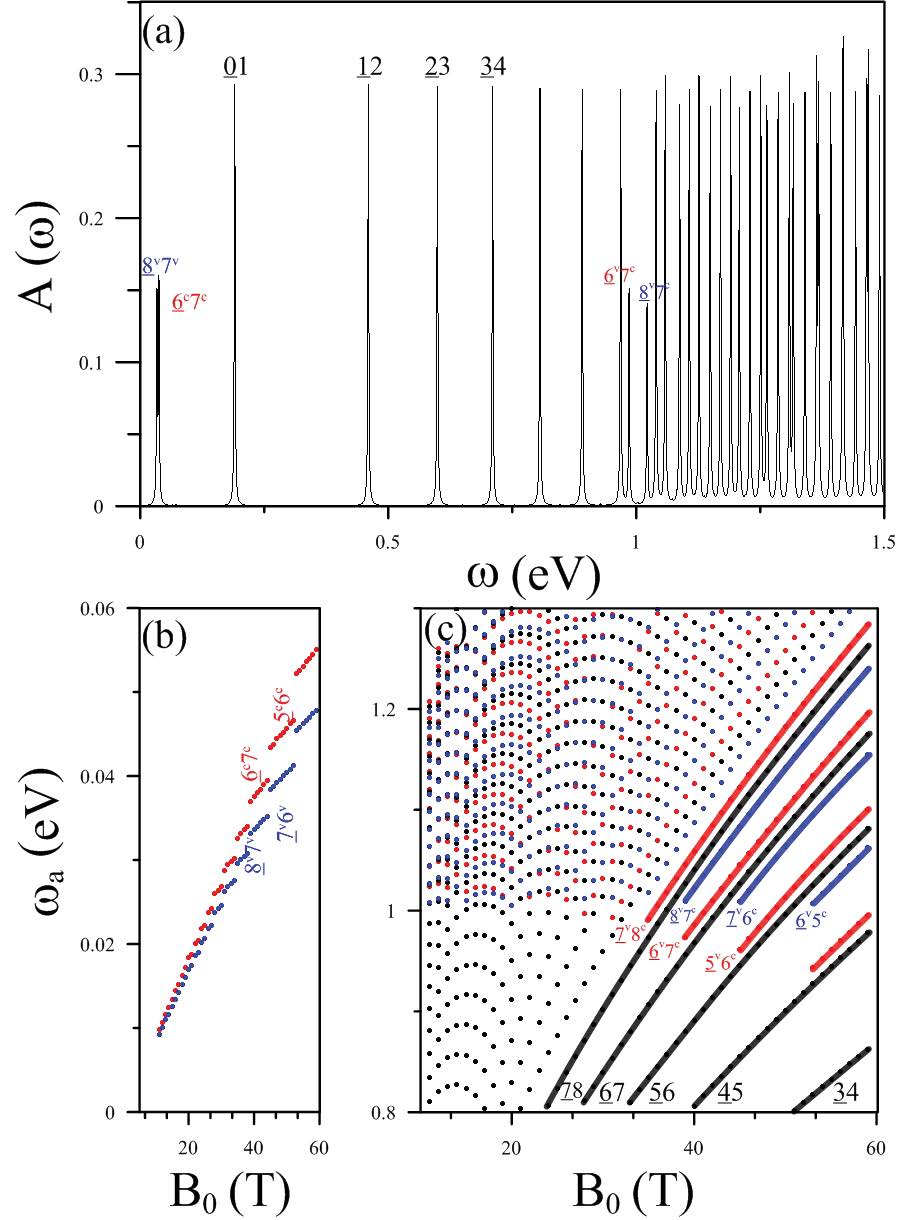}
\caption{(a) Magneto-absorption spectrum of AA-stacked trilayer graphene under $B_0$ = 40 T. The three kinds of intragroup excitation channels are indicated by different colors.
The $B_0$-dependent frequencies of threshold and full channels are shown in (b) and (c), respectively.}
\label{fig:graph}
\end{figure}

\begin{figure}
\centering
\includegraphics[width=0.9\linewidth]{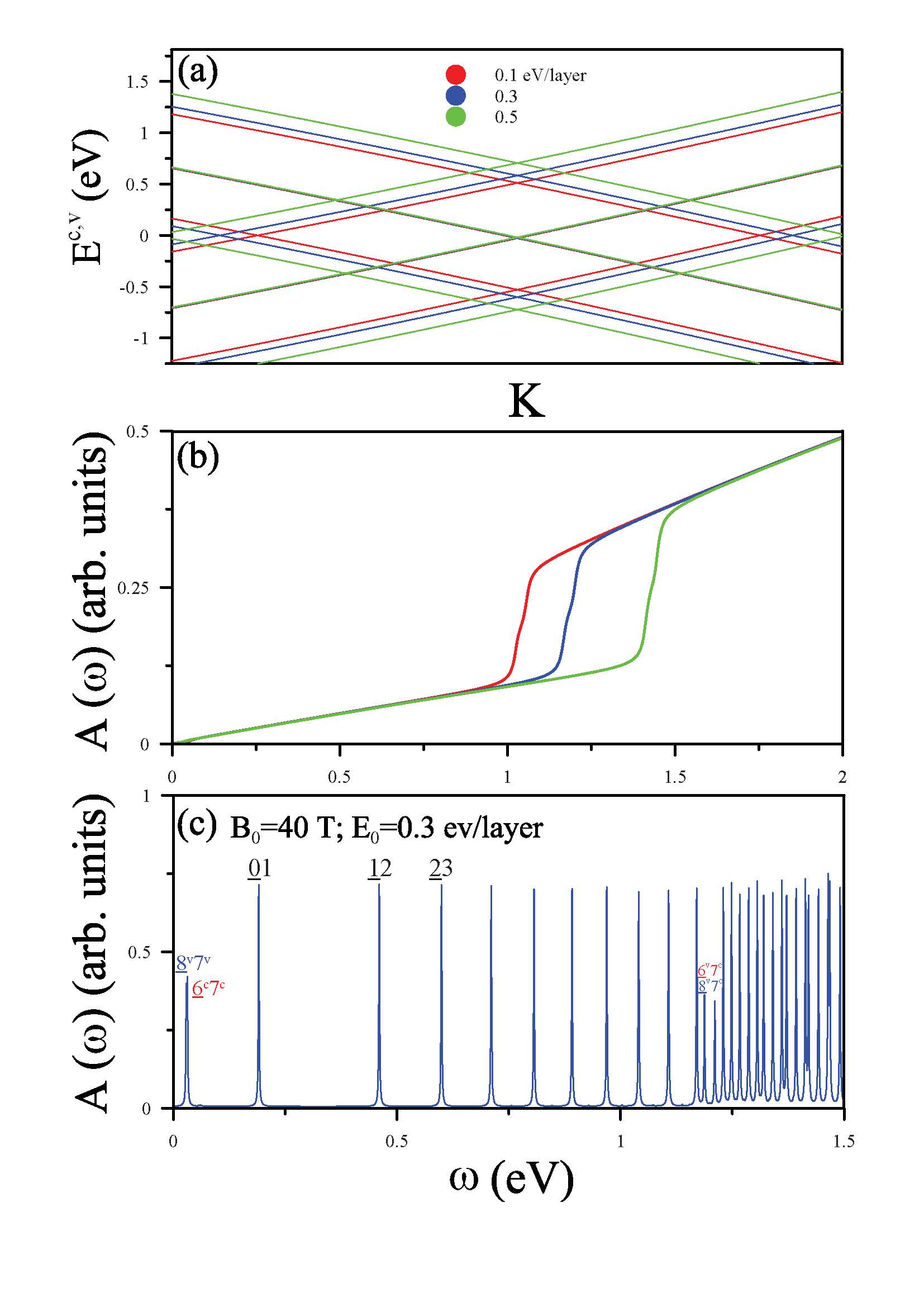}
\caption{The $E_0$-enriched (a) band structure, (b) absorption spectrum and (c) magneto-absorption spectrum of AA-stacked trilayer graphene.}
\label{fig:graph}
\end{figure}

\begin{figure}
\centering
\includegraphics[width=0.9\linewidth]{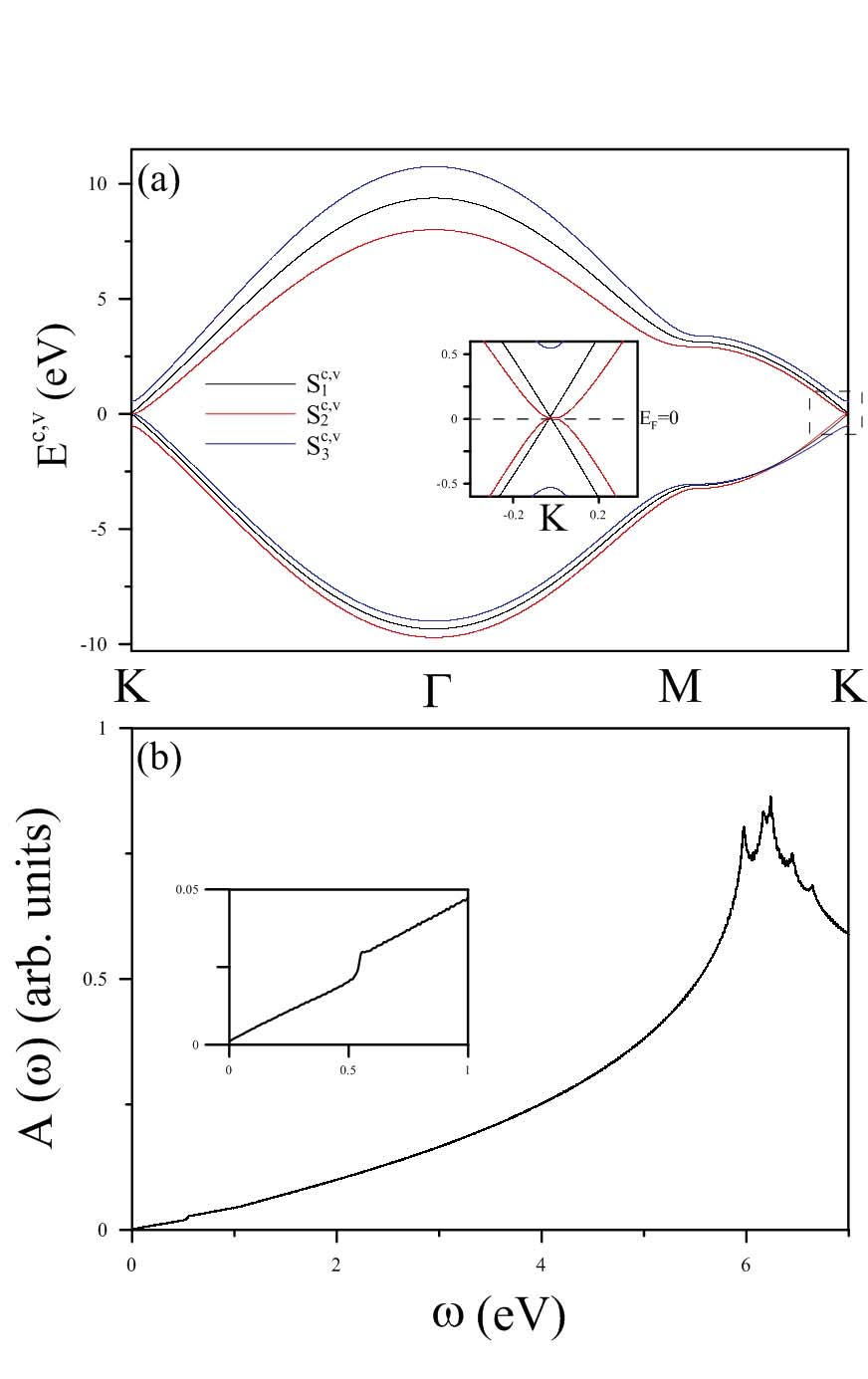}
\caption{(a) Zero-field band structure and (b) absorption spectrum of AB-stacked trilayer graphene. In (a ), the three subbands are labeled as $S_1^{c,v}$ (black), $S_2^{c,v}$ (red) and $S_3^{c,v}$ (blue). The insets of (a) and (b) show the zoomed-in view at low energies.}
\label{fig:graph}
\end{figure}

\begin{figure}
\centering
\includegraphics[width=0.9\linewidth]{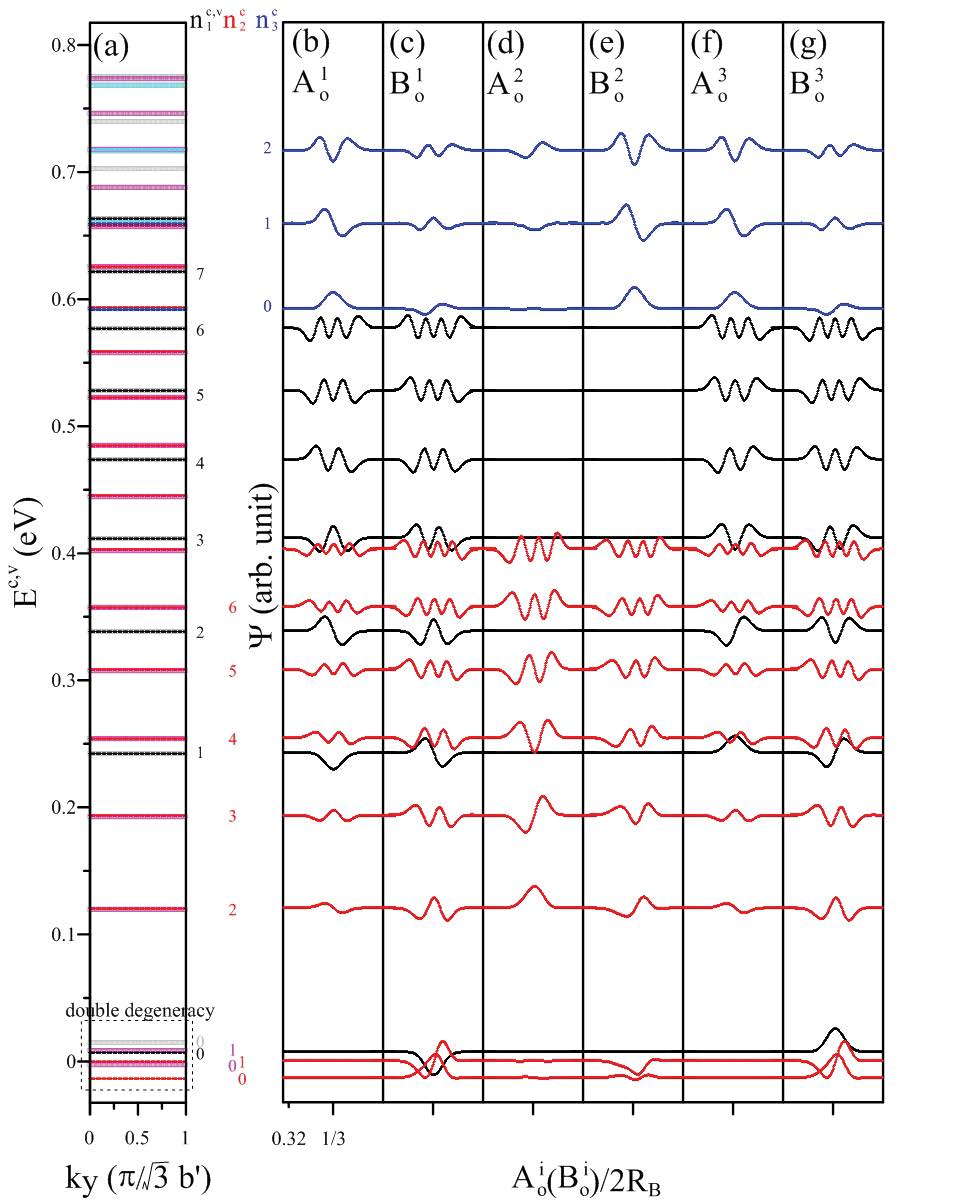}
\caption{Landau levels and subenvelope functions of 2/6 localized Landau states for AB-stacked trilayer graphene at $B_0$ = 40 T. Energies of 1/6 localized Landau states are plotted in (a) for comparison (light colors).}
\label{fig:graph}
\end{figure}

\begin{figure}
\centering
\includegraphics[width=0.9\linewidth]{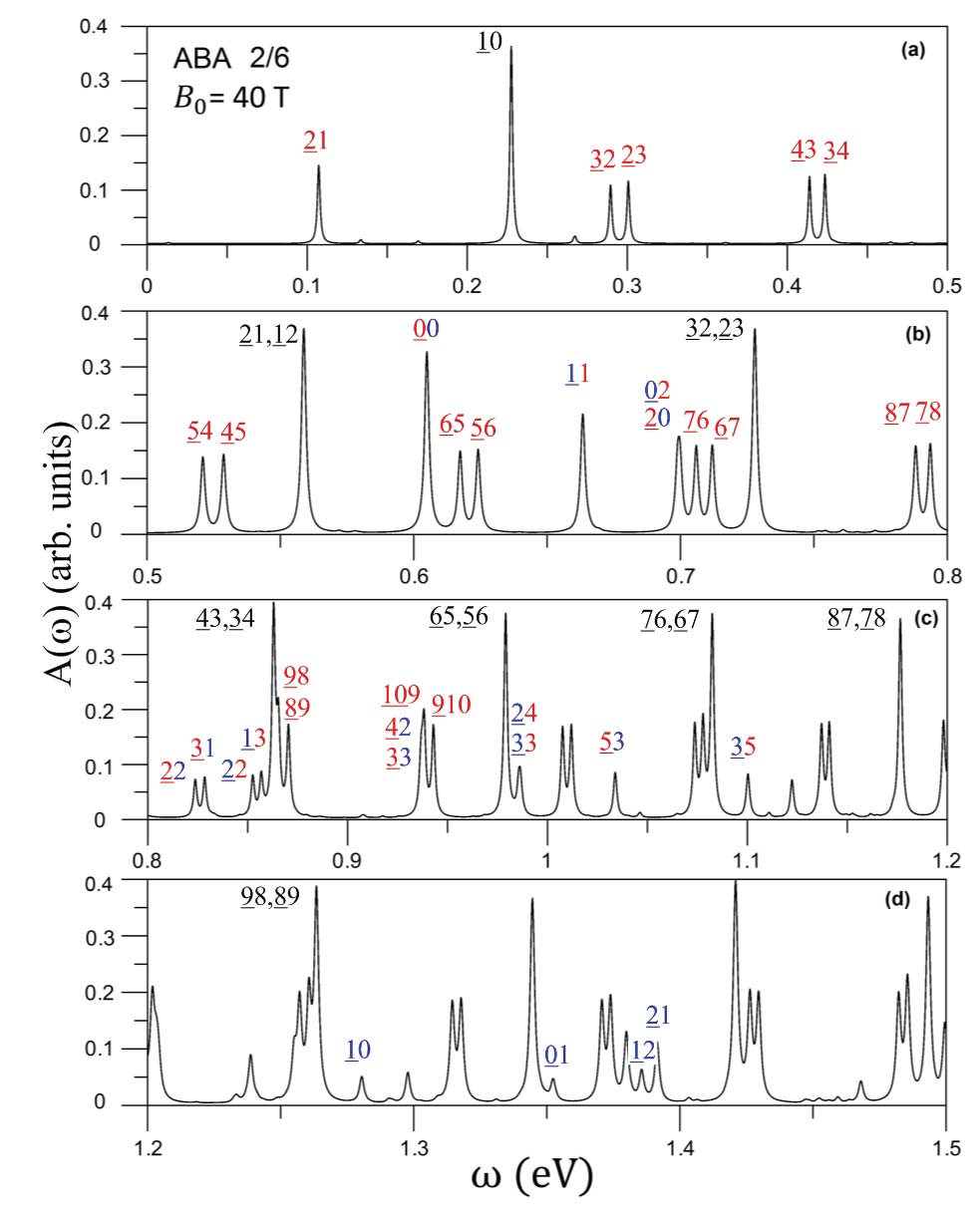}
\caption{The magneto-absorption spectrum of AB-stacked trilayer graphene with three intragroup and two intergroup categories of LL excitations under $B_0$ = 40 T. The black, red, and blue color numbers, respectively, indicate the excitations from the first, second, and third LL groups.}
\label{fig:graph}
\end{figure}

\begin{figure}
\centering
\includegraphics[width=0.9\linewidth]{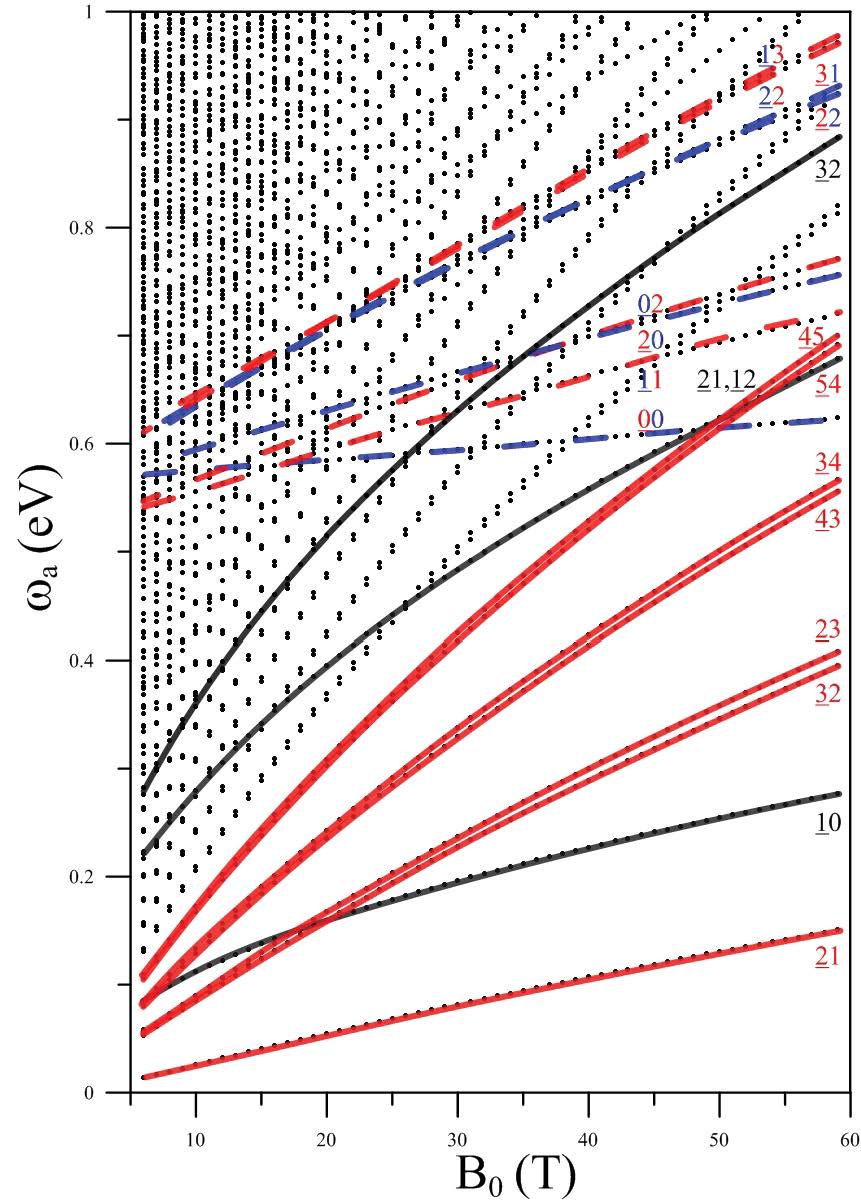}
\caption{$B_0$-dependent magneto-absorption frequencies of AB-stacked trilayer graphene. The solid and dashed curves indicate the excitation channels of the intragroup and intergroup categories, respectively.}
\label{fig:graph}
\end{figure}

\begin{figure}
\centering
\includegraphics[width=0.9\linewidth]{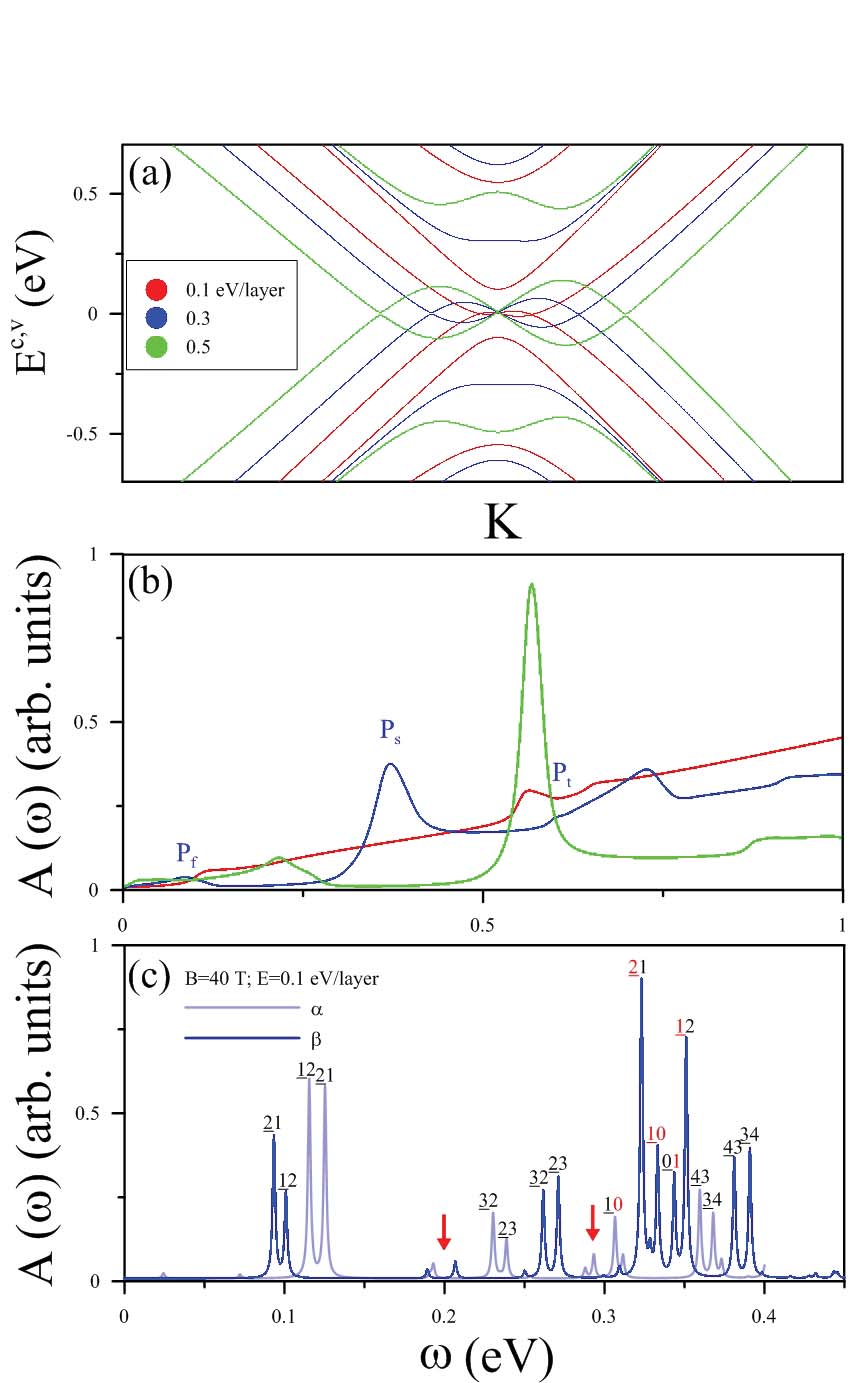}
\caption{The $E_0$-enriched (a) band structure, (b) absorption spectrum and (c) magneto-absorption spectrum of AB-stacked trilayer graphene. $\alpha$- and $\beta$-type spectra, respectively, stand for the excitations from the two valley-splitting LLs, 1/6 and 2/6 localized states in Fig. 10.}
\label{fig:graph}
\end{figure}

\begin{figure}
\centering
\includegraphics[width=0.9\linewidth]{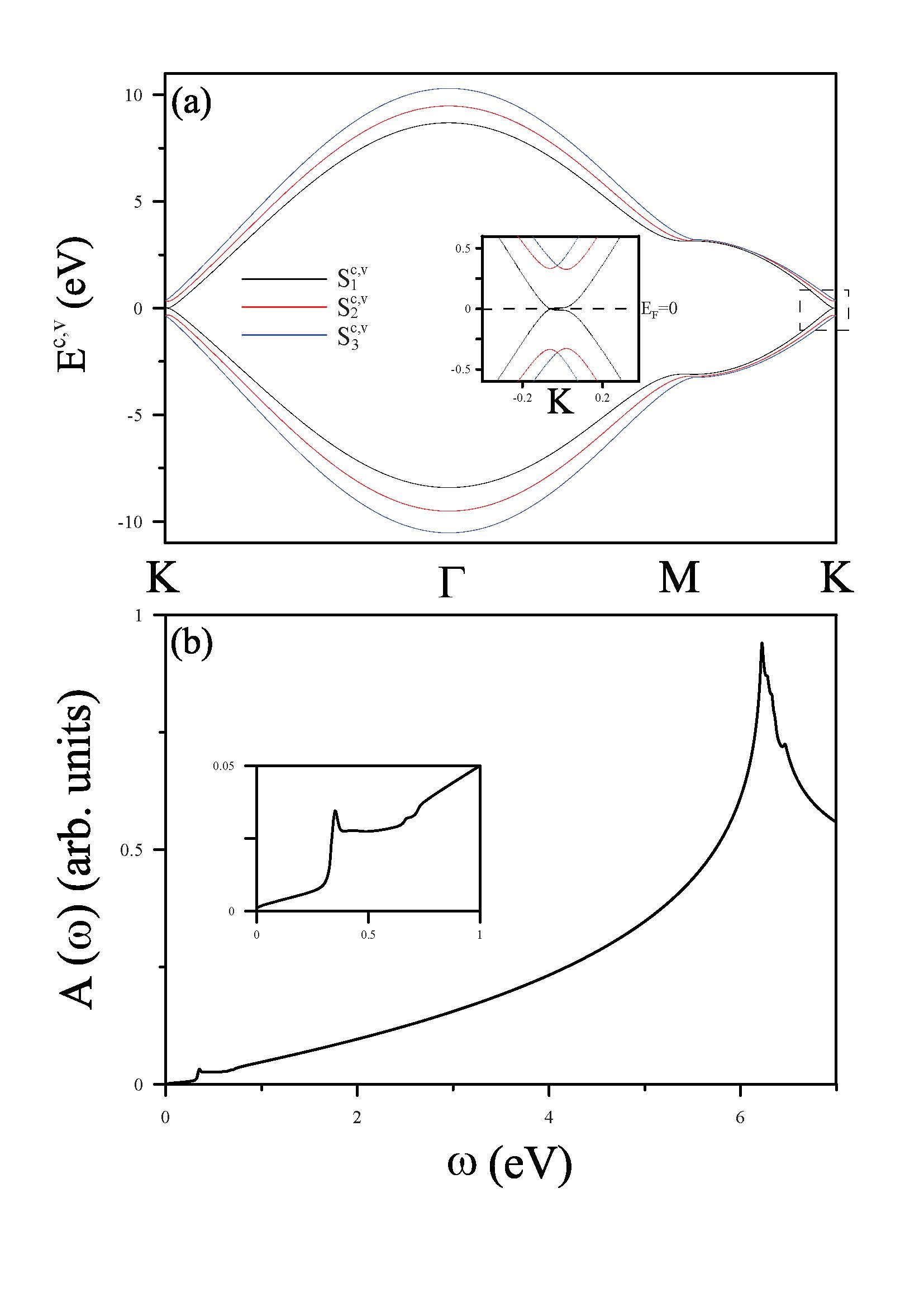}
\caption{(a) Zero-field band structure and (b) absorption spectrum of ABC-stacked trilayer graphene. In (a ), the three subbands are labeled as $S_1^{c,v}$ (black), $S_2^{c,v}$ (red) and $S_3^{c,v}$ (blue). The insets of (a) and (b) show the zoomed-in view at low energies.}
\label{fig:graph}
\end{figure}

\begin{figure}
\centering
\includegraphics[width=0.9\linewidth]{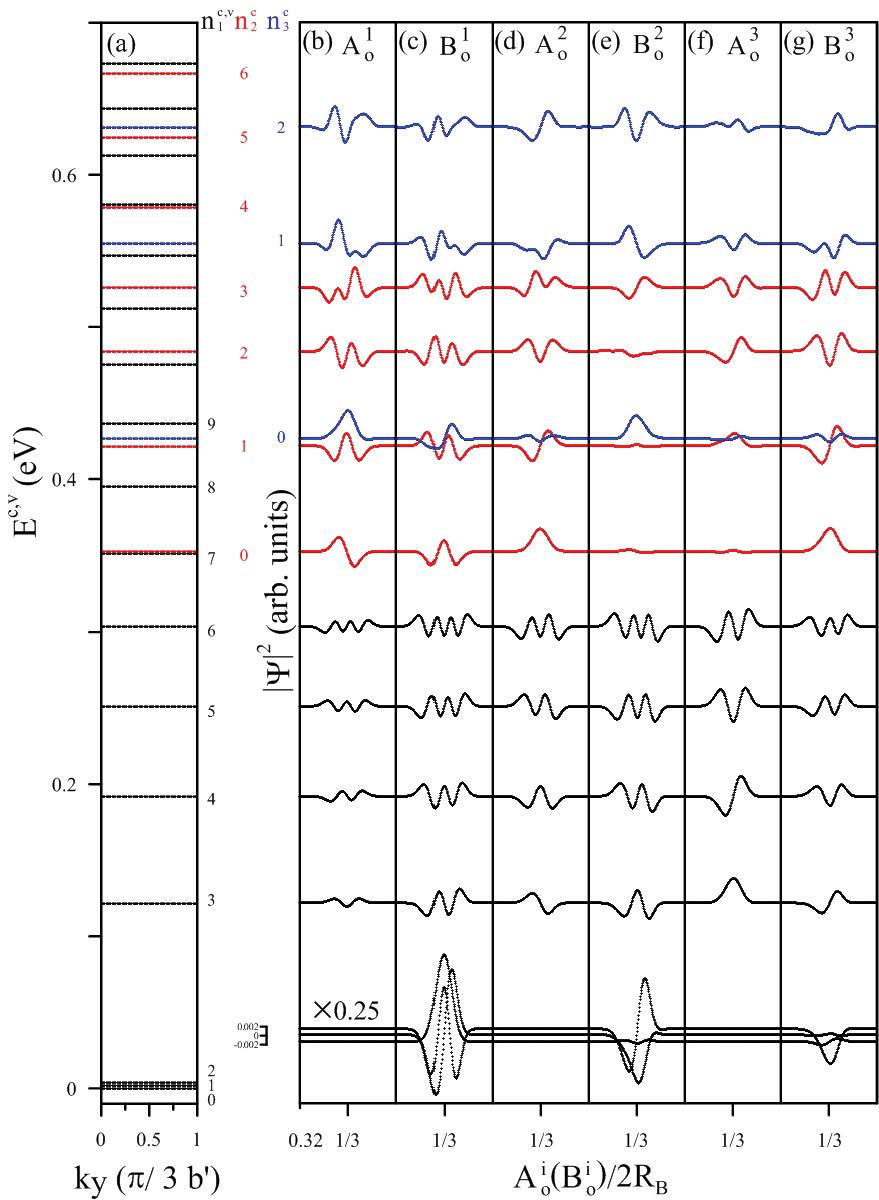}
\caption{Landau levels and subenvelope functions of 2/6 localization Landau states for ABC-stacked trilayer graphene at $B_0$ = 40 T.}
\label{fig:graph}
\end{figure}

\begin{figure}
\centering
\includegraphics[width=0.9\linewidth]{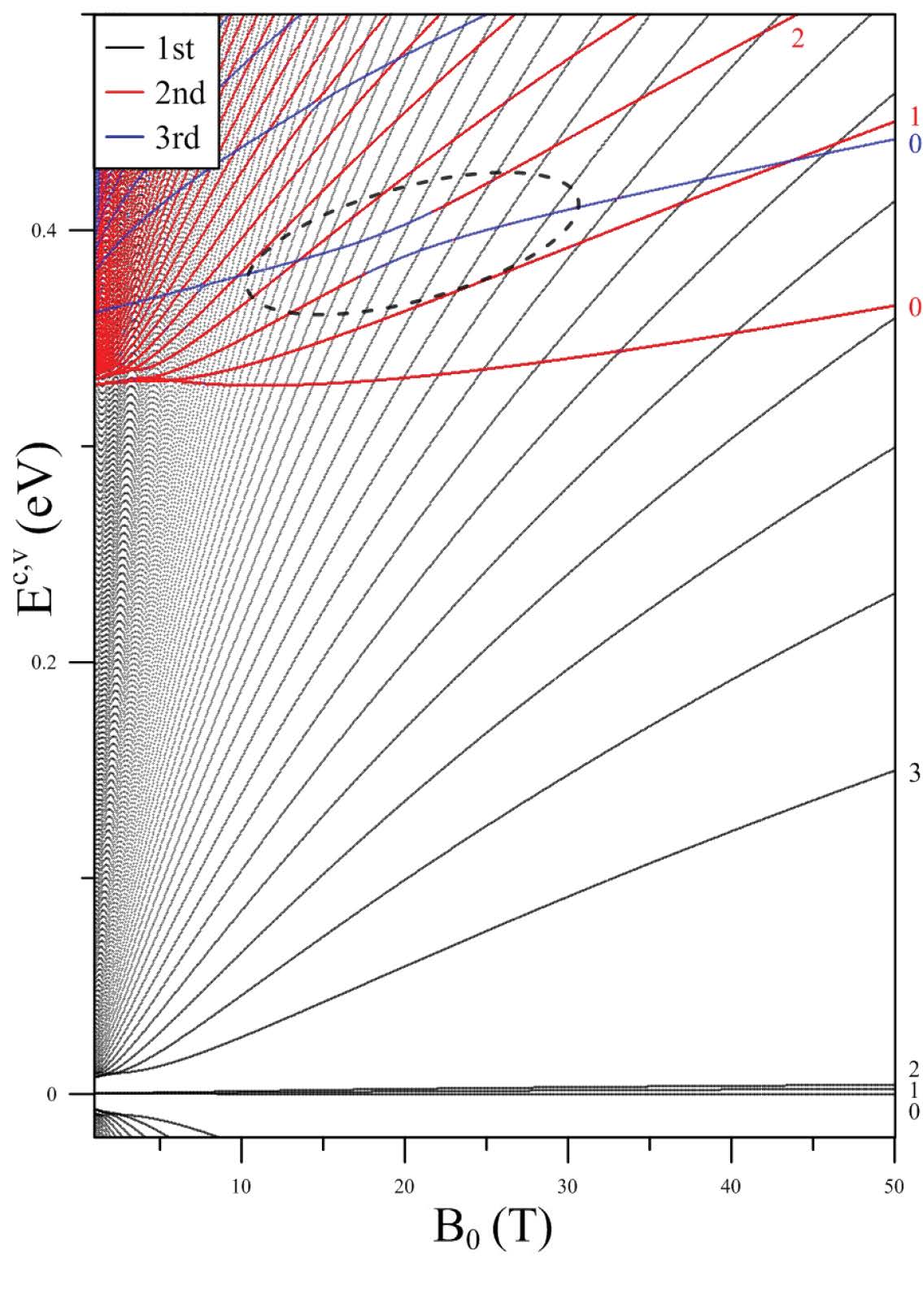}
\caption{Landau-level spectrum of ABC-stacked trilayer graphene. The intergroup anticrossing of the $n^{c}_{2}=2$ and $n^{c}_{3}=0$ LLs is shown in the dashed circle.}
\label{fig:graph}
\end{figure}

\begin{figure}
\centering
\includegraphics[width=0.9\linewidth]{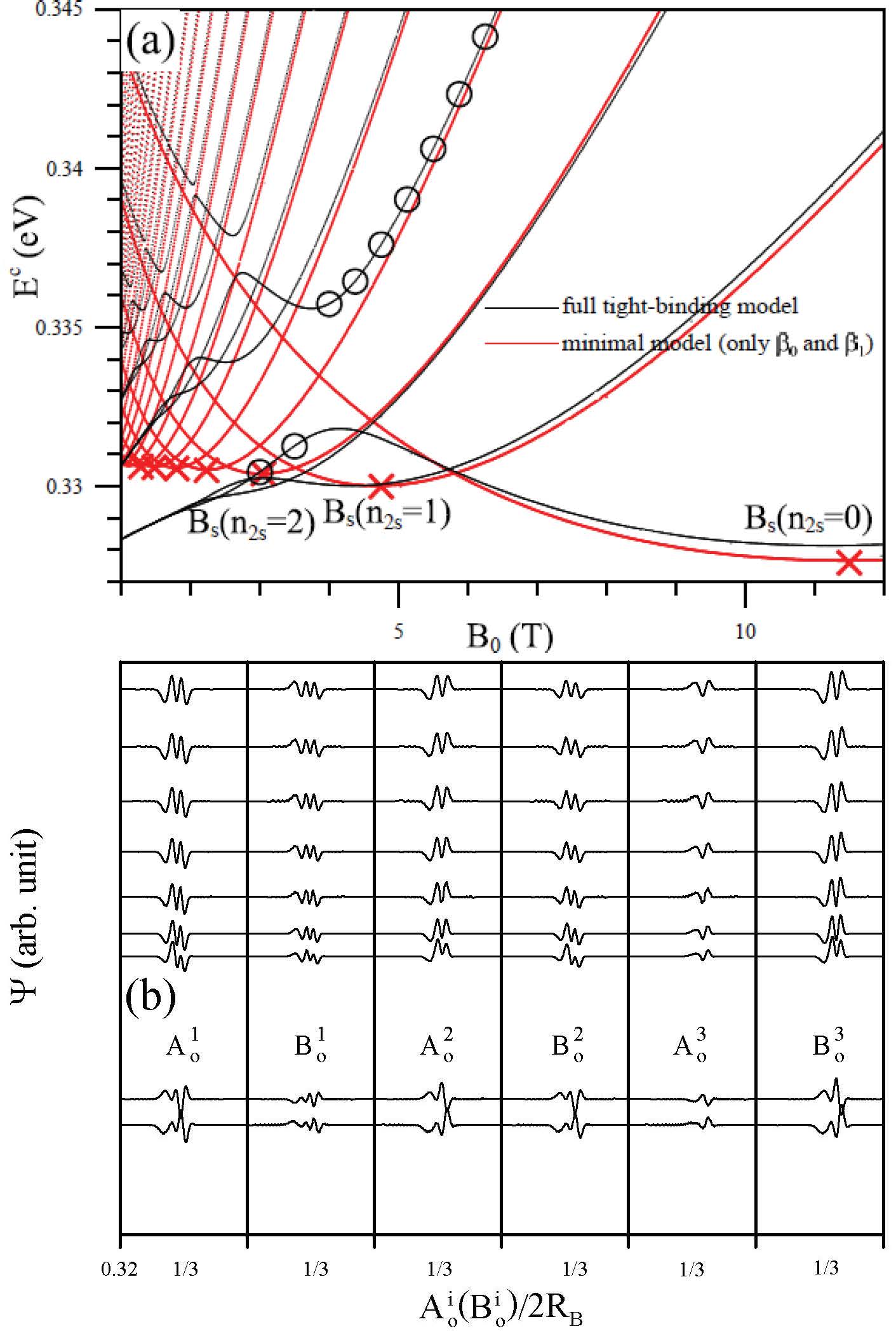}
\caption{Landau levels of the second group of ABC-stacked trilayer graphene based on the minimum model (red) and the full tight-binding model (black).
Red cross symbols indicate the critical fields $B_{s} (n_{2s})$ for specific Landau levels.
(b) Evolution of the subenvelope functions for the anticrossing Landau states marked in (a).}
\label{fig:graph}
\end{figure}

\begin{figure}
\centering
\includegraphics[width=0.9\linewidth]{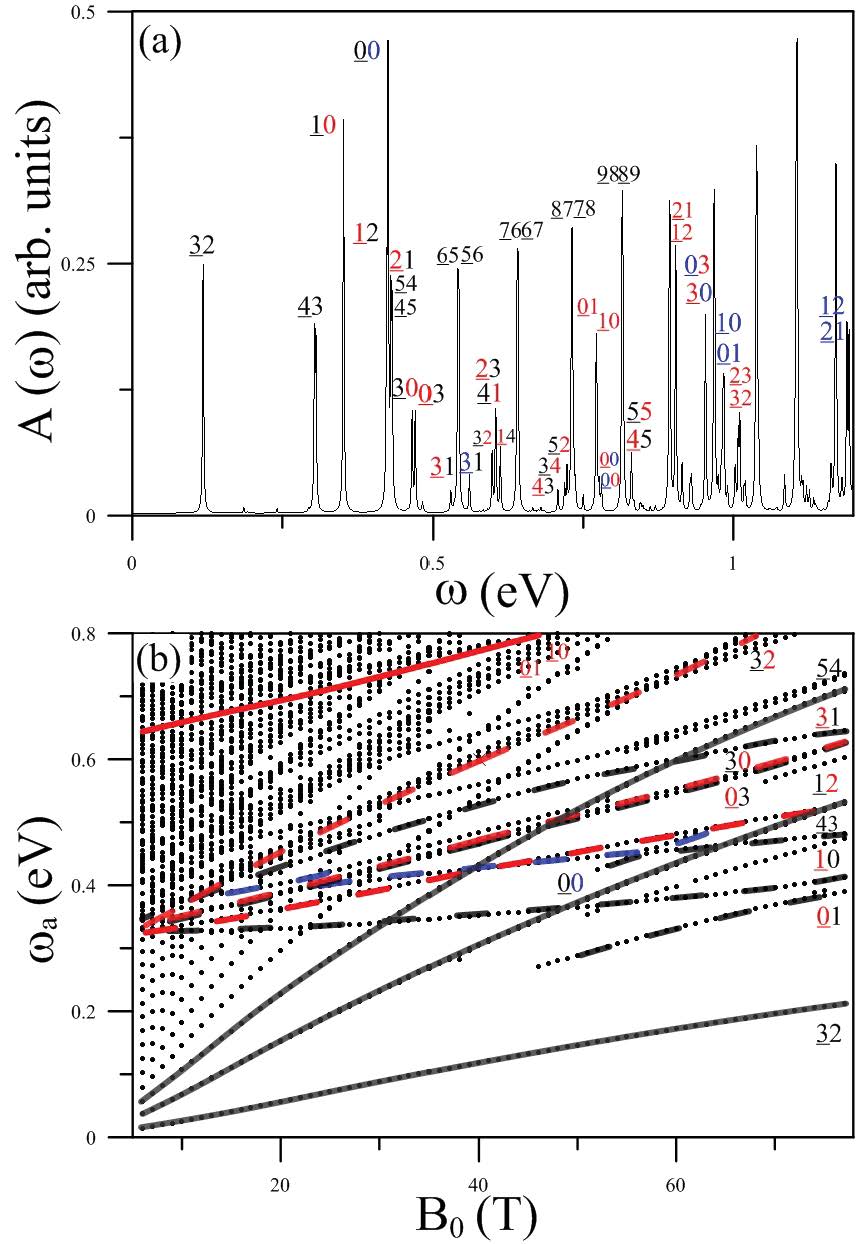}
\caption{(a) Magneto-absorption spectrum of ABC-stacked trilayer graphene under $B_0$ = 40 T. The black, red, and green color numbers, respectively, indicate the excitations from the first, second, and third LL groups. (b) $B_0$-dependent magneto-absorption frequencies. The solid and dashed curves indicate the excitation channels of the intragroup and intergroup categories, respectively.}
\label{fig:graph}
\end{figure}

\clearpage

\begin{figure}
\centering
\includegraphics[width=0.9\linewidth]{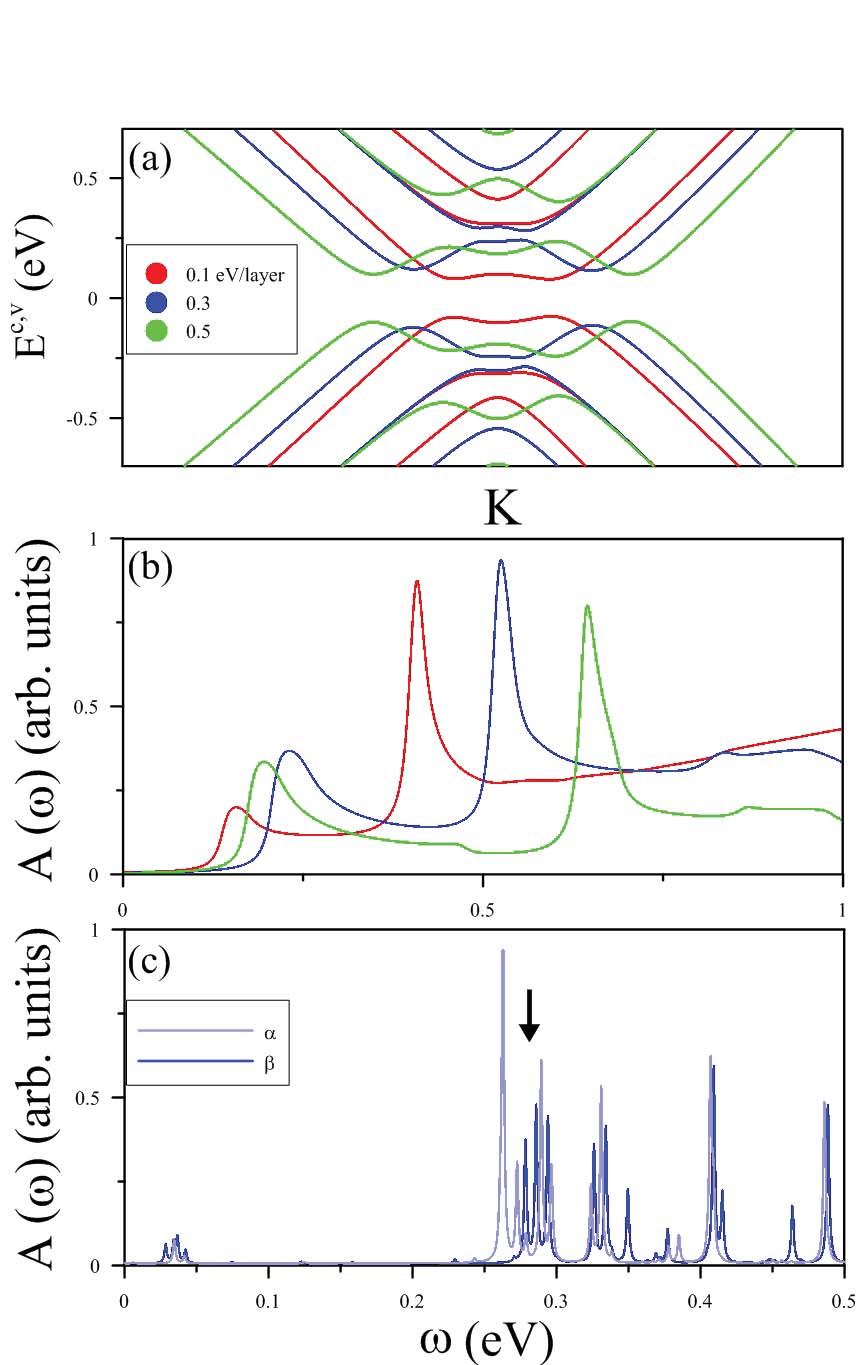}
\caption{The $E_0$-enriched (a) band structure, (b) absorption spectrum and (c) magneto-absorption spectrum of ABC-stacked trilayer graphene.}
\label{fig:graph}
\end{figure}

\begin{figure}
\centering
\includegraphics[width=0.9\linewidth]{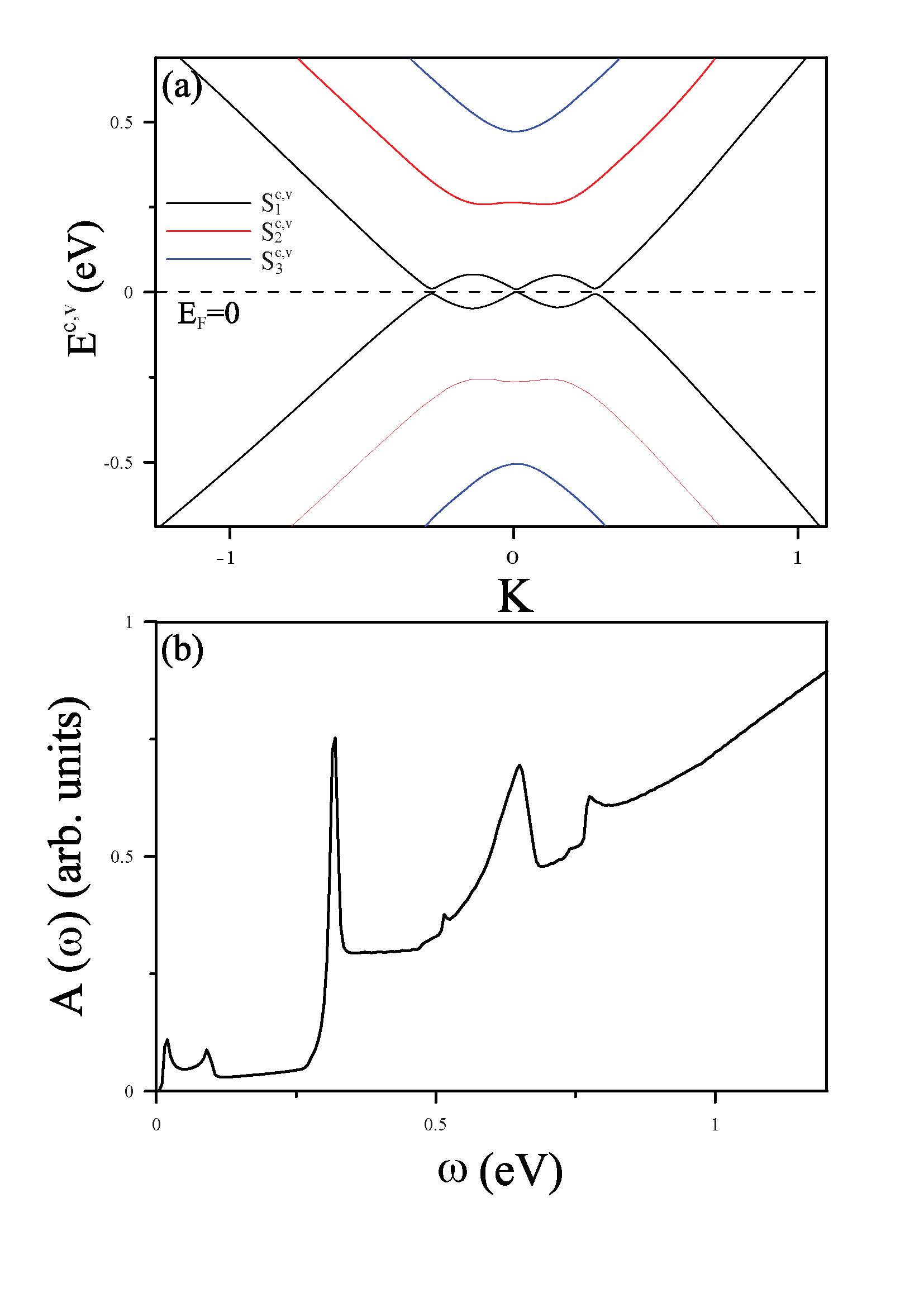}
\caption{(a) Zero-field band structure and (b) absorption spectrum of AAB-stacked trilayer graphene.}
\label{fig:graph}
\end{figure}

\begin{figure}
\centering
\includegraphics[width=0.9\linewidth]{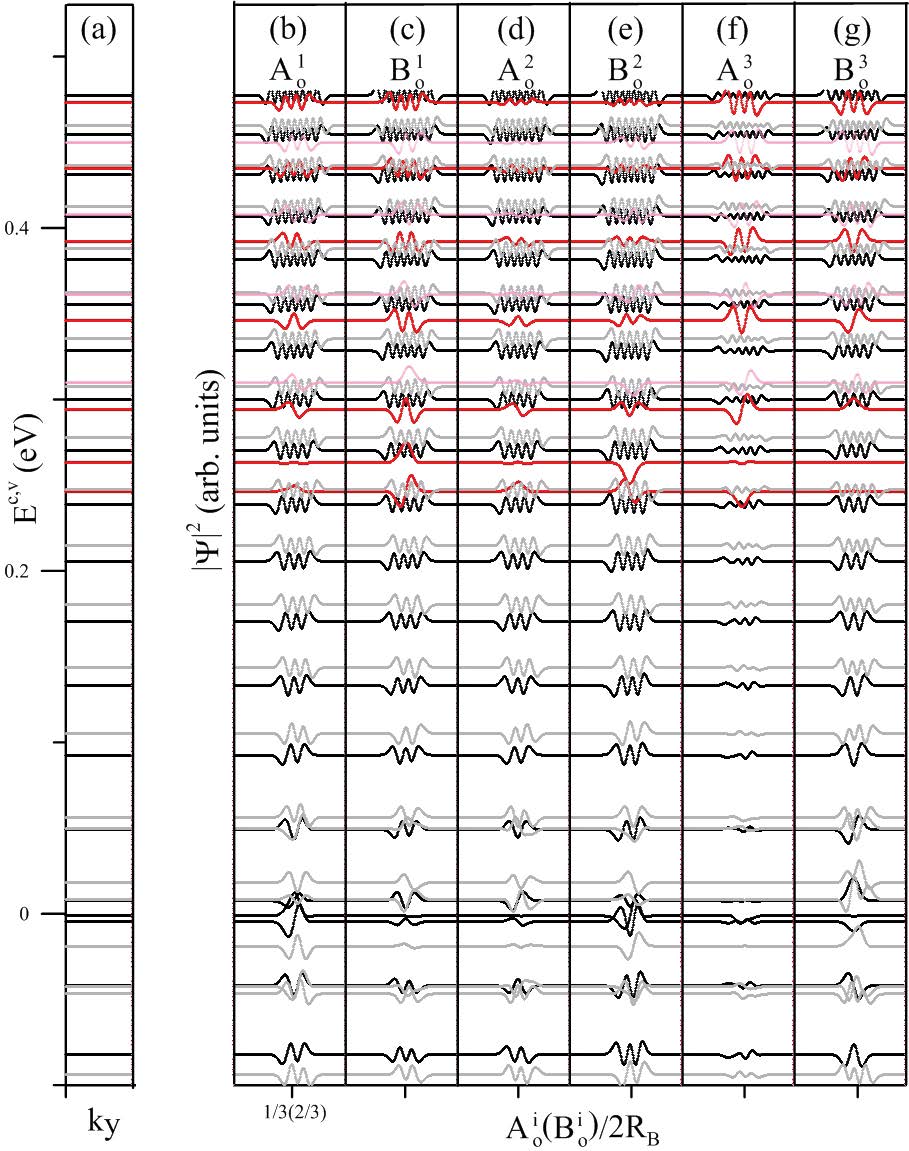}
\caption{The low-lying LL wavefunctions with six distinct sublattices localized at 2/6 and 4/6 centers of AAB-stacked trilayer graphene under $B_0$ = 40 T.}
\label{fig:graph}
\end{figure}

\begin{figure}
\centering
\includegraphics[width=0.9\linewidth]{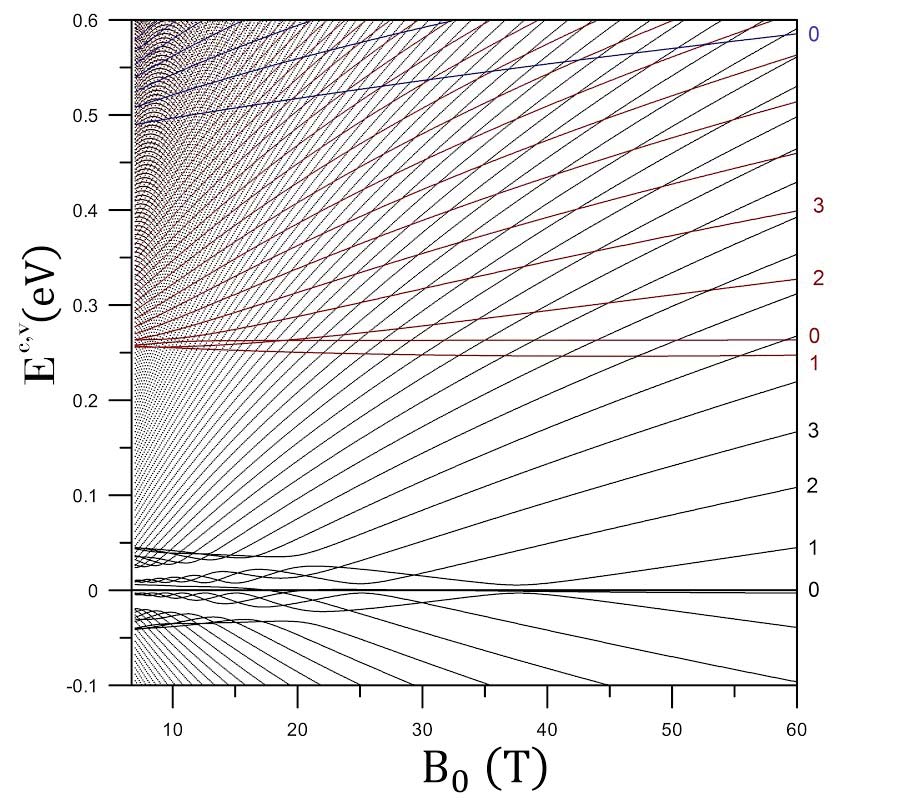}
\caption{Landau-level spectrum of AAB-stacked trilayer graphene.}
\label{fig:graph}
\end{figure}

\begin{figure}
\centering
\includegraphics[width=0.9\linewidth]{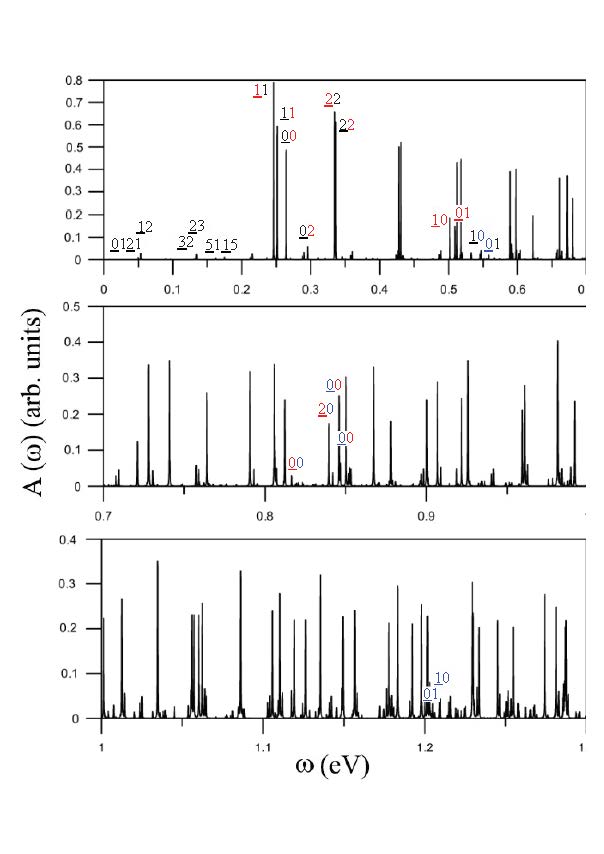}
\caption{The magneto-absorption spectrum of AAB-stacked trilayer graphene localized at 1/6 center with three intragroup and six intergroup categories of LL excitations under $B_0$ = 40 T. The black, red, and blue color numbers, respectively, indicate the excitations from the first, second, and third LL groups.}
\label{fig:graph}
\end{figure}

\begin{figure}
\centering
\includegraphics[width=0.9\linewidth]{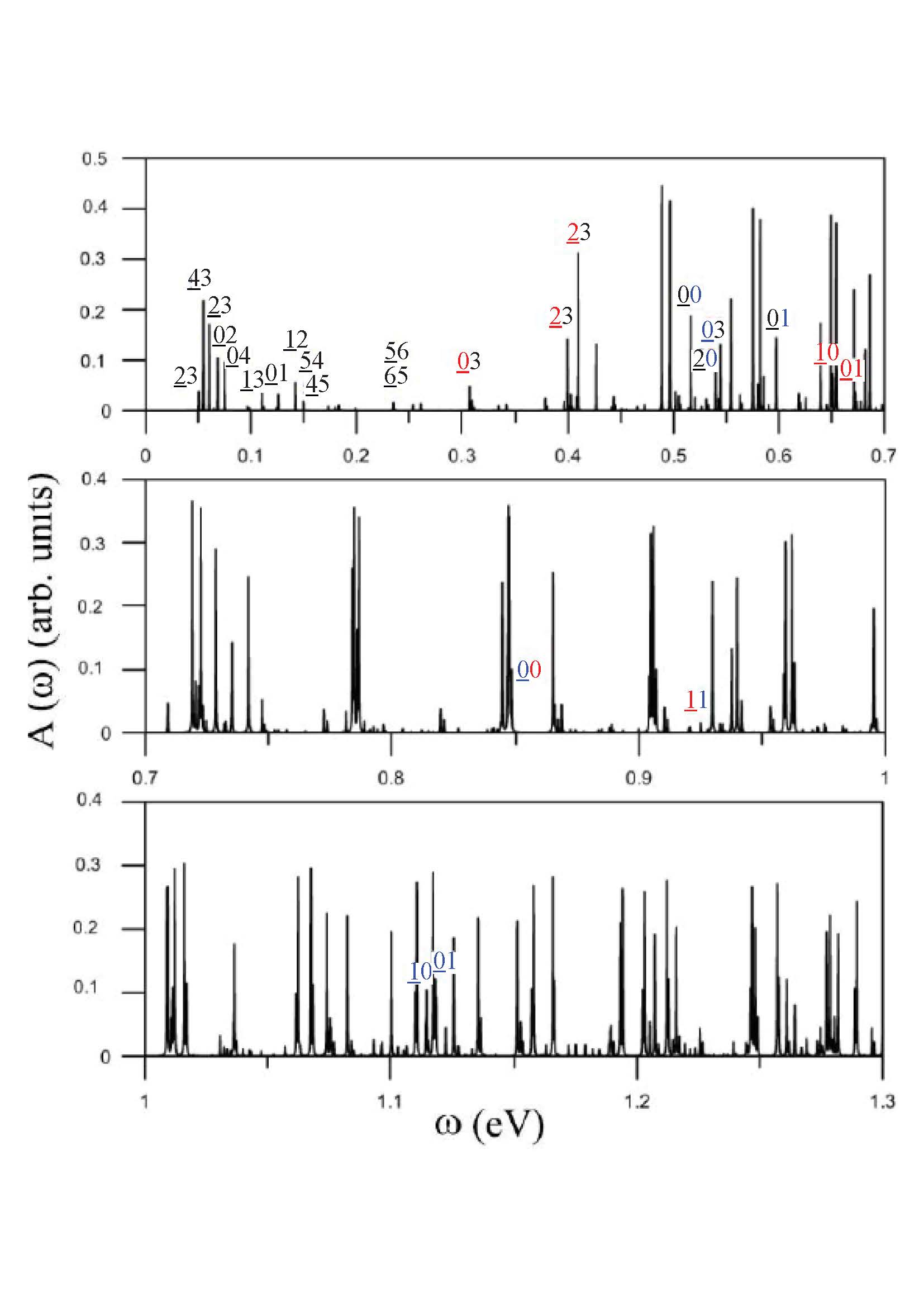}
\caption{The magneto-absorption spectrum of AAB-stacked trilayer graphene localized at 2/6 center with three intragroup and six intergroup categories of LL excitations under $B_0$ = 40 T. The black, red, and blue color numbers, respectively, indicate the excitations from the first, second, and third LL groups.}
\label{fig:graph}
\end{figure}

\begin{figure}
\centering
\includegraphics[width=0.9\linewidth]{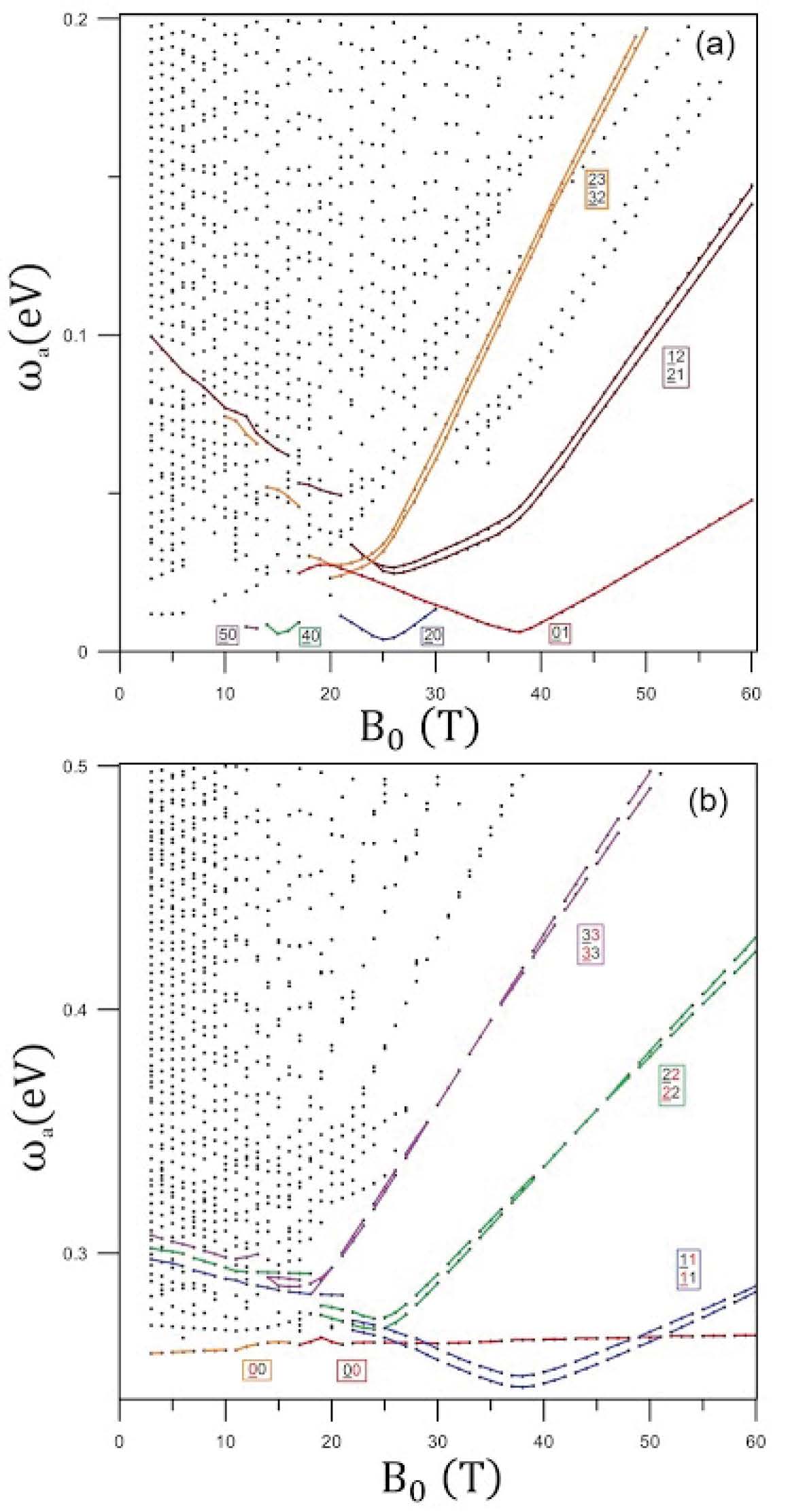}
\caption{The magneto-absorption spectrum of AAB-stacked trilayer graphene localized at 1/6 center with three intragroup and six intergroup categories of LL excitations under $B_0$ = 40 T. The black, red, and blue color numbers, respectively, indicate the excitations from the first, second, and third LL groups.}
\label{fig:graph}
\end{figure}

\begin{figure}
\centering
\includegraphics[width=0.9\linewidth]{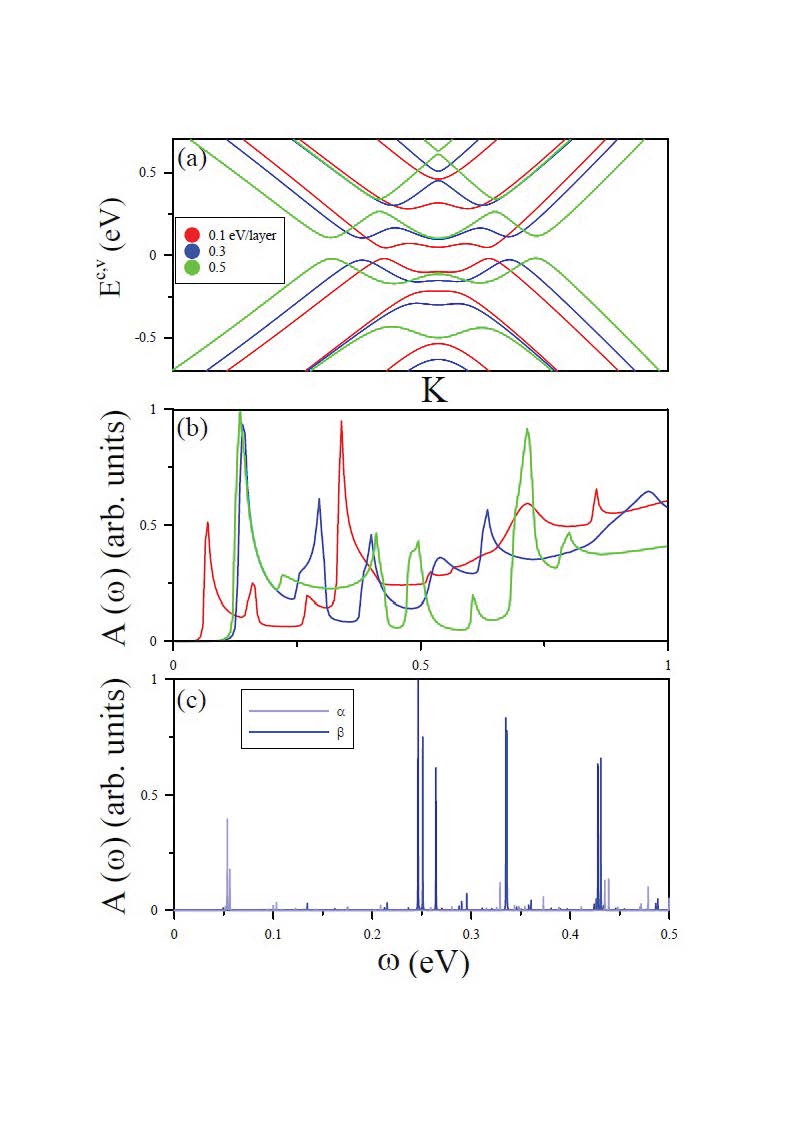}
\caption{The $E_0$-enriched (a) band structure, (b) absorption and (c) magneto-absorption spectra of AAB-stacked trilayer graphene.}
\label{fig:graph}
\end{figure}

\begin{figure}
\centering
\includegraphics[width=0.9\linewidth]{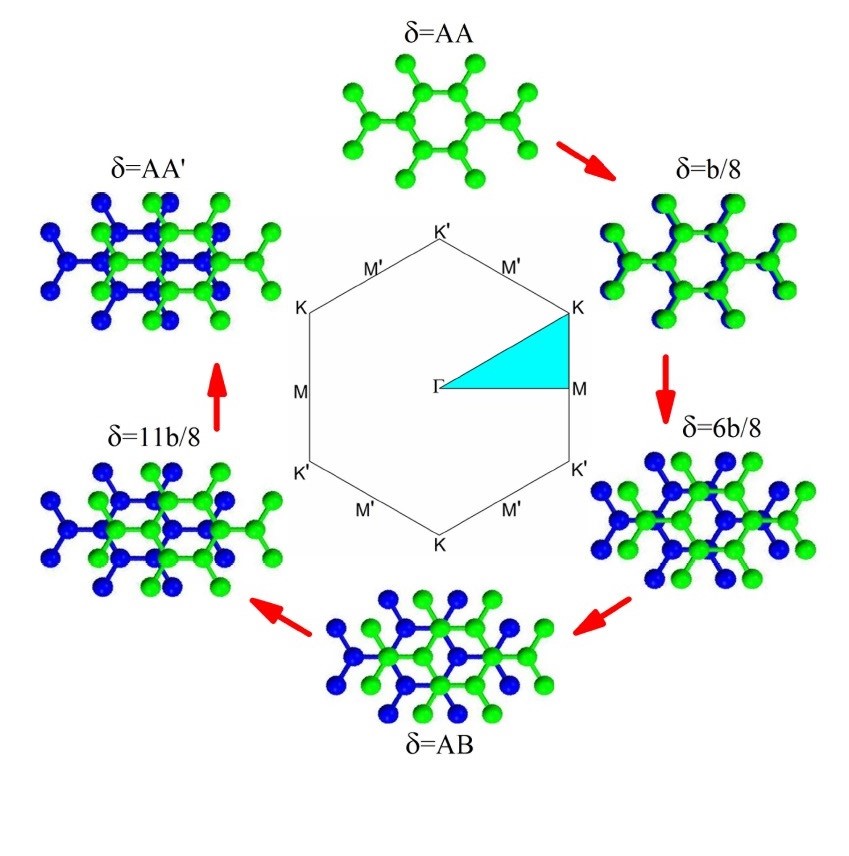}
\caption{The geometric structures of shift-dependent bilayer graphenes along the armchair
direction.}
\label{fig:graph}
\end{figure}

\begin{figure}
\centering
\includegraphics[width=0.9\linewidth]{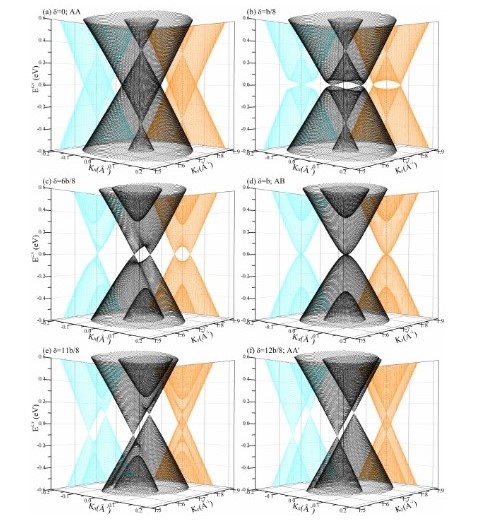}
\caption{The low-energy band structures near the K point for various stacking configurations: (a) $\delta$=0, (b) b/8, (c) 6b/8, (d) b, (e) 11b/ 8; (f) 12b/8.}
\label{fig:graph}
\end{figure}

\begin{figure}
\centering
\includegraphics[width=0.9\linewidth]{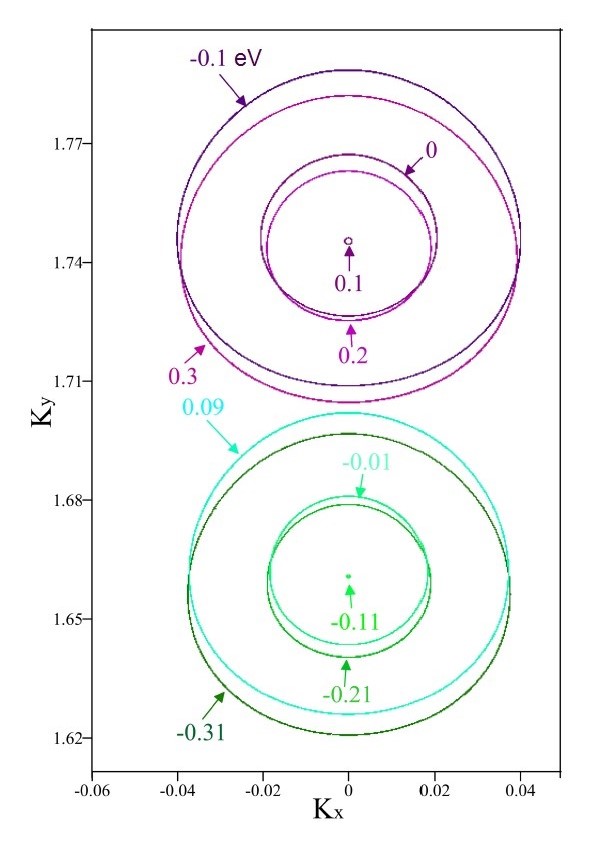}
\caption{The constant-energy contours of $\delta$=12b/8 related to the two tilted Dirac-cone structures.}
\label{fig:graph}
\end{figure}

\begin{figure}
\centering
\includegraphics[width=0.9\linewidth]{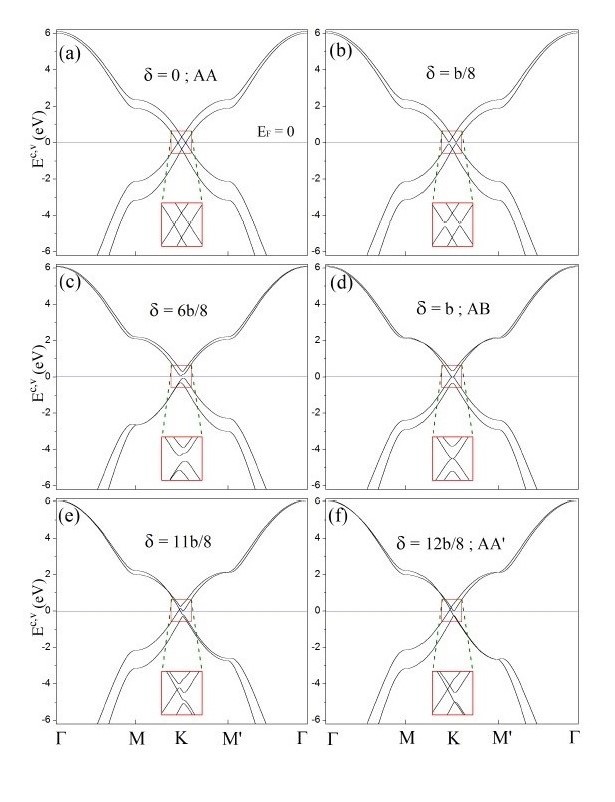}
\caption{Two-dimensional band structures of various stackings: (a) $\delta$=0, (b) b/8, (c)
6b/8, (d)b, (e) 11b/ 8; (f) 12b/8. Also shown in insets are low-lying ones.}
\label{fig:graph}
\end{figure}

\clearpage

\begin{figure}
\centering
\includegraphics[width=0.9\linewidth]{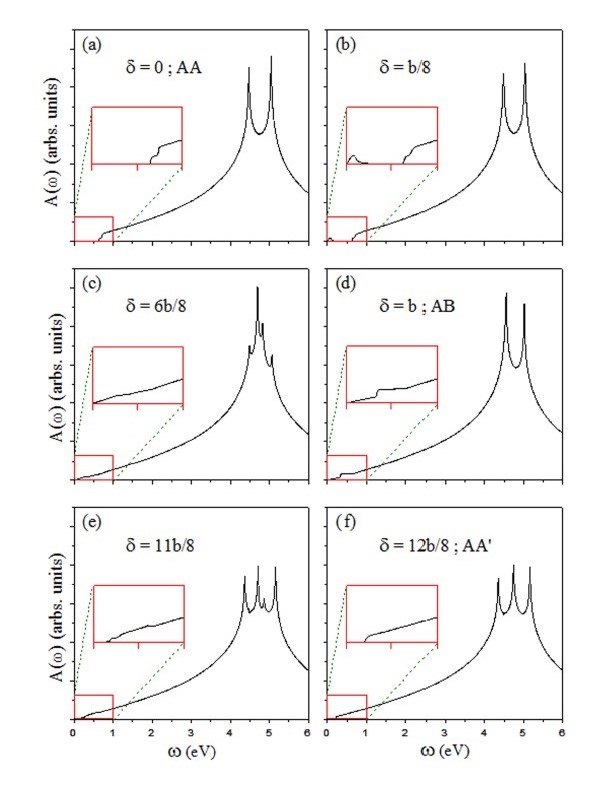}
\caption{Optical absorption spectra of (a) $\delta$=0, (b) b/8, (c) 6b/8, (d) b, (e) 11b/ 8;
(f) 12b/8. Insets show low-frequency ones.}
\label{fig:graph}
\end{figure}

\begin{figure}
\centering
\includegraphics[width=0.9\linewidth]{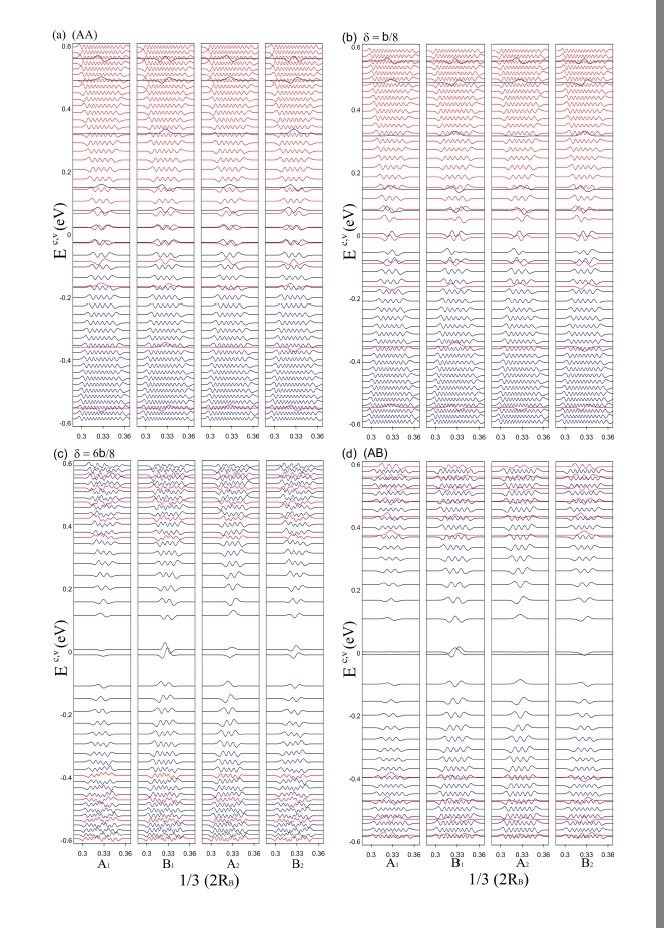}
\caption{The LL energies and wave functions at B$_{0}$=40 T for (a) $\delta$=0, (b) b/8, (c)
6b/8, (d) b, (e) 11b/ 8; (f) 12b/8.}
\label{fig:graph}
\end{figure}

\begin{figure}
\centering
\includegraphics[width=0.9\linewidth]{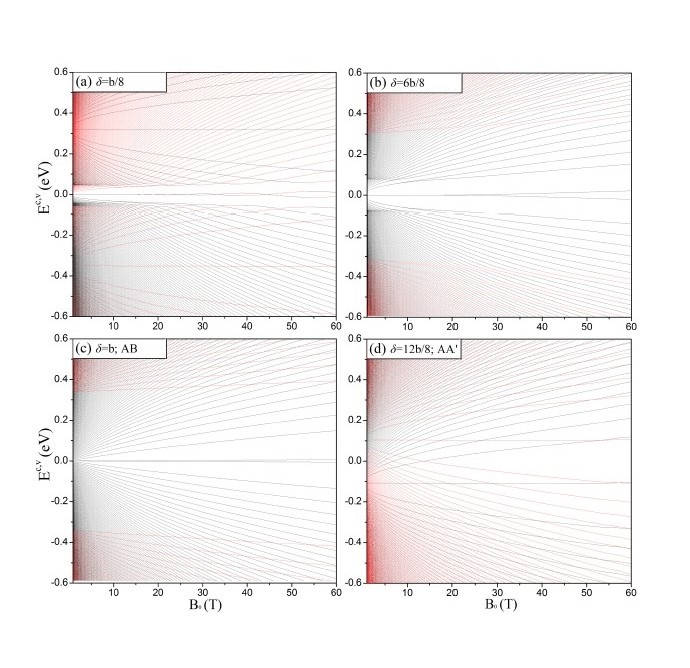}
\caption{The B$_{0}$-dependent LL energy spectra of (a) b/8, (b) 6b/8, (c) b; (d) 12b/8.}
\label{fig:graph}
\end{figure}

\begin{figure}
\centering
\includegraphics[width=0.9\linewidth]{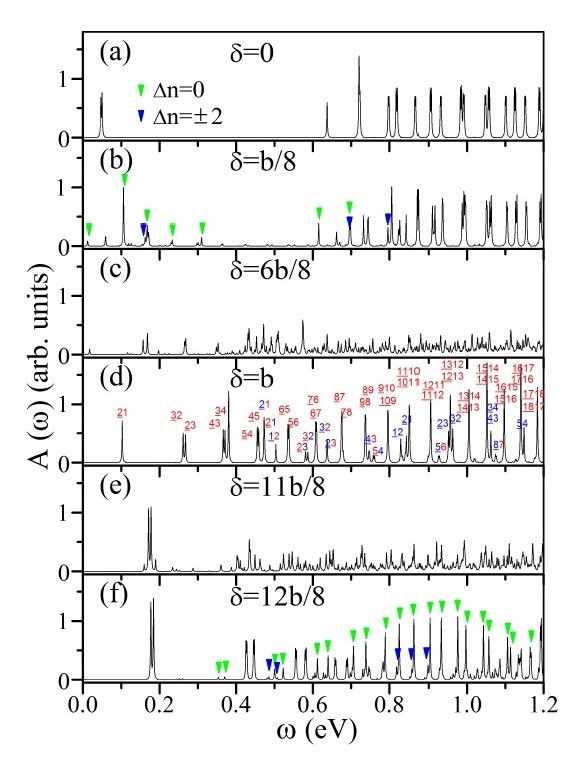}
\caption{Magneto-absorption spectra at B$_{0}$=40 T for (a) $\delta$=0, (b) b/8, (c) 6b/8,
(d) b, (e) 11b/ 8; (f) 12b/8.}
\label{fig:graph}
\end{figure}

\begin{figure}
\centering
\includegraphics[width=0.9\linewidth]{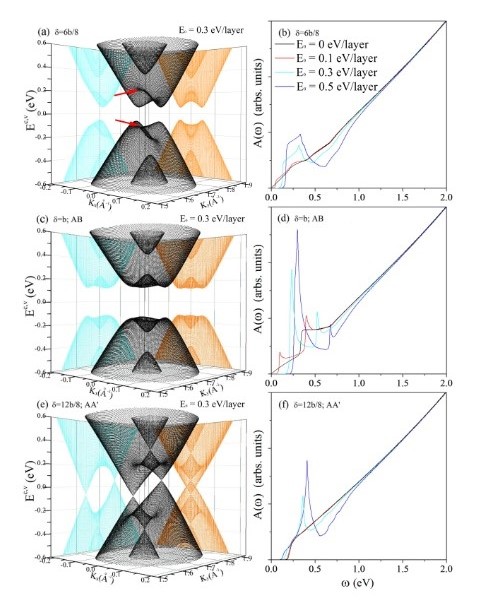}
\caption{Low-energy electronic structures and optical spectra at distinct electric fields
for bilayer graphenes with (a)-(b) $\delta$=6b/8, (c)-(d) b; (e)-(f) 12b/8.}
\label{fig:graph}
\end{figure}

\begin{figure}
\centering
\includegraphics[width=0.9\linewidth]{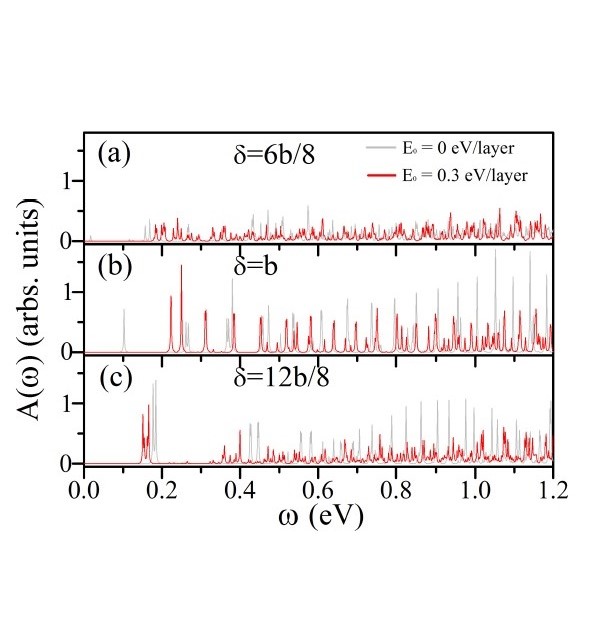}
\caption{Absorption spectra under a composite field for bilayer graphenes with (a)
$\delta$=6b/8, (b) b; (c) 12b/8.}
\label{fig:graph}
\end{figure}

\end{document}